\documentclass[twocolumn]{aastex61}

\newcommand{\titlerunning}[1]{\shorttitle{#1}}
\newcommand{\authorrunning}[1]{\shortauthors{#1}}

\newcommand*\inst[1]{\unskip\hbox{\@textsuperscript{\normalfont$#1$}}}

\newcommand*\institute[1]{
  \begingroup
    \let\and\relax
    \renewcommand*\inst[1]{}%
    \renewcommand*\thanks[1]{}%
    \renewcommand*\email[1]{}%
  \endgroup
  \newcommand{\institutions}{#1}
}%

\renewcommand{\ion}[2]{\textup{#1\,\textsc{\lowercase{#2}}}}

\let\oldarcsec\arcsec
\let\oldarcmin\arcmin
\renewcommand{\arcsec}{\oldarcsec\xspace}
\renewcommand{\arcmin}{\oldarcmin\xspace}

\usepackage{latexsym}		
\usepackage{graphicx}		
\usepackage{rotating}		
\usepackage{natbib}  
\usepackage{savesym}
\usepackage{amssymb}
\usepackage{morefloats}
\savesymbol{doublespace}
\usepackage{xspace}
\usepackage{color}
\usepackage{mdframed}
\usepackage{url}
\usepackage{subfigure}
\usepackage{grffile}
\usepackage{import}
\usepackage[utf8]{inputenc}
\usepackage{booktabs}
\usepackage{fancyhdr}

\usepackage[hang,flushmargin]{footmisc}
\usepackage{ifpdf}

\newcommand{\msun}{\ensuremath{M_{\odot}}\xspace}			

\newcommand{\lsun}{\ensuremath{L_{\odot}}\xspace}			
\newcommand{\rsun}{\ensuremath{R_{\odot}}\xspace}			
\newcommand{\hh}{\ensuremath{\textrm{H}_{2}}\xspace}			

\newcommand{\formaldehyde}{\ensuremath{\textrm{H}_2\textrm{CO}}\xspace}
\newcommand{\formamide}{\ensuremath{\textrm{NH}_2\textrm{CHO}}\xspace}

\newcommand{\methanol}{\ensuremath{\textrm{CH}_3\textrm{OH}}\xspace}

\newcommand{\fourtwotwo}{\ensuremath{4_{2,2}-3_{1,2}}\xspace} 

\newcommand{\ketene}{\ensuremath{\textrm{H}_{2}\textrm{CCO}}\xspace}
\newcommand{\ethylcyanide}{\ensuremath{\textrm{CH}_3\textrm{CH}_2\textrm{CN}}\xspace}
\newcommand{\cyanoacetylene}{\ensuremath{\textrm{HC}_{3}\textrm{N}}\xspace}
\newcommand{\methylformate}{\ensuremath{\textrm{CH}_{3}\textrm{OCHO}}\xspace}
\newcommand{\dimethylether}{\ensuremath{\textrm{CH}_{3}\textrm{OCH}_{3}}\xspace}
\newcommand{\gaucheethanol}{\ensuremath{\textrm{g-CH}_3\textrm{CH}_2\textrm{OH}}\xspace}
\newcommand{\acetone}{\ensuremath{\left[\textrm{CH}_{3}\right]_2\textrm{CO}}\xspace}
\newcommand{\methyleneamidogen}{\ensuremath{\textrm{H}_{2}\textrm{CN}}\xspace}

\newcommand{\uchii}{\ion{UCH}{ii}\xspace}

\newcommand{\hchii}{\ion{HCH}{ii}\xspace}

\newcommand{\hii}{\ion{H}{ii}\xspace}

\newcommand{\kms}{\textrm{km~s}\ensuremath{^{-1}}\xspace}	

\newcommand{\pers}{\ensuremath{\mathrm{s}^{-1}}\xspace}

\newcommand{\percc}{\ensuremath{\textrm{cm}^{-3}}\xspace}

\newcommand{\persc}{\ensuremath{\textrm{cm}^{-2}}\xspace}

\newcommand{\peryr}{\ensuremath{\textrm{yr}^{-1}}\xspace}
\newcommand{\um}{\ensuremath{\mu \textrm{m}}\xspace}    

\newcommand{\ammonia}{NH\ensuremath{_3}\xspace}
\newcommand{\twelveco}{\ensuremath{^{12}\textrm{CO}}\xspace}

\newcommand{\ceighteeno}{\ensuremath{\textrm{C}^{18}\textrm{O}}\xspace}
\def\ee#1{\ensuremath{\times10^{#1}}}

\newcommand{\perbeam}{\ensuremath{\textrm{beam}^{-1}}\xspace}

\def\eqref#1{Equation \ref{#1}}

\def\Figure#1#2#3#4#5{
\begin{figure*}[!htp]
\includegraphics[scale=#4,width=#5]{#1}
\caption{#2}
\label{#3}
\end{figure*}
}

\def\FigureOneCol#1#2#3#4#5{
\begin{figure}[!htp]
\includegraphics[scale=#4,width=#5]{#1}
\caption{#2}
\label{#3}
\end{figure}
}

\def\FigureTwo#1#2#3#4#5#6{
\begin{figure*}[!htp]
\subfigure[]{ \includegraphics[scale=#5,width=#6]{#1} }
\subfigure[]{ \includegraphics[scale=#5,width=#6]{#2} }
\caption{#3}
\label{#4}
\end{figure*}
}

\newenvironment{rotatepage}%
{}{}

\def\FigureFour#1#2#3#4#5#6{
\begin{figure*}[!htp]
\subfigure[]{ \includegraphics[width=0.5\textwidth]{#1} }
\subfigure[]{ \includegraphics[width=0.5\textwidth]{#2} }
\subfigure[]{ \includegraphics[width=0.5\textwidth]{#3} }
\subfigure[]{ \includegraphics[width=0.5\textwidth]{#4} }
\caption{#5}
\label{#6}
\end{figure*}
}

\begin{document}
\title{Thermal Feedback in the high-mass star and cluster forming region W51}
\titlerunning{Thermal Feedback in W51}
\authorrunning{Ginsburg et al}
\newcommand{\nraojansky}{\affiliation{\it{Jansky fellow of the National Radio Astronomy Observatory, Socorro, NM 87801 USA }}}
\newcommand{\nrao}{\affiliation{\it{National Radio Astronomy Observatory, Socorro, NM 87801 USA }}}
\newcommand{\eso}{ \affiliation{\it{ European Southern Observatory, Karl-Schwarzschild-Stra{\ss}e 2, D-85748 Garching bei München, Germany } } }
\newcommand{\radboud}{\affiliation{\it{Department of Astrophysics/IMAPP, Radboud University Nijmegen, PO Box 9010, 6500 GL Nijmegen, the Netherlands}}}
\newcommand{\allegro}{\affiliation{\it{ALLEGRO/Leiden Observatory, Leiden University, PO Box 9513, 2300 RA Leiden, the Netherlands}}}
\newcommand{\zah}{\affiliation{\it{Astronomisches Rechen-Institut, Zentrum f{\"u}r Astronomie der Universit{\"a}t Heidelberg, M{\"o}nchhofstra{\ss}e 12-14, 69120 Heidelberg, Germany}}}
\newcommand{\casa}{\affiliation{\it{CASA, University of Colorado, 389-UCB, Boulder, CO 80309}} }
\newcommand{\jodrell}{\affiliation{\it{Jodrell Bank Centre for Astrophysics, School of Physics and Astronomy, University of Manchester, Oxford Road, Manchester M13 9PL, UK}}}
\newcommand{\morelia}{\affiliation{\it{Instituto de Radioastronom{\'i}a y Astrof{\'i}sica, UNAM, A.P. 3-72, Xangari, Morelia, 58089, Mexico}}}
\newcommand{\sjsu}{\affiliation{\it{{San Jose State University, One Washington Square, San Jose, CA 95192}}}}
\newcommand{\herts}{\affiliation{\it{Centre for Astrophysics Research, University of Hertfordshire, College Lane, Hatfield, AL10 9AB, UK}}}
\newcommand{\uofa}{\affiliation{\it{Dept. of Physics, University of Alberta, Edmonton, Alberta, Canada}}}
\newcommand{\arcetri}{\affiliation{\it{INAF-Osservatorio Astrofisico di Arcetri, Largo E. Fermi 5, I-50125, Florence, Italy } } }
\newcommand{\exclus}{\affiliation{\it{Excellence Cluster Universe, Boltzman str. 2, D-85748 Garching bei M\"unchen, Germany } }}
\newcommand{\ljmu}{\affiliation{\it{Astrophysics Research Institute, Liverpool John Moores University, 146 Brownlow Hill, Liverpool L3 5RF, UK }}}

\author[0000-0001-6431-9633]{Adam Ginsburg}
\nraojansky
\eso

\author{Ciriaco Goddi}
\radboud
\allegro

\author{J.~M.~Diederik Kruijssen}
\zah

\author{John Bally}
\casa

\author{Rowan Smith}
\jodrell

\author{Roberto Galv{\'a}n-Madrid}
\morelia

\author{Elisabeth A.C. Mills }
\sjsu

\author{Ke Wang }
\eso

\author{James E. Dale}
\herts

\author{Jeremy Darling}
\casa

\author{Erik Rosolowsky }
\uofa

\author{Robert Loughnane}
\morelia

\author{Leonardo Testi }
\eso
\arcetri
\exclus

\author{Nate Bastian }
\ljmu

\correspondingauthor{Adam Ginsburg}
\email{aginsbur@nrao.edu; adam.g.ginsburg@gmail.com}


\begin{abstract}
    High-mass stars have generally been assumed to accrete most of their
    mass while already contracted onto the main sequence, but this
    hypothesis has not been observationally tested.
    We present ALMA observations of a $3\times1.5$ pc area in the W51
    high-mass star-forming complex.  We identify dust continuum sources and
    measure the gas and dust temperature through both rotational diagram
    modeling of \methanol and brightness-temperature-based limits.  The
    observed region contains three high-mass YSOs that appear to be at the
    earliest stages of their formation, with no signs of ionizing radiation
    from their central sources.
    The data reveal high gas and dust
    temperatures ($T > 100$ K) extending out to about 5000 AU from each of
    these sources.  There are no clear signs of disks or rotating
    structures down to our 1000 AU resolution.
    The extended warm gas provides evidence that, during the process of
    forming, these high-mass stars heat a large volume and correspondingly
    large mass of gas in their surroundings, inhibiting fragmentation and
    therefore keeping a large reservoir available to feed from.  By contrast,
    the more mature massive stars that illuminate compact \hii regions have
    little effect on their surrounding dense gas, suggesting that these main
    sequence stars have completed most or all of their accretion.  
    The high luminosity of the massive protostars ($L>10^4$ \lsun), combined
    with a lack of centimeter continuum emission from these sources, implies
    that they are not on the main sequence while they accrete the majority of
    their mass; instead, they may be bloated and cool.


    


\end{abstract}

\section{Introduction}
High-mass stars are the drivers of galaxy evolution, cycling enriched materials
into the interstellar medium (ISM) and illuminating it.  During their formation
process, however, these stars are nearly undetectable because of their rarity
and their opaque surroundings.  We therefore know relatively little about how
massive stars acquire their mass and what their immediate surroundings look
like at this early time.  We expect, though, that the physical conditions
should be changing rapidly.

The stellar initial mass function (IMF) appears to be a universal distribution
\citep{Bastian2010a}.  However, massive  O-stars (with $M>50 \msun$)
almost always form in a clustered fashion \citep[in proto-clusters or
proto-associations;][]{de-Wit2004a,de-Wit2005a,Parker2007a}. 
Their presence, and the strong feedback they produce, may directly influence
how the IMF around them is formed.  If feedback from these stars is relevant
while most of the  mass surrounding them is still in gas (not yet in stars),
the mass function in such clusters cannot be determined by ISM properties
(initial conditions) alone. 

Models of high-mass star formation universally have difficulty collapsing enough
material to a stellar radius to form very massive stars.  Generally, these models
produce a high-mass star with enough luminosity to halt further
\emph{spherical} accretion at a very early stage, with $M_* \sim 10-20\msun$.
Radiation pressure provides a fundamental limit on how much mass can be
accreted \citep{Wolfire1987a,Osorio1999a}, but geometric effects can circumvent
this limit and allow further accretion
\citep{Yorke2002a,Krumholz2005b,Krumholz2009a,Krumholz2009b,Kuiper2012a,Kuiper2013c,Rosen2016a}.
Additionally, fragmentation-induced starvation can limit the amount of mass
available to the most massive star, instead breaking up massive cores into many
lower-mass fragments \citep{Peters2010a,Girichidis2012b}, though other simulations suggest
that feedback should suppress this fragmentation
\citep{Myers2013a,Krumholz2016a}.  The simulations used to demonstrate that disk
accretion can form massive stars still have limited physics and can only
produce stars up to $M\sim80$ \msun even in the current
best 3D cases  \citep{Kuiper2015a,Kuiper2016a}.  The question of how massive
stars acquire their mass, and especially whether they ever form Keplerian
disks, remains open \citep{Beltran2016b}.

Nature is clearly capable of producing massive stars larger than those produced
in simulations.  Within the LMC, stars up to $M\sim300$ \msun have been
spectroscopically identified \citep{Crowther2016a}.  Within our own Galaxy,
very massive stars have been found in compact, high-mass clusters such as NGC
3603 and the Arches \citep{Crowther2010a}.  While it is difficult to identify
and characterize the most massive stars in our own galaxy because the UV
features best capable of establishing their spectral types are 
extinguished,
it is still possible to find examples of very massive stars close to their
birth environments using infrared lines.  \citet{Barbosa2008a} identified an O3
and an O4 star ($M\gtrsim50$ \msun) within the W51 IRS2 region, demonstrating
that this region has at some time formed stars on the high
end tail of the IMF.  It remains to be seen whether W51 will form any very
massive stars ($M>100$ \msun), but it is  an appropriate environment to
investigate the process.

The W51 cloud contains two protocluster regions, IRS2 and e1/e2, which each
contain $M\gtrsim10^4$ \msun of gas and have large far-infrared luminosities
that
indicate the presence of embedded, recently-formed or forming massive stars
\citep[Figure
\ref{fig:overview}][]{Harvey1986a,Sievers1991a,Ginsburg2012a,Ginsburg2016b}.
Previous
millimeter and centimeter observations have revealed the gas reservoir
that is forming new stars and, because of the high masses of the individual
cores detected, indicated that these new stars are likely to be massive
\citep{Zhang1997a,Eisner2002a,Zapata2009a,Tang2009a,
Zapata2010a,Shi2010b,Shi2010a,Koch2010a,Koch2012a,Koch2012b,Tang2013b,Goddi2016a}.  The
W51 protoclusters, while distant \citep[5.4 kpc;][]{Sato2010a}, therefore
provide a powerful laboratory for studying high-mass star formation in an
environment where feedback from massive stars is already evident, but 
formation is still ongoing.

The protocluster region within W51 exhibits many signs of strong feedback.  In
particular, there are many giant \hii regions detected in the infrared through
radio \citep{Mehringer1994a,Ginsburg2015a}.  These \hii region bubbles exist
on many scales, and the driving populations of OB stars have been identified
\citep{Kumar2004a,Ginsburg2016a}.  While the larger W51 cloud, which stretches
about 100 pc along Galactic longitude, shows some signs of interaction with a
supernova remnant \citep{Brogan2013a,Ginsburg2015a}, there is as yet no
sign that supernovae have occurred within the W51 IRS2 or e1/e2 protocluster
regions.  They are in the relatively short stage after  high mass stars
have formed but before the gas has been exhausted or expelled.

This combination of feedback and ongoing formation is essential for testing
components of high mass star formation theory that are relatively inaccessible
to simulations.  While simulations have verified the conclusion that
early-stage accretion heating can control the mass scale within low-mass star
forming regions \citep{Krumholz2007c, Offner2011b, Bate2012a,Bate2014b,
Guszejnov2015c, Guszejnov2016a,Krumholz2016a}, there
have been neither theoretical nor observational tests of this model for
high-mass stars.   For example, \citet{Krumholz2006a} suggests that accretion
heating during the formation of high-mass stars can heat massive cores to
$\gtrsim100$ K and therefore suppress fragmentation into smaller stars, which
would be expected for cold cores, though these models have $T>100$ K out to
only $R\lesssim100$ AU.  

\Figure{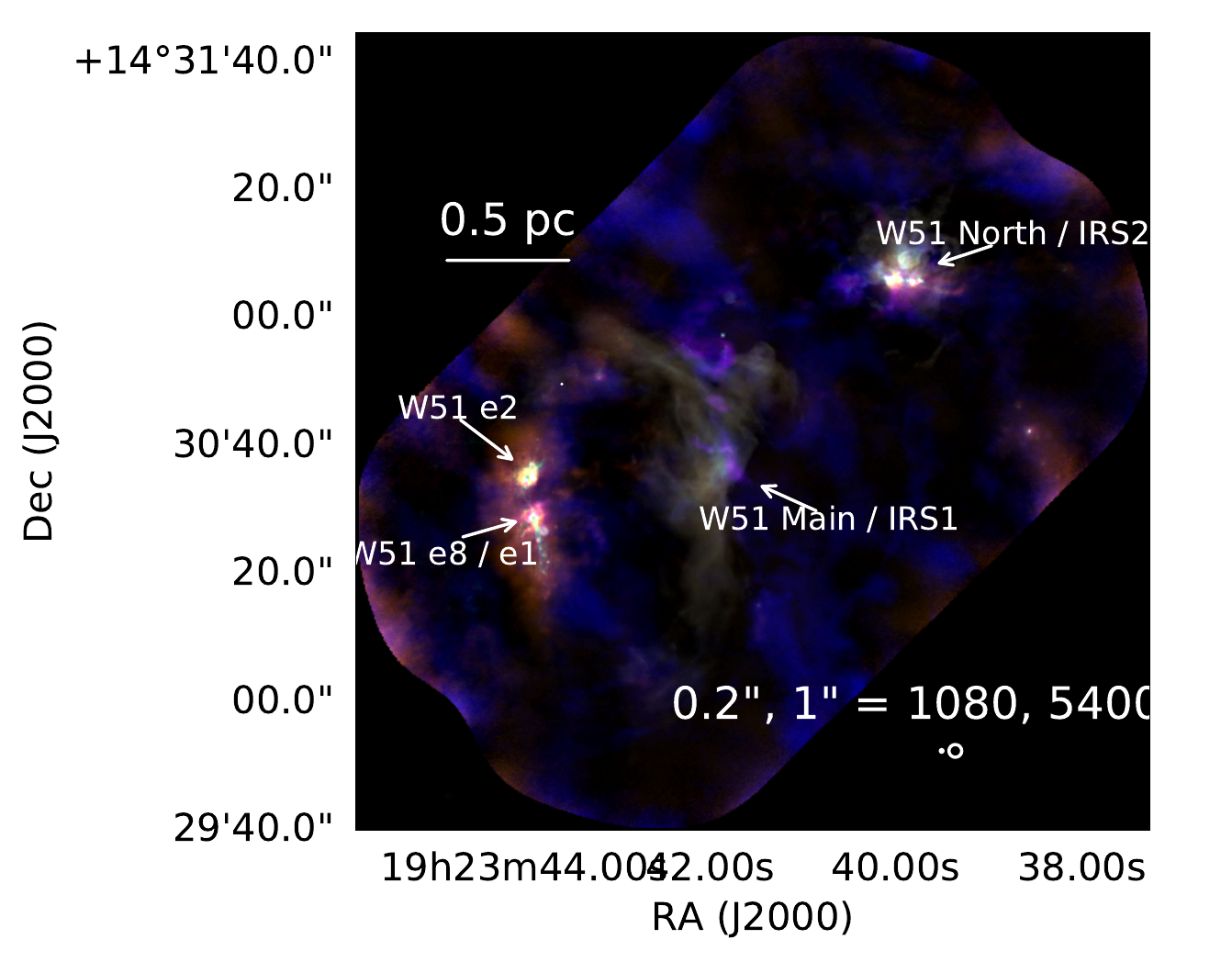}
{An overview of the W51A region as seen by ALMA and the VLA.  The main regions
discussed in this paper are labeled.  W51 e8 is a mm dust source, while W51 e1
is the neighboring \hii region.  Similarly, W51 IRS2 is the \hii region, and
W51 North is the brightest mm source in that area.  The colors are a composite
of millimeter emission lines: C$^{18}$O 2-1 in blue, \methanol \fourtwotwo in
orange, and HC$_3$N 24-23 in
purple.  The 1.3mm continuum is shown in green.  The white hazy emission
shows VLA Ku-band free-free continuum emission
\citep{Ginsburg2016b}.}
{fig:overview}{1}{18cm}

We present an observational study of the high-mass star-forming region W51,
showing that the actively forming massive stars significantly affect their
surrounding dense gas, while stars that are not accreting have little effect.
 In Section
\ref{sec:observations}, we describe the observations and data reduction
process.  Section \ref{sec:results} describes the analysis:
We discuss source identification (\S \ref{sec:sourceid}),
the mass and flux recovered on different
spatial scales (\S \ref{sec:massbudget}),
the observed chemical distribution (\S \ref{sec:chemistrymaps}), 
temperatures inferred from \methanol lines (\S \ref{sec:methanol}),
the radial mass profiles (\S \ref{sec:radialmass}), the gas kinematics
(\S \ref{sec:kinematics}), nondetection of disks (\S \ref{sec:disks}),
the signatures of ionizing and non-ionizing feedback around MYSOs
(\S \ref{sec:nonionizingradiation}), and finally a brief note about outflows
(\S \ref{sec:mainoutflows}).
Section \ref{sec:discussion} discusses scales and types of feedback (\S
\ref{sec:feedbackscales}), outflows (\S \ref{sec:outflowdiscussion}),
the implications of these outflows for accretion (\S \ref{sec:accretionandoutflows}),
and fragmentation (\S \ref{sec:fragmentation}).  
Section \ref{sec:cooperative} discusses implications
of the fragmentation analysis and the existence of these cores on star
formation theory.  Section \ref{sec:faintsrcs_discussion} discusses the
low-mass cores and protostars.
We conclude in Section \ref{sec:conclusion}.
Additional interesting features in the W51 data not directly relevant
to our main topic, the formation of high-mass stars,
are discussed in the Appendices, including
some remarkable outflows 
(\S \ref{sec:outflows}),  a characterization of the lower-mass sources
(\S \ref{sec:contsrcs}),
and an interesting bubble (\S \ref{sec:e5bubble}).

\section{Observations}
\label{sec:observations}
As part of ALMA Cycle 2 program 2013.1.00308.S, we observed a
$\sim2\arcmin\times1\arcmin$ region centered between W51 IRS2 and W51 e1/e2
with a 37-pointing mosaic.  Two configurations of the 12m array were used,
achieving a resolution of 0.2\arcsec.  Additionally, a 12-pointing mosaic was
performed using the 7m array, theoretically probing scales up to
$\sim28$\arcsec.  The full UV coverage included baselines over
the range $\sim12$ to $\sim1500$ m.
The spectral windows (SPWs) covered are listed in Table \ref{tab:spw},
and the lines they cover described in Section \ref{sec:obslines}.

\begin{table*}[htp]
\caption{Spectral Setup}
\begin{tabular}{lllll}
\label{tab:spw}
SpwID & Minimum Frequency & Maximum Frequency & Channel Width [$\nu$] & Channel Width [$v$] \\
 & $\mathrm{GHz}$ & $\mathrm{GHz}$ & $\mathrm{kHz}$ & $\mathrm{km\,s^{-1}}$ \\
\hline
0 & 218.11930228 & 218.619301 & -122.07 & 0.17 \\
1 & 218.36288652 & 220.355073 & -488.281 & 0.67 \\
2 & 230.376575 & 232.36876148 & 488.281 & 0.64 \\
3 & 232.981075 & 234.97326148 & 488.281 & 0.63 \\
\hline
\end{tabular}
\end{table*}

\subsection{Data Reduction}
Data reduction was performed using CASA 4.5.2-REL (r36115), including
reprocessing of datasets that were delivered with earlier versions.  The
QA2-produced visibility data products were combined using the standard inverse
variance weighting.  Two sets of images were produced for different aspects of
the analysis, one including the 7m array data and one including only 12m data.
Except where otherwise noted, the 12m-only data were used in order to focus on
the compact structures.  The conversion from
flux density to brightness temperature is $T_B \approx 220
\mathrm{\,K}/(\mathrm{Jy}\,\perbeam)$ for a 0.33\arcsec beam (most of the spectral
line data) or $T_B=590\mathrm{\,K}/(\mathrm{Jy}\,\perbeam)$ for a 0.2\arcsec
beam (for the higher-resolution images of the continuum) assuming a central
frequency 226.6 GHz (see below).

Full details of the data reduction, including all scripts used, can be found on
the project's github
repository\footnote{\url{https://github.com/adamginsburg/W51_ALMA_2013.1.00308.S}}.

\subsubsection{Continuum}
A continuum image combining all 4 spectral windows was produced using
\texttt{tclean}.  We identified line-rich channels from a spectrum of source e8
and flagged them out prior to imaging\footnote{The velocity
range of e8, e2, and North is similar enough that a common range was acceptable
for this process.  Note also that, while the sources are line-rich,
failure to flag out the data results in a $<10\%$ error in the continuum
estimates \citep[see][showing that even the richest sources in the Galaxy have
$<40\%$ line contribution.]{Sanchez-Monge2017a} }.  We then phase
self-calibrated the data on baselines longer than 100m to increase the dynamic
range.  The final image
was cleaned to a threshold of 5 mJy.  The lowest noise
level in the image, away from bright sources, is $\sim0.2$ mJy/beam
($M\sim0.14$ \msun at $T=20$ K using the extrapolation of
\citet{Ossenkopf1994a} opacity from \citet{Aguirre2011a} with $\beta=1.75$),
but near the bright sources e2 and IRS2, the noise reached as high as $\sim2$
mJy/beam.  Deeper cleaning was attempted, but these attempts produced
instabilities that resulted in divergent maps.  The combined image has a central
frequency of about 226.6 GHz assuming a flat spectrum source; a steep-spectrum
source, with $\alpha=4$, would have a central frequency closer to 227 GHz, a
difference that is negligible for all further analysis.

\subsubsection{Lines}
\label{sec:obslines}
We produced spectral image cubes of the lines listed in Tables
\ref{tab:linesspw0}, \ref{tab:linesspw1}, \ref{tab:linesspw2}, and
\ref{tab:linesspw3}.  For kinematic and moment analysis, the median value over
the spectral range [25,30],[80,95] \kms was used to estimate and subtract the
local continuum.

\section{Results \& Analysis}
\label{sec:results}

\subsection{Continuum Sources}
\label{sec:contsources}
In this section, we describe our overall catalogue of continuum sources,
then examine in detail the three most prominant hot cores that contain
massive young stellar objects (MYSOs), W51 e2e, W51 e8, and W51 North.
We also discuss W51 d2, which appears to be somewhat older and less massive
than these three dominant objects.

\subsubsection{Source Identification and Catalogue}
\label{sec:sourceid}
We used the \texttt{dendrogram} method described by \citet{Rosolowsky2008c} and
implemented in \texttt{astrodendro} to identify sources.  We used a minimum
value of 1 mJy/beam ($\sim5\sigma$) and a minimum $\Delta=0.4$ mJy/beam
($\sim2\sigma$) with minimum 10 pixels (each pixel is 0.05\arcsec).  This
cataloging yielded over 8000 candidate sources, of which the majority are noise
or artifacts around the brightest sources.  To filter out these bad sources,
we created a noise map taking the local RMS of the \texttt{tclean}-produced
residual map, using a weighted RMS over a $\sigma=30$ pixel (1.5\arcsec)
gaussian.  We then removed
all sources with peak S/N $< 8$, mean S/N per pixel $< 5$, or minimum S/N per
pixel $ < 1$.  We also only included the smallest sources in the dendrogram,
the ``leaves''.  These parameters were tuned by checking against ``real''
sources identified by eye and selected using \texttt{ds9}: most real sources are
recovered and few spurious sources ($<10$) are
included.  The resulting catalog includes 113 sources.

The `by-eye' core extraction approach, in which we placed ds9 regions on all
sources that look `real', produced a more reliable but less complete (and less
quantifiable) catalog containing 75 sources.  This catalog is more useful in
the regions around the bright sources e2 and North, since these regions are
affected by substantial uncleaned PSF sidelobe artifacts.  In particular, the
dendrogram catalog includes a number of sources around e2/e8 that, by eye,
appear to be parts of continuous extended emission rather than local peaks;
``streaking'' artifacts in the reduced data result in their identification
despite our threshold criteria.  The dendrogram extraction also identified
sources within the IRS 2 \hii region that are not dust sources.  Dendrogram
extraction missed a few clear sources in the low-noise regions away from
W51 Main and IRS 2 because the identification criteria were too conservative.

When extracting properties of the `by-eye' sources, we used variable sized
circular apertures, where the apertures were selected to include all of the
detectable symmetric emission around a central peak up to a maximum radius
$r\sim0.6$\arcsec.  This approach is necessary because some of the sources are
not centrally peaked and are therefore likely to be spatially resolved starless
cores.

Further information about and general discussion of the continuum sources is in
Appendix \ref{sec:contsrcs}.  For the rest of this section, we focus on only
the few brightest sources.  The general point source population is briefly
revisited in Section \ref{sec:faintsrcs_discussion}.

\subsubsection{W51e2e mass and temperature estimates from continuum}
\label{sec:W51e2e}

In a $0.21\arcsec\times0.19\arcsec$ beam ($1100\times1000$ au), the peak flux
density toward W51 e2e is 0.38 Jy, which corresponds to a brightness
temperature $T_B=225$ K.  This is a lower limit to the surface brightness of
the millimeter core, since an optical depth $\tau<1$ or a filling factor of the
emission $ff<1$ would both imply higher intrinsic temperatures.  The implied
luminosity, assuming blackbody emission from a spherical beam-filling source,
is $L = 4\pi r^2 \sigma_{sb} T^4 = 2.3\ee{4} \lsun$, where
$\sigma_{sb}=5.670373\ee{-5} \mathrm{\,g\,s^{-3}\,K^{-4}}$ is the
Stefan-Boltzmann constant.  Since any systematic
uncertainties imply a higher temperature, this estimate is a lower limit on the
source luminosity.  Such a luminosity corresponds to a B0.5V, 15 \msun main
sequence star with effective temperature $4\ee{4}$ K \citep[][see
Section \ref{sec:stellarproperties} for further discussion of
stellar types]{Pecaut2013a}\footnote{For the B-star parameters, we used
\url{http://www.pas.rochester.edu/~emamajek/EEM_dwarf_UBVIJHK_colors_Teff.txt},
which primarily comes from \citet{Pecaut2013a}.
}.

If we assume that the dust is optically thick throughout our beam, and assume
an opacity constant $\kappa(227 \mathrm{GHz})=0.0083$ cm$^2$ g$^{-1}$  (which
incorporates and assumed gas-to-dust ratio of 100), the minimum mass
per beam to achieve $\tau\geq1$ is $M=18$ \msun beam$^{-1}$.  This  mass is not
a strict limit in either direction: if the dust is indeed optically thick,
there may be substantial hidden or undetected gas, while if the filling factor
is lower than 1, the dust may be much hotter and therefore optically thin and
lower  mass.  However, simulations and models both predict that the dust will
become highly optically thick at radii $r\lesssim1000$ au
\citep{Forgan2016a,Klassen2016a}, so it is likely that this measurement
provides  a lower limit on the total gas mass surrounding the protostar.
Therefore, unless the stars are extremely efficient at removing material or the
gas fragments significantly on $<1000$ AU scales, the stellar mass is likely to
at least double before accretion halts.

For an independent measurement of the temperature that is not limited to the
optically thick regions, we use the \methanol lines in band, calculating an LTE
temperature that is $200 < T < 600$ K out to $r<2$\arcsec ($r<10^4$ AU; Section
\ref{sec:methanol}).  As noted in Section \ref{sec:methanol}, these
temperatures may be overestimates when the low-J lines of \methanol are
optically thick, but for now they are the best measurements we have available.
If the dust temperature matches the methanol temperature, it would be optically
thin ($\tau \lesssim 1/3$) and the central source dust mass would be only
$\sim6$ \msun.  However, this latter estimate discounts any substructure at
scales $<1000$ AU. 

An upper limit on the radio continuum emission from W51e2e is $S_{14.5 GHz} <
0.6$ mJy/beam (2-$\sigma$) in a FWHM=$0.34\arcsec$ beam, or $T_{B,max} < 30$ K
\citep{Ginsburg2016b}.  Assuming emission from an optically thick \hii region
with $T_e = 8500$ K \citep{Ginsburg2015a}, the upper limit on the emitting
radius is $R(\hii) < 110 AU$.  Similar limits are obtained from other
frequencies in those data.  The free-free contribution to the millimeter
flux is therefore negligible, and the central source is unlikely to be
ionizing.  Limits on the stellar properties are further discussed in
Section \ref{sec:stellarproperties}.

\subsubsection{W51 e8 and North mass and temperature from continuum}
\label{sec:w51e8andnorth}
We repeat the above analysis for e8 and North.  They have peak intensities
of 0.35 and 0.44 Jy/beam respectively, corresponding to peak brightness
temperatures of 205 and 256 K.  The North source was detected at 25 \um, but
not at shorter wavelengths, by \citet{Barbosa2016a}, confirming the presence
of warm dust.
The lower limit luminosities of W51 e8 and North in a single beam, assuming the
brightest detected beam is optically thick, are 1.6\ee{4} and 3.9\ee{4} \lsun,
respectively.

W51 North has a free-free upper limit similar to that of W51e2e, but somewhat less
restrictive because the noise in that region is substantially higher.  W51 e8,
by contrast with the others, has a clear detection at cm wavelengths.  The
source e8n, which is offset from the peak mm emission by 0.13\arcsec\ (700 AU),
has $S_{25 GHz}=4.7$ mJy/beam, corresponding to $T_B=135$ K, which implies an
\hii region size $R=180$ AU if the emission is produced by  optically thick
free-free emission.  This could be part of an ionized jet or an ionizing binary
companion, but its offset from the central mm source
suggests that it is not a simple spherically symmetric HC\hii region.

The apparent dust masses in the central beams of e8 and North are the same
as in e2e, $M\sim18$ \msun, but these measurements are subject to the same
limits discussed in Section \ref{sec:W51e2e}.

\subsubsection{W51 d2: a smaller, likely older hot core}
\label{sec:w51d2}
The source W51 d2 is something of an outlier in our sample.  Like the three
main hot cores, e2e, e8, and North, d2 has a small extended molecular hot core
around it, with $R\lesssim3000$ AU.  However, unlike these cores, d2 is a very
bright centimeter continuum source, $\sim17$ mJy at 15 GHz
\citep{Ginsburg2016a}.  Its millimeter continuum emission can readily be
explained as free-free emission, requiring a spectral index of only
$\alpha\sim0.6-0.7$ from the cm to account for all of its mm emission.  There
is little doubt that it contains a compact \hii region.  Because of this free-free
contamination, we cannot estimate the central core's dust mass.  If we assume
the free-free is optically thin at 36 GHz \citep[the highest-frequency cm-wave
measurement
we have available;][]{Goddi2015a}, with $S_{36 \mathrm{GHz}} = 29$ mJy and
$S_{227 \mathrm{GHz}}=110$
mJy, the dust-produced flux would be $S_{227 \mathrm{GHz}} = 86$ mJy, or about
$\sim20-25\%$ as bright as the other three cores ($T_{B}=65$ K).  With such a modest
lower limit brightness temperature, the dust source is likely to be optically
thin or less than beam-filling, making its upper limit dust mass 
$M\ll18\msun$; if we assume $T_{dust} = T_{line,max} = 220 K$, the upper limit
dust mass is $M<7$ \msun.  If d2 were a purely dust source, its lower limit
luminosity is a meager 160 \lsun.  Since the lowest-luminosity 
stars with ionizing photospheres have $L>10^4$ \lsun, d2 is unlikely to be a
dust-only source.

Additionally, unlike the three hot cores, d2 does not drive an outflow.  It
does, however, power a unique set of ammonia (\ammonia) masers \citep[][Wootten
\& Wilson in prep]{Gaume1993a,Wilson1990a,Zhang1995a,Henkel2013a,Goddi2015a}.  These
features imply it is in an intermediate evolutionary state between the larger
compact \hii regions and the hot cores that exhibit no centimeter continuum.

\citet{Barbosa2016a} reported W51 d2 (OKYM 6) as ``just a ridge of emission''
because it appears only in their 25 \um images and is invisible at shorter
wavelengths.  Our clear detection of both the known \hchii region and a
surrounding molecular core indicate instead that it is just extremely embedded.

\subsection{The mass and light budget on different spatial scales}
\label{sec:massbudget}
An evolutionary indicator used for star-forming regions is the amount of mass
at a given density; a more evolved (or more efficiently star-forming) region will
have more mass at high densities.  We cannot measure the dense gas fraction
directly, but the amount of flux density recovered by an interferometer
provides an approximation.

For the ``total'' flux density in the region, we use the Bolocam Galactic Plane
Survey observations \citep{Aguirre2011a,Ginsburg2013a}, which are the closest
in frequency single-dish millimeter data available.  We assume a spectral index
$\alpha=3.5$ to convert the BGPS flux density measurements at 271.4 GHz to the
mean ALMA frequency of 226.6 GHz.  The ALMA data (specifically, the
0.2\arcsec resolution 12m-only data) have a total flux 23.2 Jy above a  conservative
threshold of 10 mJy/beam in our
mosaic; in the same area, the BGPS data have a flux of 144 Jy, which scales down to
76.5 Jy.  The recovery fraction is 30$\pm3$\%, where the error bar accounts
for a change in $\alpha\pm0.5$.  The threshold of 10 mJy/beam corresponds to a
column threshold $N>1.3\ee{25}$ \percc for 20 K dust. This threshold also
corresponds to an optical depth of $\tau\approx0.5$, implying that a substantial
fraction of the cloud is either approaching optically thick or is warmer than 20
K.  For an unresolved spherical source in the $\sim0.2\arcsec$ beam, this
column density corresponds to a volume density $n>10^{8.1}$ \percc.
Of the area with significant emission, 23\% has $T_B>20$ K (34 mJy \perbeam)
and must have $T_{dust}>20$ K, guaranteeing that a substantial fraction of all
of the detected continuum emission is coming from warmer dust.

Even more impressive is the amount of the total flux density concentrated
into the three massive cores, W51 e2e, e8, and North.  These three contain
12.3 Jy (within 1\arcsec or 5400 AU apertures) of the total 23.2 Jy in the
observed field - more than half of the total ALMA flux density, or 15\% of the
BGPS flux density.  In a \citet{Kroupa2001a} IMF, massive stars ($M>20$ \msun)
account for only 0.15\% of the mass, so in order for the gas mass distribution
to produce a `normal' stellar distribution, the high-mass-star-producing gas
must be much brighter (hotter) than that making low-mass stars, or the gas 
in these cores must be substantially redistributed and fragmented into a
mixture of high- and low-mass stars as the region evolves.

\subsection{Chemically Distinct Regions}
\label{sec:chemistrymaps}
\label{sec:chemistrymapsobs}
The large ``hot cores'' in W51 (e2, e8, and North) are spatially well-resolved
and multi-layered.  These cores are detected in lines of many different species
spanning areas $\sim5\ee{3}-1\ee{4}$ AU across.  We describe some of the
specific notable chemical features in this section, but the overall point that
the three biggest hot cores have extended chemical structure is highlighted in
Figures \ref{fig:chemmapse2}, \ref{fig:chemmapse8}, and
\ref{fig:chemmapsnorth}, with a fainter hot core shown for contrast in Figure
\ref{fig:chemmapsALMAmm14}.

Surrounding W51e2e, there are relatively sharp-edged and uniform-brightness
regions in a few spectral lines over the range 51-60 \kms (Figure
\ref{fig:chemmapse2}, especially the \methanol and \methylformate lines).  Some
of these features are elongated in the direction
of the outflow, but most have significant extents orthogonal to the outflow.
The circularly symmetric features are prominent in \methanol, OCS, and
\dimethylether, weak but present in \formaldehyde and SO, and absent in
\cyanoacetylene and HNCO.

Around e8, a similar chemically enhanced region is observed, but in this case
\dimethylether is absent.  Toward W51 North, \methanol, \formaldehyde, and SO
exhibit the sharp-edged enhancement feature, while the other species do not.

By contrast, along the south end of the e8 filament, no such enhanced features
are seen; only \formaldehyde and the lowest transition of methanol, \methanol
\fourtwotwo, are evident.

The relative chemical structures of e2, e8, and North are  similar.
The same species are detected in all of the central cores.  However, in e2,
\dimethylether, \methylformate, \ethylcyanide, and Acetone (\acetone) are
significantly more extended than in the other sources.
\gaucheethanol is detected in W51 North, but is weak in e8 and almost absent
in e2 (Figures \ref{fig:chemmapse2}, \ref{fig:chemmapse8},
\ref{fig:chemmapsnorth}, \ref{fig:chemmapsALMAmm14}).

Different chemical groups exhibit different morphologies around e2, and this
approximate grouping is also seen around the other cores.  Species that are
elongated in the NW/SE direction are associated primarily with the outflow
(\cyanoacetylene, \ethylcyanide).  Other species are associated primarily with
the extended circular core (\methylformate, \dimethylether, \acetone).  Some
are only seen in the compact core ($R<0.4\arcsec\sim2000$ AU;
\methyleneamidogen, HNCO, \formamide, and vibrationally excited
\cyanoacetylene).  Only \methanol and OCS are associated with both the extended
core and the outflow, but not the greater extended emission.  \ketene seems to
be associated with only the extended core, but not the compact core. Finally,
there are the species that trace the broader ISM in addition to the cores and
outflows: \formaldehyde, $^{13}$CS, OCS, \ceighteeno and SO.  Both HCOOH and
N$_2$D$^+$ are weak and associated only with the innermost e2e core.

The presence of these complex species symmetrically distributed at large
distances ($r\sim5000$ AU) from the central sources is an independent
indication of the gas heating provided by these sources.  The abundance
increase most likely corresponds to $T\gtrsim85$ K, the approximate
sublimation temperature of \methanol ice \citep{Green2009a}.

While we focused on the three main hot cores, which all have radii $\sim5000$
AU, there are a few others that have similar chemical enhancements, but
significantly smaller extents.  The sources d2 and ALMAmm31 can be seen in
Figure \ref{fig:chemmapsnorth} on the right (west) side of the map.  These both
have resolved chemical structure, but the structures are smaller than in the
main hot cores.  d2 is also unique in having a central ionizing source detected
in H30$\alpha$ and a (moderately) extended chemical envelope.

\Figure{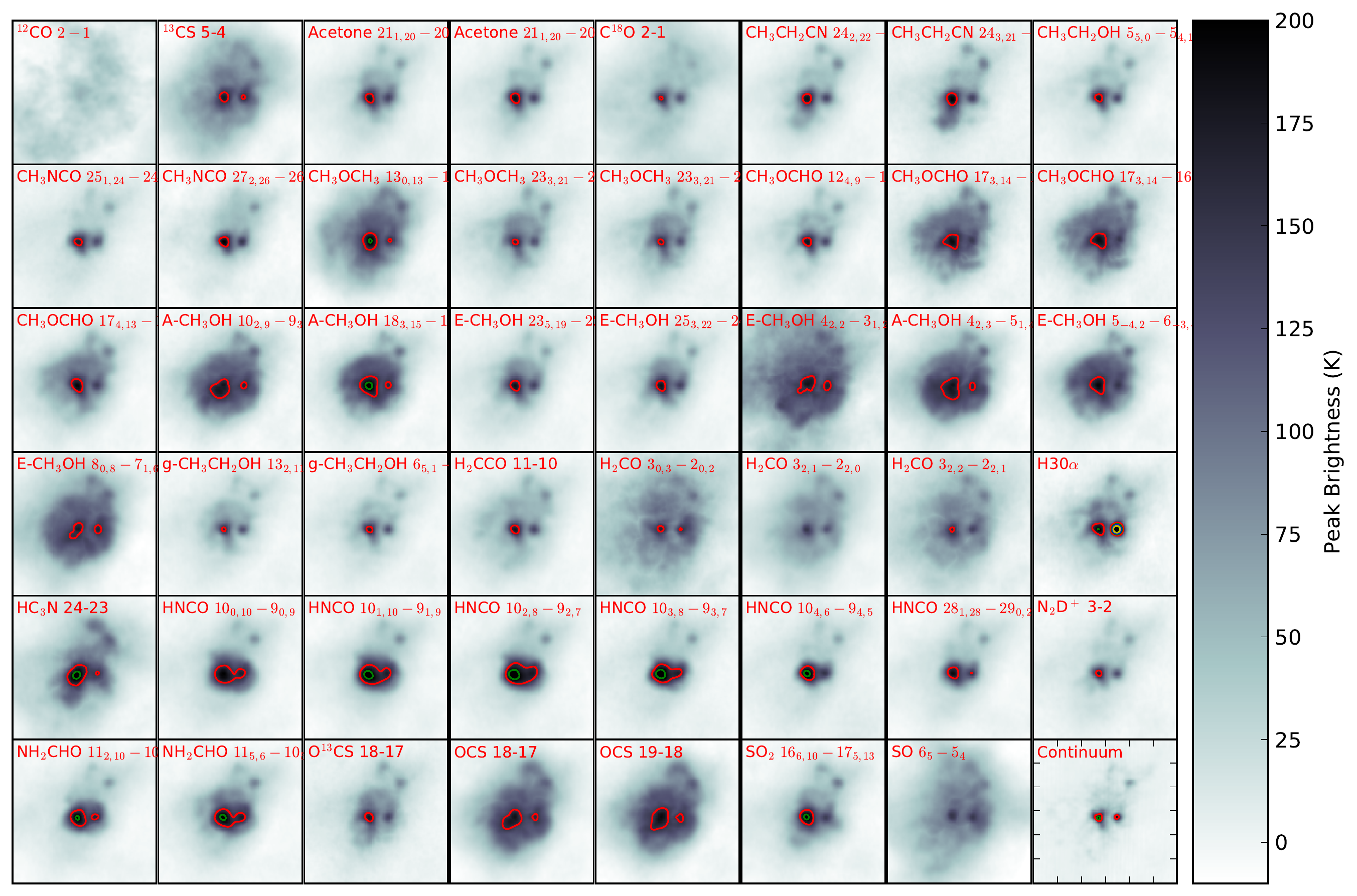}
{Peak brightness maps of the e2 region in 47 different lines over the range 51 to 60
\kms.
The cutouts are $6\arcsec\times6\arcsec$ ($3.2\ee{4}\times3.2\ee{4}$ AU).
To illustrate the lower limit temperature implied by the observed brightness,
the maps are not continuum subtracted.  
For additional contrast, contours are shown at 150, 200, 250, and 300 K
(red, green, blue, yellow).
There
is a strong `halo' of emission seen in the CH$_3$Ox lines and OCS.  Extended
emission is also clearly seen in SO, $^{13}$CS, and \formaldehyde, though these
lines more smoothly blend into their surroundings.  HNCO and \formamide have
smaller but substantial regions of enhancement with a sharp contrast to their
surroundings.  HC$_3$N traces the e2e outflow.  The bright H30$\alpha$ emission
marks the position of e2w, the hypercompact HII region that dominates the
centimeter emission in e2.
}{fig:chemmapse2}{1}{18cm}

\Figure{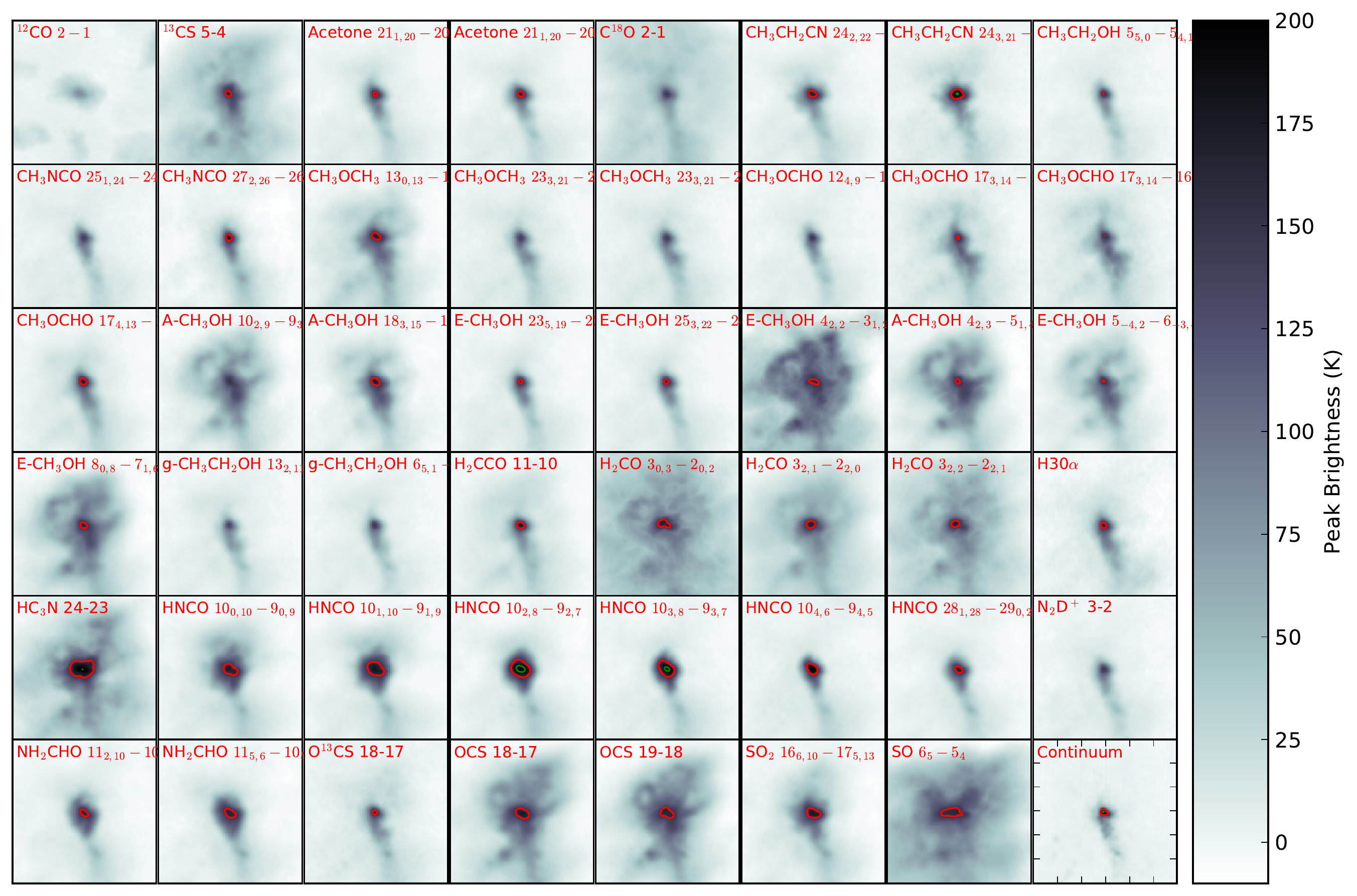}
{Peak brightness maps of the e8 region in 47 different lines over the range 52 to 63
\kms.
The cutouts are $6\arcsec\times6\arcsec$ ($3.2\ee{4}\mathrm{~AU}\times3.2\ee{4}$ AU).
To illustrate the lower limit temperature implied by the observed brightness,
the maps are not continuum subtracted.  
For additional contrast, contours are shown at 150 and 200 K
(red and green, respectively).
As in e2 (Figure \ref{fig:chemmapse2}),
there is extended emission in the CH$_3$OH and OCS lines, but in contrast with e2,
the other CH$_3$Ox lines are more compact. SO is brighter than OCS in e8,
whereas the opposite is true in e2.
}{fig:chemmapse8}{1}{18cm}

\Figure{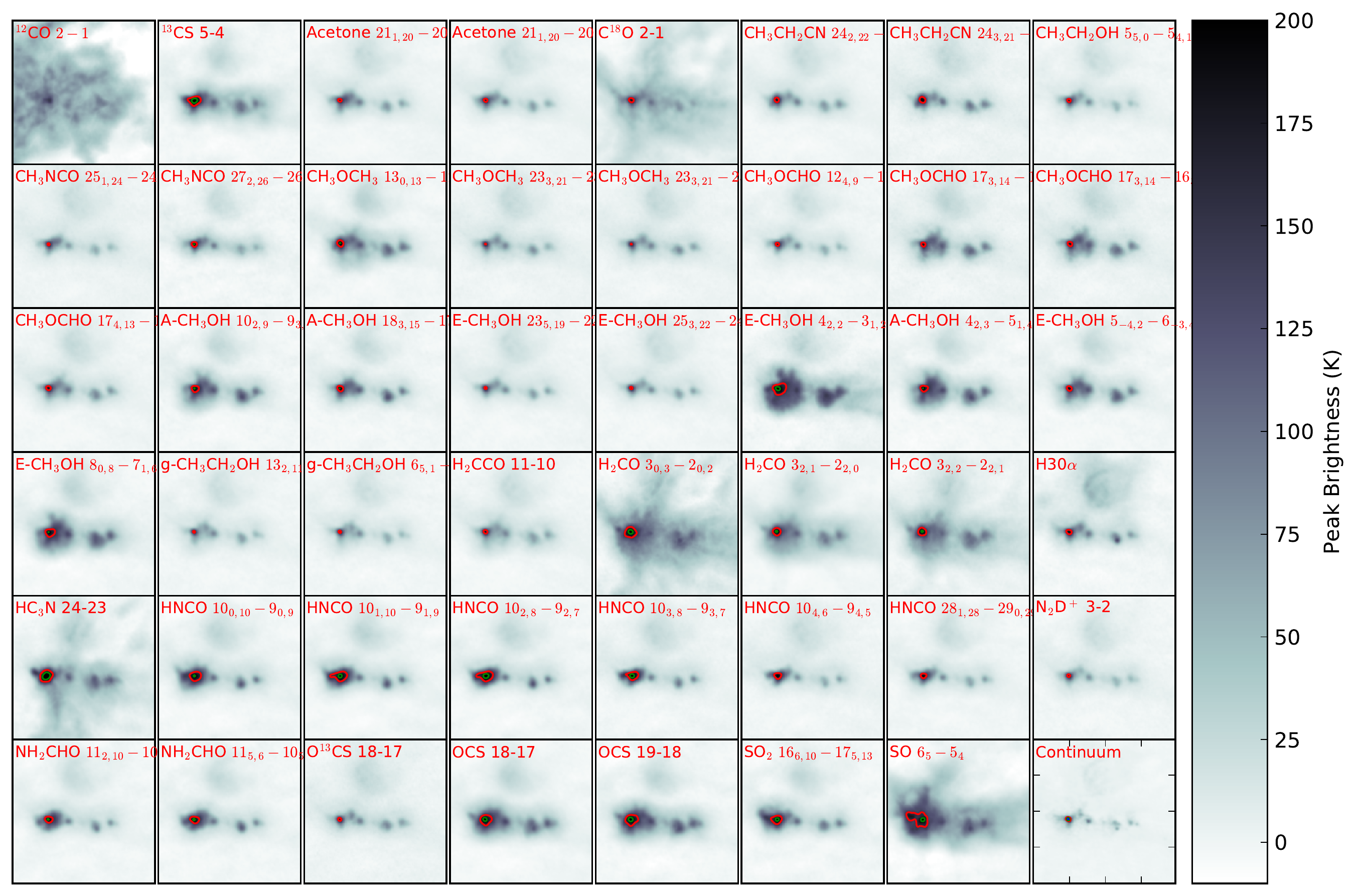}
{Peak brightness maps of the W51 IRS2 region containing the North core in 47
different lines over the range 54 to 64 \kms.  The cutouts
are $10\arcsec\times10\arcsec$ ($5.4\ee{4}\times5.4\ee{4}$ AU).  To illustrate
the lower limit temperature implied by the observed brightness, the maps are
not continuum subtracted.  For additional contrast, contours are shown at 150
and 200 K (red and green, respectively).  Qualitatively, the relative extents
of species seem comparable to e8 (Figure \ref{fig:chemmapse8}).  
The W51 North core is the brightest region highlighted by the contours in some
frames.  W51 d2 is right of center and slightly south of the other cores.
}{fig:chemmapsnorth}{1}{18cm}

\Figure{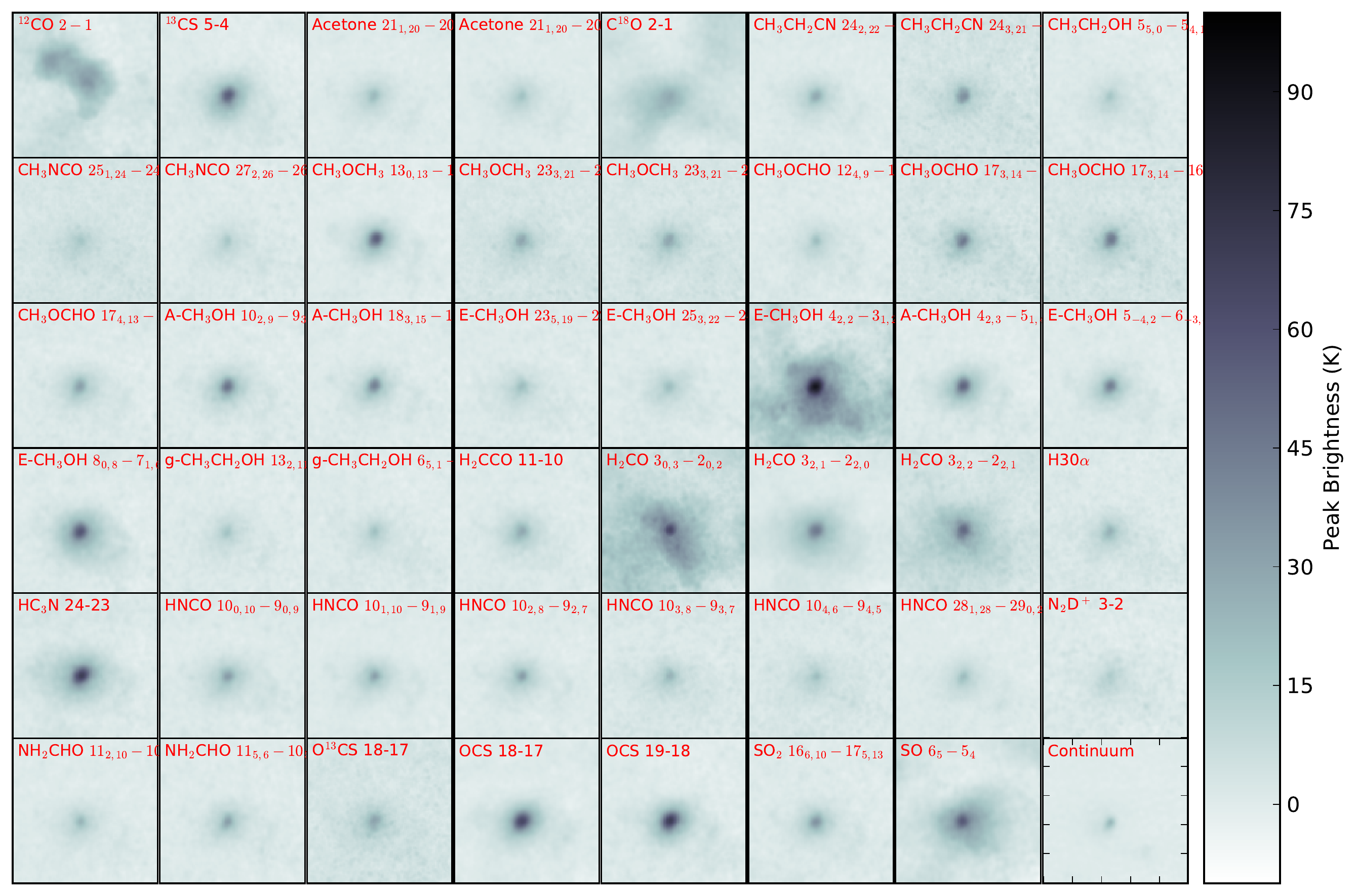}
{Peak brightness maps of the ALMAmm14 region in 47 different lines over the
range 58 to 67 \kms.
The cutouts are $5\arcsec\times5\arcsec$ ($2.7\ee{4}\times2.7\ee{4}$ AU).
ALMAmm14 is one of the brightest sources outside of
e2/e8/IRS2, but it is substantially fainter than those regions.  Still, it has
a notably rich chemistry.
}{fig:chemmapsALMAmm14}{1}{18cm}

\subsection{\methanol temperatures \& columns in the hot cores}
\label{sec:ch3ohtem}
\label{sec:methanol}
The  chemically enhanced regions appear to be associated with regions of
elevated gas temperature.  We examine the temperature structure directly by
analyzing the excitation of lines for which we have detected multiple
transitions with significant energy differences.  We do not use \formaldehyde
for this analysis despite its usefulness as a thermometer because it is clearly
optically thick (self-absorbed) in all lines in the hot cores.
This section presents the details of the temperature determination, while 
the implications of the temperature measurements will be discussed later,
throughout Section \ref{sec:discussion}.

We produce rotational diagrams for each spatial pixel covering all \methanol
lines detected at high significance toward at least one position\footnote{We
observe both A- and E-type \methanol, but assume the ratio $E/A=1$, 
as expected if the molecules have an even moderately high formation 
temperature $T\gtrsim20$ K \citep{Wirstrom2011a}.}.  The detected
lines span a range $45 < E_U < 800$ K, allowing robust measurements of the
temperature
assuming the lines are optically thin, in LTE, and the gas temperature is high
enough to excite the lines.  These conditions are likely to be satisfied in the
e2e, e8, and North cores, except for the optically thin requirement; the lower-J
lines in particular are optically thick across much of the extent of the cores. 

\FigureTwo
{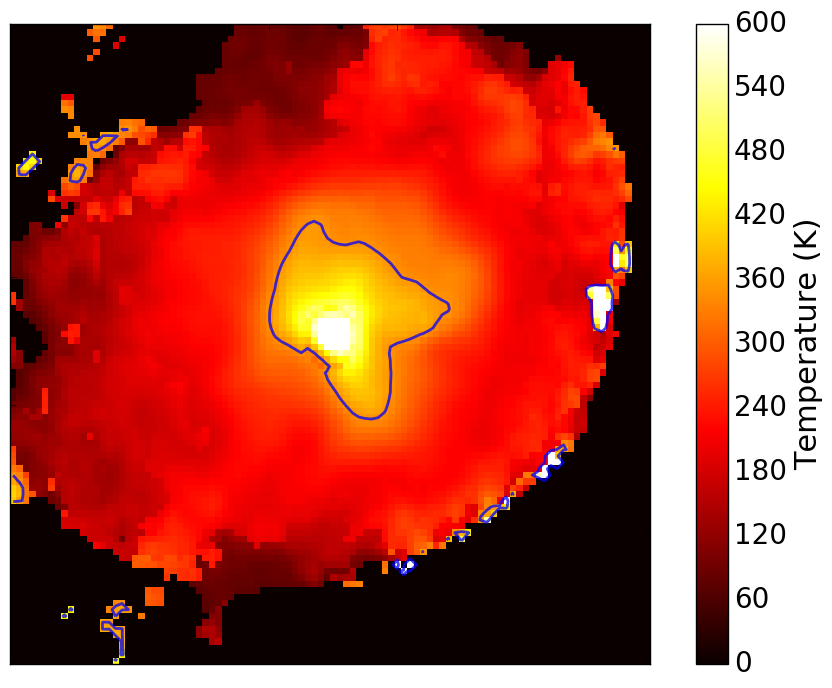}
{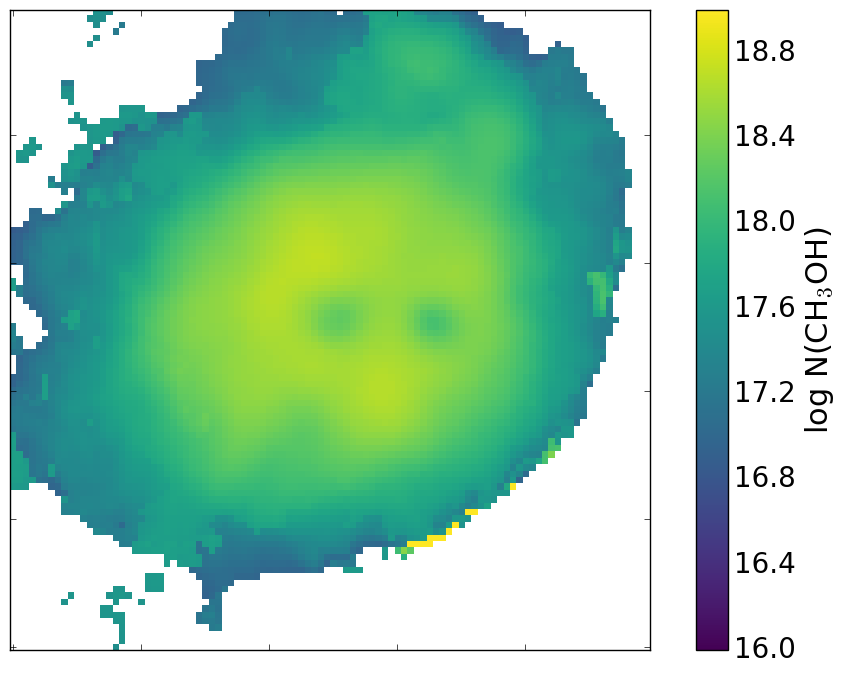}
{Methanol temperature and column density maps around e2. 
The maps are $5\arcsec\times5\arcsec$ ($2.7\ee{4}\times2.7\ee{4}$ AU).
The central regions around
the cores appear to have lower column densities because the lines become
optically thick and self-absorbed.  The contour in the temperature map is at
350 K, where red meat is typically considered ``well-done''.}
{fig:ch3ohe2}{1}{9cm}

The fitted temperature and \methanol column maps are shown in Figure 
\ref{fig:ch3ohe2}.
Sample fitted rotational diagrams are displayed in Figure \ref{fig:ch3ohe2rot}.
The line intensities are computed from moment maps integrating over the range
(51, 60) \kms in continuum subtracted spectral cubes, where the continuum
was estimated as the median over the ranges (25-35,85-95) \kms, except
for the J$_u$=25 lines, which had a continuum estimated from the 10th percentile
over the same range to exclude contamination from the SO outflow line wings.
The fitted species are listed in the order plotted in Table \ref{tab:methanol}.
Note that A- and E-type methanol can only interchange in chemical reactions,
but barring peculiar excitation processes, they should be governed by the same
partition function \citep{Rabli2010c}.

\Figure{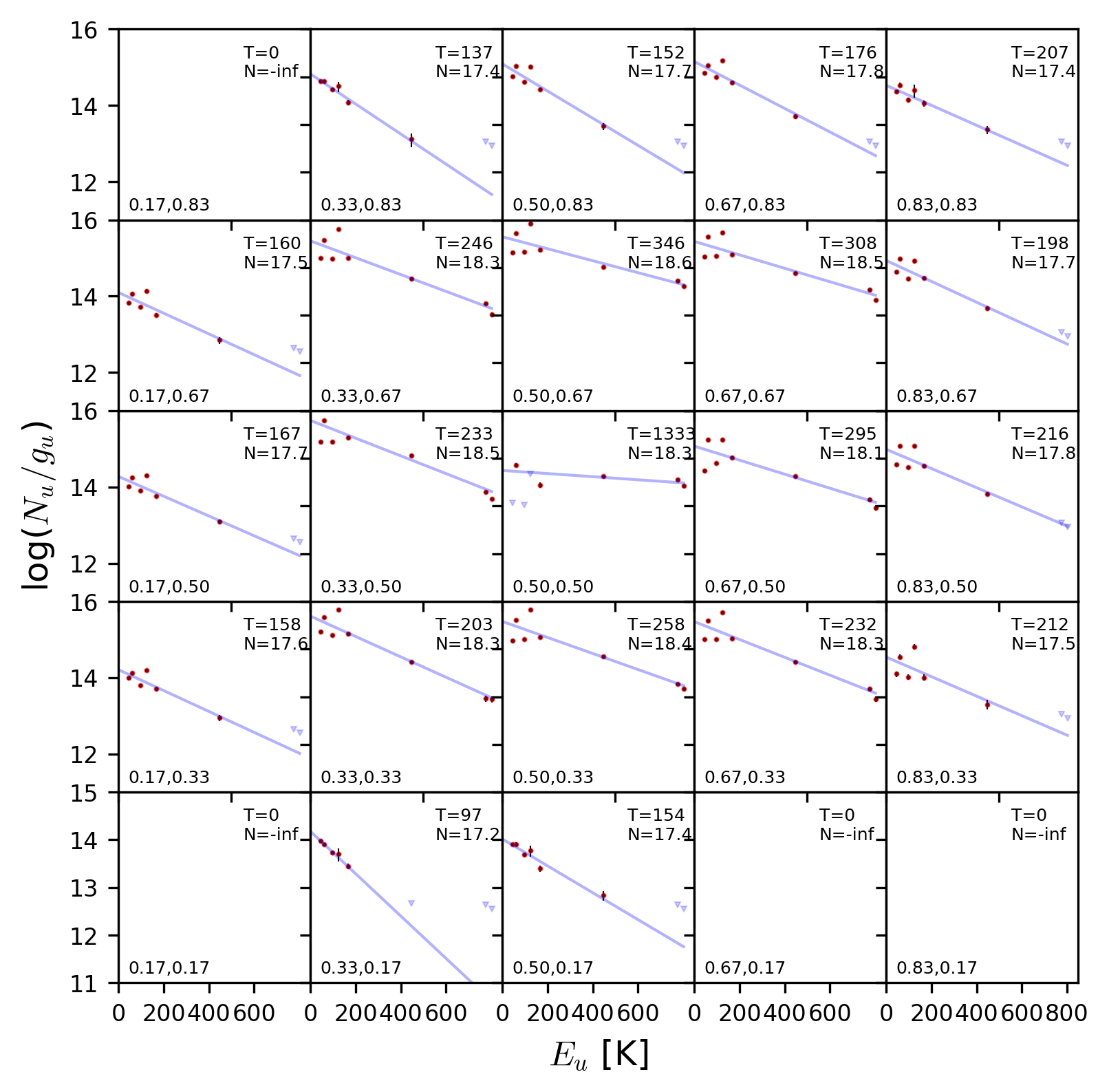}
{A sampling of fitted rotation diagrams of the detected \methanol transitions.
These are shown to provide validation of the temperatures and column densities
derived and shown in Figure \ref{fig:ch3ohe2}.  The lower-left corner of each
panel shows the position from which the data were extracted in that figure in
units of figure fraction.  Error bars show the measurement error on each point;
because these are plotted on a log scale, the errors are often smaller than
the plotted points.  Pixels with nondetections at the 3-$\sigma$ level are plotted
as triangles indicating the 1-$\sigma$ error at that position; they are included
in the fit as zero-column values with the appropriate error.  The fitted
temperature and column are shown in the top right of each plot.
The central position is severely affected by absorption and can be ignored.
The corners do not have enough line detections to be fit.
}{fig:ch3ohe2rot}{1}{18cm}

\begin{table*}[htp]
\caption{\methanol lines used to determine temperature}
\begin{tabular}{lll}
\label{tab:methanol}
Line Name & Frequency & E$_U$ \\
 & $\mathrm{GHz}$ & $\mathrm{K}$ \\
\hline
E-CH$_3$OH $4_{2,2}-3_{1,2}$ & 218.44005 & 45.45988 \\
A-CH$_3$OH $4_{2,3}-5_{1,4}$ & 234.68345 & 60.9235 \\
E-CH$_3$OH $8_{0,8}-7_{1,6}$ & 220.07849 & 96.61336 \\
E-CH$_3$OH $5_{-4,2}-6_{-3,4}$ & 234.69847 & 122.72222 \\
A-CH$_3$OH $10_{2,9}-9_{3,6}$ & 231.28115 & 165.34719 \\
A-CH$_3$OH $18_{3,15}-17_{4,14}$ & 233.7958 & 446.58025 \\
E-CH$_3$OH $23_{5,19}-22_{6,17}$ & 219.99394 & 775.89371 \\
E-CH$_3$OH $25_{3,22}-24_{4,20}$ & 219.98399 & 802.17378 \\
\hline
\end{tabular}

\end{table*}

To validate some of the rotational diagram fits, we examined the modeled
spectra overlaid on the real (Figure \ref{fig:ch3ohe2epeaks}).  These generally
display significant discrepancies, especially at low J where self-absorption is
evident.  In Figure \ref{fig:ch3ohe2epeaks}, there is clearly a low-temperature
component slightly redshifted from the high-J peak that can be seen as a dip
within the line profile.  The presence of this unmodeled low-temperature
component renders our \methanol temperature measurements uncertain, biasing
them to be slightly high.  Nevertheless,
the general trend exhibited by \methanol temperatures matches expectations
if there is a central heating source.

\FigureTwo
{f9}
{f10}
{Spectra of the \methanol lines toward a pair of selected pixels just outside
of the central e2e core. (a) is 0.55\arcsec  and (b) is 1.33\arcsec from e2e.  The red
curves show the LTE model fitted from a rotational diagram as shown in Figure
\ref{fig:ch3ohe2rot}.  \emph{The model is not a fit to the data shown}, but is instead
a single-component LTE model fit to the integrated intensity of the lines
shown.  As such, the fit is not convincing, and it is evident that a
single-temperature, single-velocity model does not explain the observed lines.
Nonetheless, a component with the modeled temperature is likely to be present
in addition to a cooler component responsible for the self-absorption in the
low-J lines.  (a) shows a pixel close to the center of e2e, which is probably
optically thick in most of the shown transitions, while (b) shows a better case
where the highest-$A_{ij}$ (highest critical density) lines are overpredicted
but many of the others are well-fit.}
{fig:ch3ohe2epeaks}{1}{8cm}

Figure \ref{fig:ch3ohvscont} shows a comparison between the \methanol
$10_{2,9}-9_{3,6}$ line and the 225 GHz continuum.  While the brightest regions
in \methanol mostly have corresponding dust emission, the dust morphology
traces the \methanol morphology very poorly.  This difference suggests that the
enhanced brightness is not simply because of higher total column density.
We examine the dust-\methanol correspondence more quantitatively in Figure
\ref{fig:ch3ohtemX}; Figure \ref{fig:ch3ohtemX}d shows the poor correlation.

\FigureTwo{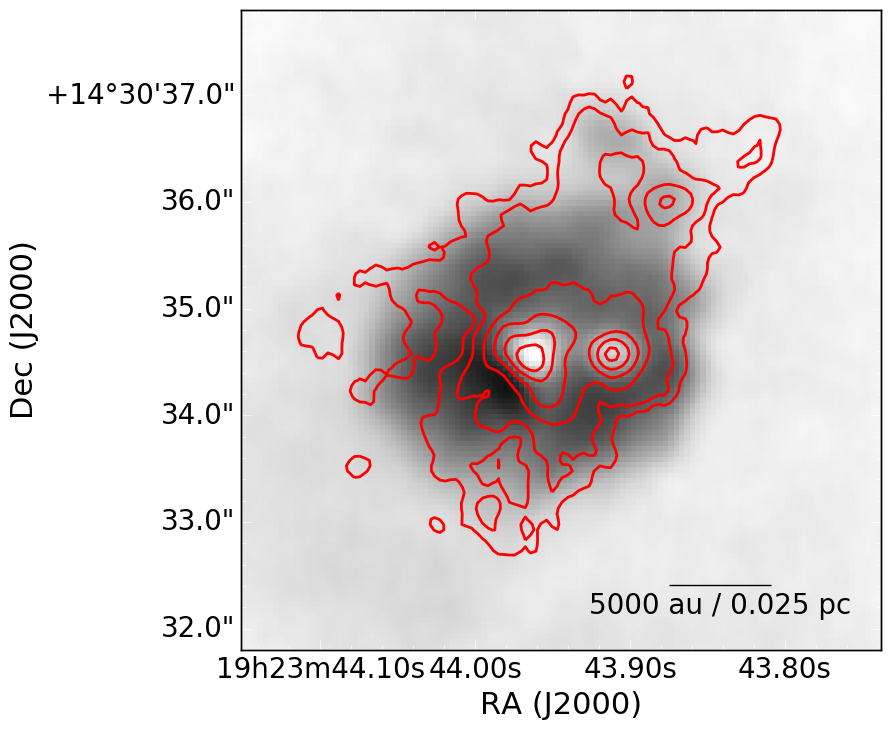}
          {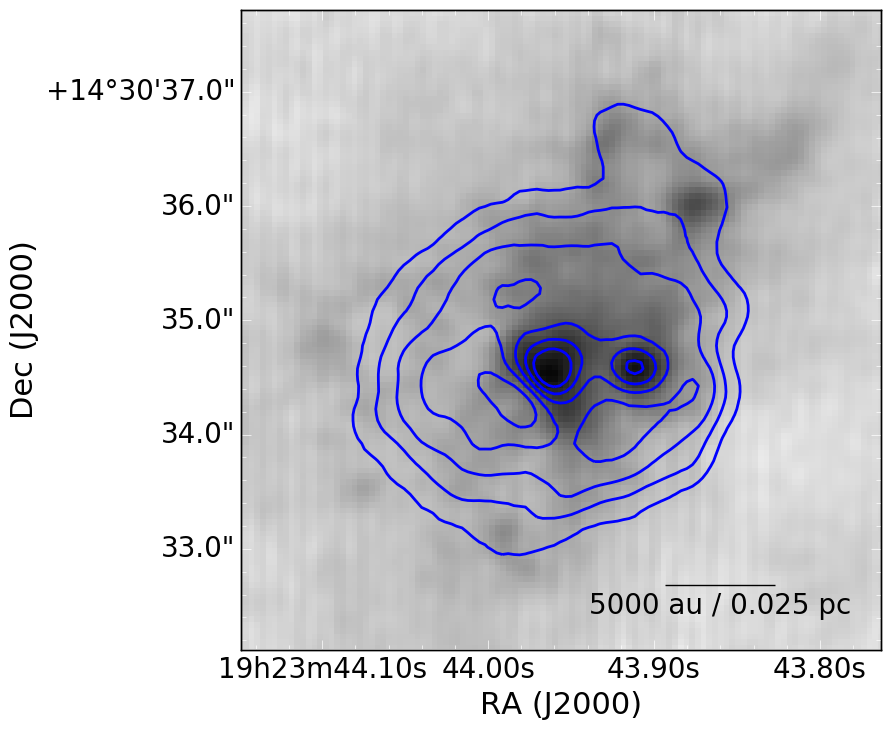}
{Images showing \methanol $10_{2,9}-9_{3,6}$ and 225 GHz continuum emission,
with \methanol in grayscale and continuum in contours (left) and continuum in
grayscale, \methanol in contours (right).  The fainter (whiter) regions in the center
of the \methanol map correspond to the bright continuum cores and show where all lines
appear to be self-absorbed.}
{fig:ch3ohvscont}{1}{8cm}

Figure \ref{fig:methanolradialprofile} shows the observed brightness profiles
of \methanol line and dust continuum emission, which gives a lower limit on
the physical temperature probed by the \methanol and continuum.
Figure \ref{fig:ch3ohtemX}a shows a comparison of the \methanol temperature and
abundance.  The \methanol abundance is derived by comparing the rotational
diagram (RTD) fitted \methanol column density to the dust column density while
using the \methanol-derived temperature as the assumed dust temperature.  The
figure shows all pixels within a 3\arcsec (16200 AU) radius of e2e, with pixels
having low column density and high temperature (i.e., pixels with bad fits) and
those near e2w (which may be heated by a different source) excluded.  We used
moment-0 (integrated intensity) maps of the \methanol lines to perform these
RTD fits, which means we have ignored the line profile entirely and in some
cases underestimated the intensity of the optically thick lower-J lines: in the
regions of highest column, the column is  underestimated and the temperature is
overestimated, as can be seen in Figure \ref{fig:ch3ohe2epeaks}.

\Figure{f14}
{Radial profiles of the azimuthally averaged peak surface brightness of the
observed \methanol transitions along with the profile of the continuum
brightness around e2e.  These profiles indicate \emph{lower limit} gas temperatures
as a function of radius; the true temperature can be substantially higher even
if the lines are optically thick because of foreground, cold, self-absorbing
layers.
The radial profiles were constructed from images with 0.2\arcsec
resolution including only 12m data.  The lines are not continuum subtracted, so
they represent the true on-sky observed brightness.  The abundance bump is
evident at $r\sim1.5\arcsec$, while the consistently increasing high-J lines
(\methanol $23_{5,19}-22_{6,17}$ and $25_{3,22}-24_{4,20}$) demonstrate that
the excitation is continuing to increase toward the center, even after the
lower-J lines become optically thick.
}
{fig:methanolradialprofile}{1}{18cm}

\FigureFour
{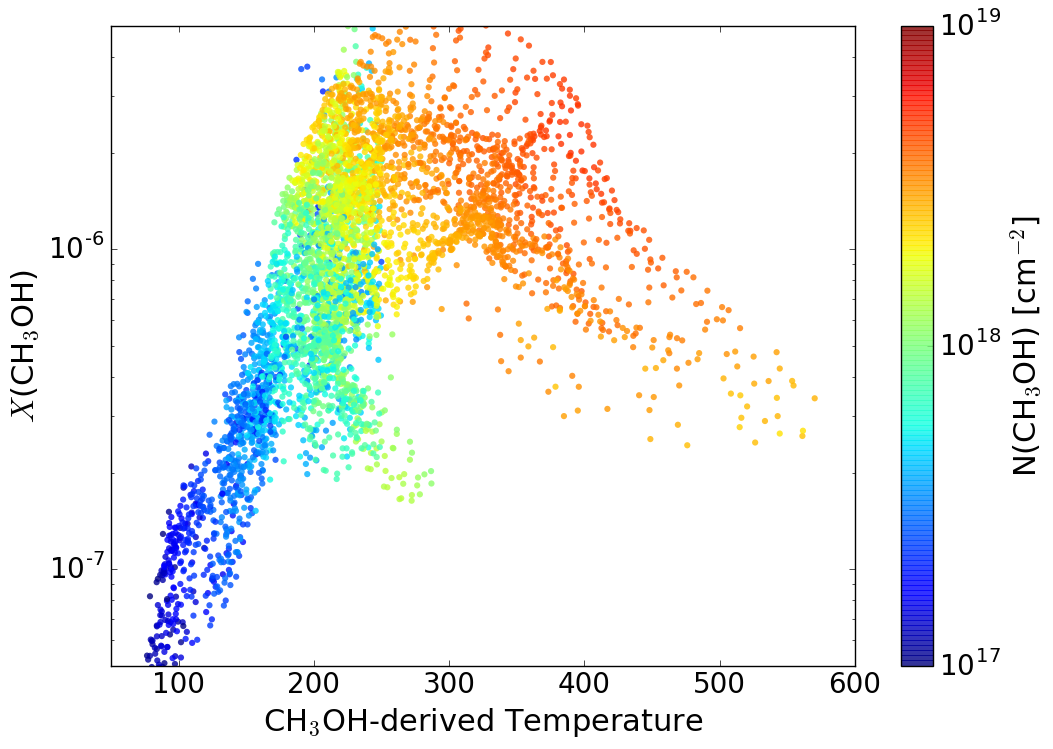}
{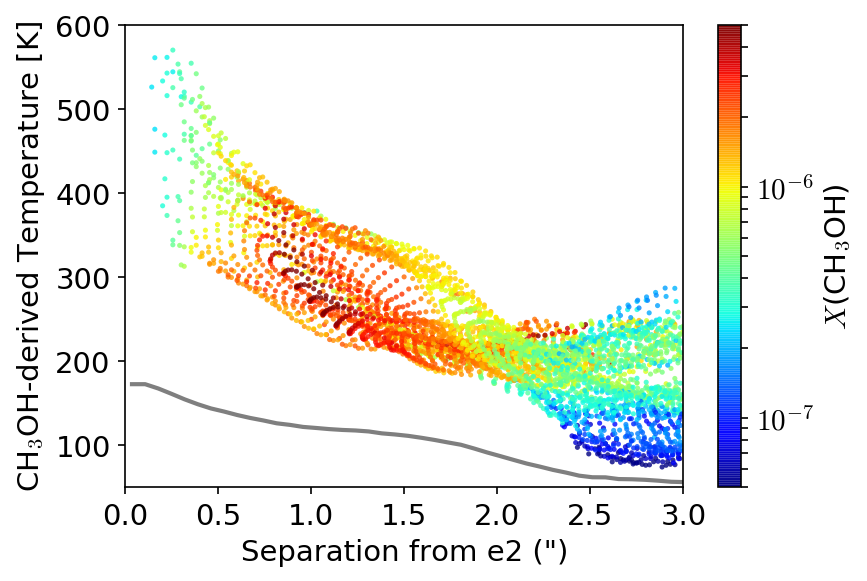}
{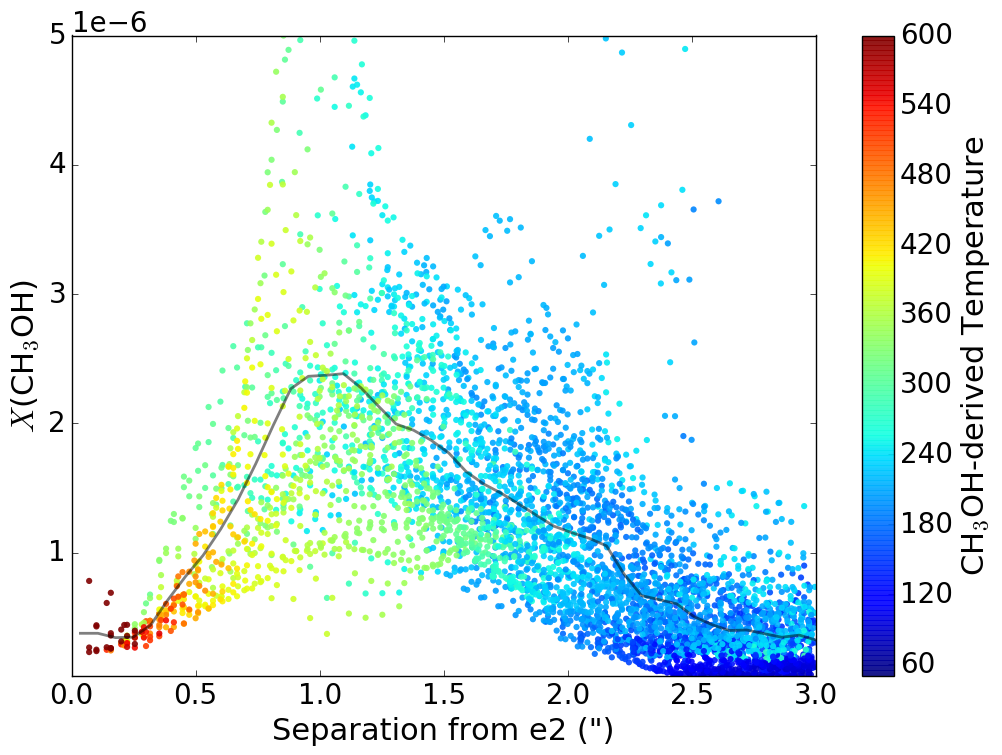}
{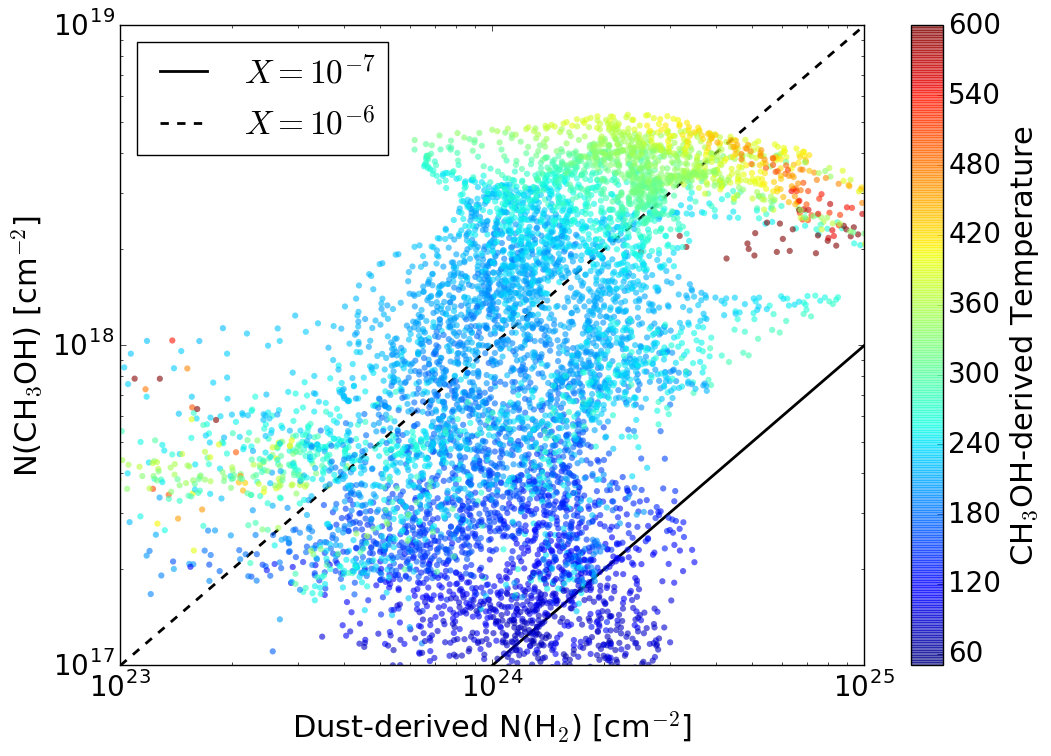}
{Comparison of the \methanol temperature, column density, and abundance.
(a) The relation between temperature and abundance.  There is a weak correlation,
but most of the high abundance regions are at high temperatures.
(b) Temperature vs distance from e2e.  There is a clear trend toward higher
temperatures closer  to the central source.
The gray line shows the azimuthally averaged peak $T_B(\methanol~
8_{0,8}-7_{1,7})$, which gives an approximate lower limit on the highest
temperature at each line of sight.
(c) Abundance vs distance from e2e.  The apparent dip at $r<1\arcsec$ is
somewhat artificial because it is driven by a rising dust emissivity that
corresponds to an increasing optical depth in the dust.  The \methanol column
in this inner region is likely to be underestimated.   The gray
line shows the azimuthally averaged abundance.
(d) \methanol vs dust column density.  }
{fig:ch3ohtemX}

A few features illustrate the effects of thermal radiative feedback on the gas.
The temperature jump starting inward of  $r\sim1.5\arcsec$ (8100 AU; Figure
\ref{fig:ch3ohtemX}b) is
substantial, though the 100-200 K floor at greater radii is likely
artificial\footnote{The low-J transitions have significant optical depth
across the whole region, but in the inner part of the core, the temperature
measurement is dominated by the high-J transitions, which give a long
energy baseline for the fit.  In the core exterior, the high-J lines are
not detected, so the (possibly optically thick) low-J lines determine
the temperature fit, which results in much lower accuracy and greater
potential bias.}.
There is an abundance enhancement at the inner radii, but in the plot it
appears to be a radial bump rather than a pure increase.  The abundance
enhancement is probably real,
and is a factor of $\sim5-10\times$.  The inner abundance dip
is caused by two coincident effects: first, the \methanol column becomes underestimated
because the low-J \methanol is \emph{self}-absorbed, and second, the dust
becomes optically thick, blocking additional \methanol emission, though this
latter effect is somewhat self-regulating since it also decreases the inferred
dust column (the denominator in the abundance expression).

\subsection{Radial mass profiles around the most massive cores}
\label{sec:radialmass}
In Figure \ref{fig:hmradprof}, we show the radial mass profiles extracted from the
three high-mass protostellar cores in W51: W51 North, W51 e2e, and W51 e8.
The plot shows the enclosed mass out to $\sim1\arcsec$ (5400 AU).  On larger
spatial scales, the enclosed mass rises more shallowly, indicating the end of the
core.

All three sources show similar radial profiles.  Figure \ref{fig:hmradprof}b
shows $M(<R)$ using $T_{dust}=T_{\scriptsize{\methanol}}$, which is a reasonable
approximation of the mass profile (though it is likely a lower limit on
the mass; see \S \ref{sec:ch3ohtem}).  Assuming $T_{dust}=40$ K, approximately
the hottest measured dust temperature in the region from Herschel SED fits,
gives a mass upper limit in each core that is up to 3000 \msun within a compact
radius of 5400 AU (0.03 pc).  If the observed dust were all at 600 K instead of
40 K, the mass would be $17\times$ lower, $\sim100-200$ \msun, which we treat
as a strict lower bound as it is unlikely that the dust at more than
$r\gtrsim1000$ AU from the central heating source is so warm.

\FigureTwo
{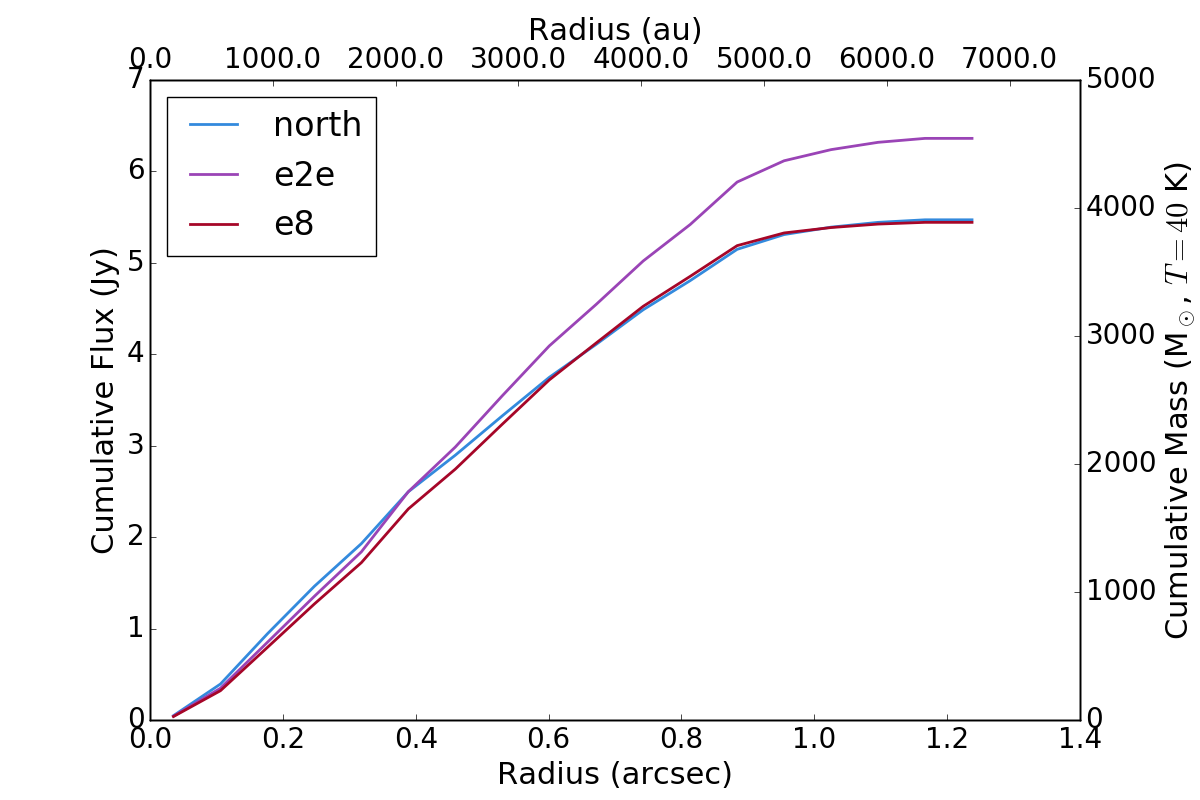}
{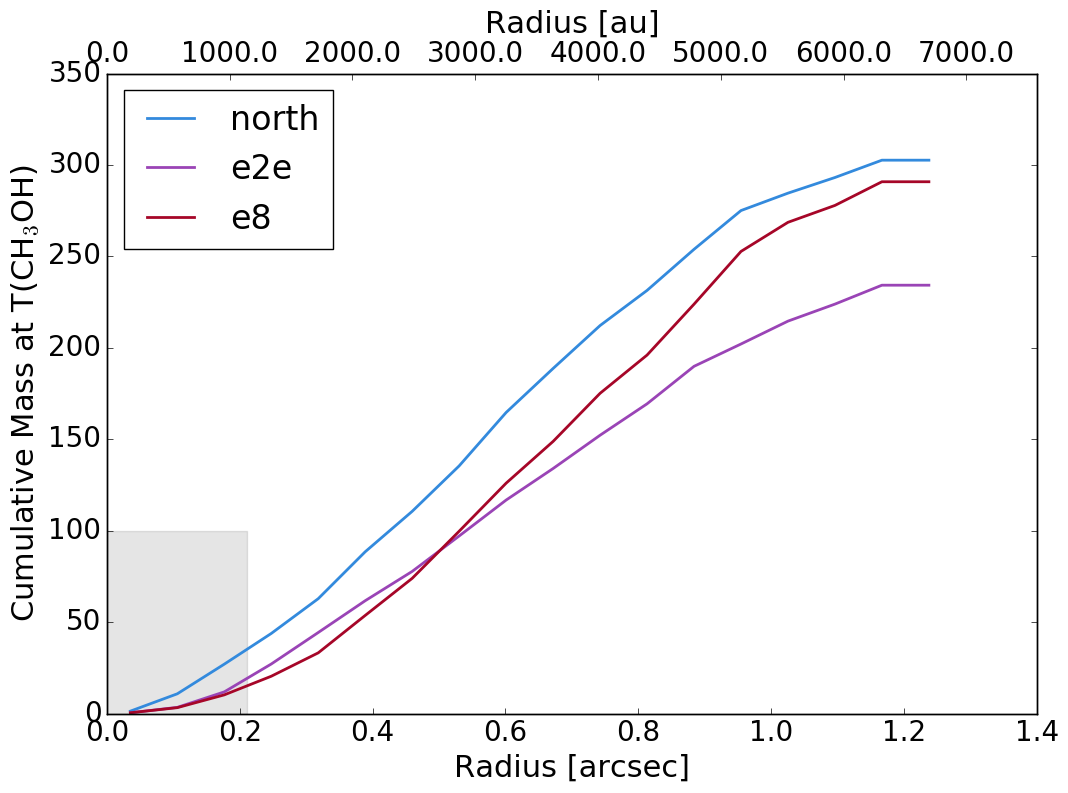}
{The cumulative (a) flux density radial profile and (b) mass radial profile
centered on three massive protostellar cores.  The cores share similar profiles
and are likely dominated by hot dust in their innermost regions, but they are
more likely to be dominated by cooler dust in their outer, more massive
regions.  The cumulative mass distribution inferred from assuming the gas is at
a constant temperature $T=40$ K (the approximate Herschel dust temperature on a
$\sim0.5$ pc scale) in (a) should be interpreted as an upper limit.  In (b), we
use the temperature map computed from \methanol in Section \ref{sec:methanol};
this plot is at least qualitatively more realistic, though it is subject to
many uncertainties discussed in \S \ref{sec:methanol}.  The grey rectangle
highlights the beam size.}
{fig:hmradprof}{1}{8cm}

\subsection{Gas kinematics around the most massive cores}
\label{sec:kinematics}
The gas motion around the massive cores is traced consistently by many species.
\methanol has some of the brightest and most isolated (i.e., not confused with
other species) lines, so we show the kinematic structure of two
moderately excited \methanol lines for the e2e MYSO core in Figure
\ref{fig:kinematicse2} (similar plots for e8 and North are showin in the
Appendix, figures \ref{fig:kinematicse8} and \ref{fig:kinematicsnorth}).

\FigureOneCol{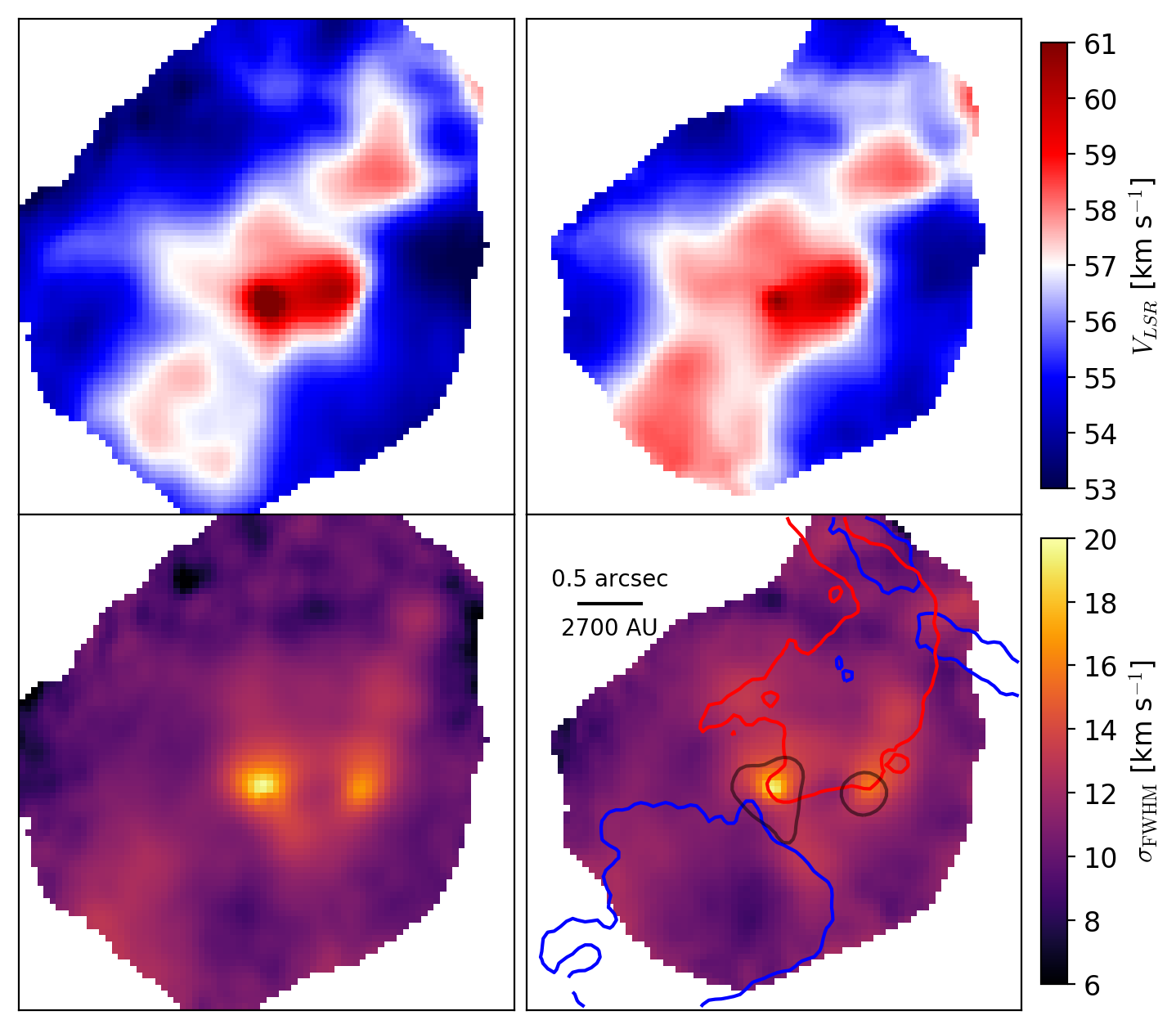}
{Moment 1 and 2 maps of the W51 e2 core over the velocity
range 45-70 \kms.  The left column shows \methanol $8_{0,8}-7_{1,6}$
and the right shows \methanol $10_{2,9}-9_{3,6}$.  We show two
lines with similar excitation but separated substantially in frequency
to demonstrate that the moment maps are not contaminated by nearby lines.
The lower-right panel has contours of high-velocity $^{12}$CO 2-1 overlaid
to show the general location of the outflows.  The black contour shows
the mm continuum at 0.15 Jy beam$^{-1}$.}
{fig:kinematicse2}{1}{9cm}

There are two notable common features in these maps. First, there is no clear
sign of systematic motion, particularly rotation, in any of them.  Second, they
have velocity dispersion uniformly much greater than the sound speed.  We
determined temperatures in Section \ref{sec:ch3ohtem}, giving
$c_s\sim0.5$ \kms.  With velocity dispersions $\sigma_{FWHM}\approx5-15$ \kms,
the gas is typically moving at Mach numbers $\mathcal{M}\approx10-30$.

In e2e, the spatial locations of both the blue and red lobes of the CO outflow
are redshifted in the dense gas, while the rest of the core is blueshifted.
The outflow axis shows some of the lowest velocity dispersion in the e2e core,
suggesting that the outflow is not responsible for driving the observed
velocity dispersion.

An increase in the velocity dispersion toward the central protostar is clearly
seen in both e2e and e8, though the opposite is seen in the north.  We caution,
though, that the high velocity dispersion toward the central source is likely
to be affected by contamination from other molecular species.  There are many
more complex species detected in the central pixel than elsewhere in these
cores.

The velocity structure around these sources is more complex than illustrated by
the moment maps alone.  For example, to the northeast of e8, there is a gap in
the emission of many lines accompanied by a double-peaked profile, hinting at
the presence of an expanding bubble.  Multiple velocity components are seen
along many lines of sight around each core.

The overall appearance of these cores suggests that many different gas flows
(both inflow and outflow) are intersecting and interacting.  While the high
velocity dispersion suggests that the gas may be highly turbulent, it remains
possible that the linewidths come from unresolved substructure in coherent
flows such as infall along a wide range of angles.

\subsubsection{Signs of infall toward e2?}
\citet{Zhang1997a}  reported a measurement of fast infall onto e2.
However, these measurements were performed with 2-3 \arcsec resolution and the
P Cygni profiles actually consist of a blend between absorption toward the
centimeter-bright e2w \hii region and emission from the \emph{extended} e2e hot
core.  

\citet{Goddi2016a} resolved the absorption toward e2w and emission
toward e2e and showed a velocity difference $v_{e2e}-v_{e2w} = \Delta v \sim
-0.9$ \kms, which is consistent with infall toward e2e of $v_{in}\sim1.2$ \kms
at $r=4000$ AU assuming an inclination of the flow $i=45\deg$.  They noted
that the lower-excitation \ammonia lines have redshifted wings relative to
the higher-excitation lines, indicating infall at up to $\lesssim 2$ \kms.

\citet{Shi2010a} measured an infall velocity toward e2e of $v=2.5$ \kms, 
but their adopted systemic velocity is inconsistent with measurements
using radio lines \citep{Goddi2016a}.  If the \methanol or \ammonia centroid
velocities from \citet{Goddi2016a} are adopted, the offset noted by
\citet{Shi2010a} is not significant and there is no clear sign of infall.

A likely reason for the inconsistent conclusions about infall in the 0.85 mm
and 1.3 cm data of \citet{Shi2010a} and \citet{Goddi2016a}, respectively, is
the optical depth of the central core in e2e.  In the presence of rapid infall,
optically thick dust would hide emission from background blueshifted material,
suppressing the inverse P Cygni profile.  Bright continuum also reduces the
line-to-continuum ratio, making the theoretically highest-velocity features
closest to the star more difficult to detect.  While cold foreground material
should still be readily detectable, such material is expected to be inflowing
at low velocities anyway.

Indeed, in our data, deep absorption is seen in the low-J lines of
\formaldehyde and \methanol, and these lines have velocity centroids
$v\sim56-57$ \kms, consistent with the centroid velocity of the central core.
The central core is at rest relative to the bulk molecular cloud.

Looking at the line profiles of some low-J lines, such as \methanol
$4_{2,2}-3_{1,2}$, it is tempting to interpret the observed double-peak
profiles as infall signatures.  However, the overall structure of the line
velocities as a function of excitation does not support this interpretation.
If material is infalling toward a central heating source and getting denser
closer to the center, the lines with the highest upper-state energy
levels and greatest critical densities should exhibit the highest
velocities, which is not observed.  Instead, we observe redshifted
wings in the lowest-excitation components (Figure \ref{fig:ch3ohe2epeaks}).
This pattern does not rule out infall, but it cannot be interpreted
so straightforwardly.

Toward both e8 and North, the same observational caveats about the optical
depth of the millimeter continuum apply.  We conclude that our ALMA data do
not provide an unambiguous signature of infall, but this nondetection is caused
by observational limitations rather than a lack of infall motion.

\subsubsection{Are there disks around the MYSOs?}
\label{sec:disks}
We find no direct evidence of disks in the gas kinematic data.  The presence of
outflows (Appendix \ref{sec:outflows}) hints that there are accretion disks,
but measurement of a Keplerian rotation curve is necessary to definitively
identify a disk.

The characteristic signature of a Keplerian disk is a velocity profile that
rises from low in the outskirts, almost certainly smaller than the turbulent
velocity dispersion, to a large value near the center.  For a 100 \msun star,
at 1000 AU, the expected circular velocity is only 9.5 \kms, which is comparable
to the velocity dispersion we observe across most of the core; any smaller
star would support a proportionately smaller orbital velocity.  Even if there
is an extraordinarily massive star at the center of each of these cores, we
would not expect a clear disk signature to be detectable anywhere except the
central pixel because of the high turbulent velocity dispersion.  As noted
above, though, the central pixel is the most chemically complex and confused
region, so the line width measurements at that location are unreliable.

Despite these limitations, many authors have reported the detection of
``rotating toroids'' or ``Keplerian-like'' rotation curves around MYSOs
\citep{Johnston2015a,Chen2016b,Ilee2016a,Zapata2015a,Hunter2014a,Sanchez-Monge2013a,Moscadelli2014a}.
Following these authors, we examine the velocity profile perpendicular to the
observed outflow direction in e2e.  Figure \ref{fig:e2epvdiagrams} shows
position-velocity diagrams of a \methanol and a \methylformate line extracted
along PA $=35^\circ$, perpendicular to the $^{12}$CO 2-1 outflow.
While there
is velocity structure, there is no obvious line broadening at the source
center, nor is there any obvious gradient indicating a rotating structure.  The
line-to-continuum ratio also drops, which could be an indication that the dust
is becoming optically thick, preventing us from detecting the
high-velocity gas.  Indeed, an optically thick inner disk at 1 mm is theoretically
expected \citep{Forgan2016a,Klassen2016a}, so it is not surprising that we fail
to detect high-velocity features associated with a disk.  Our result fits with
\citet{Maud2017a} and Cesaroni et al (in prep), who similarly failed to find
disk signatures around O-type ($>10^5$ \lsun) YSOs.

\FigureTwo
{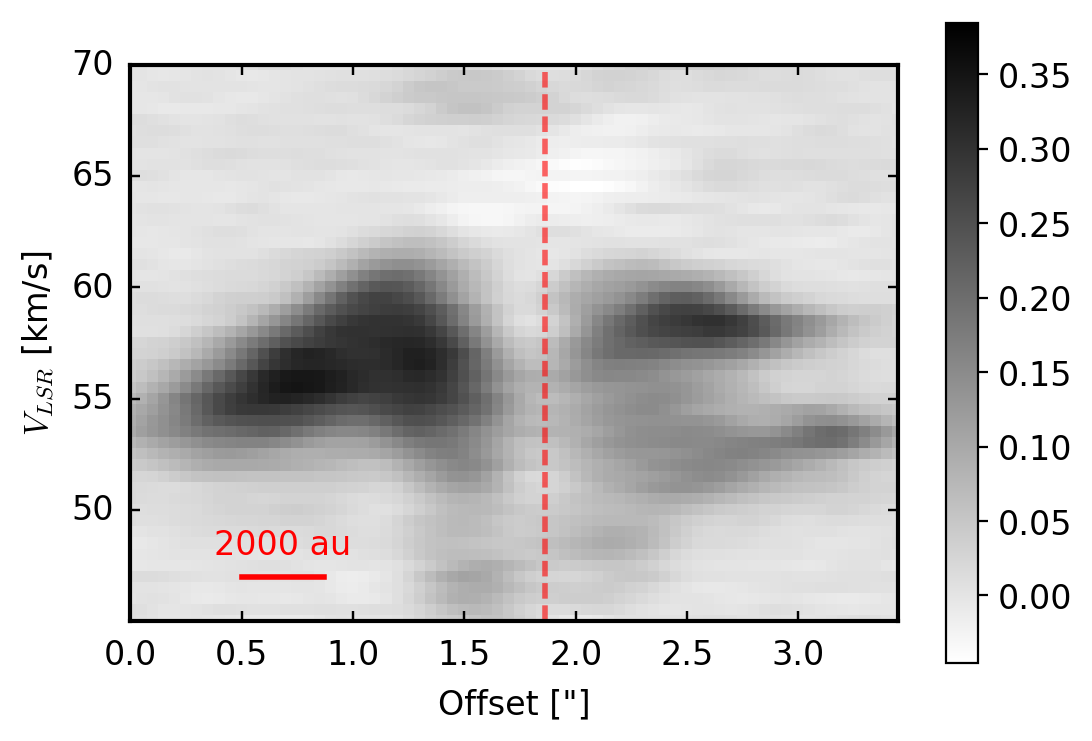}
{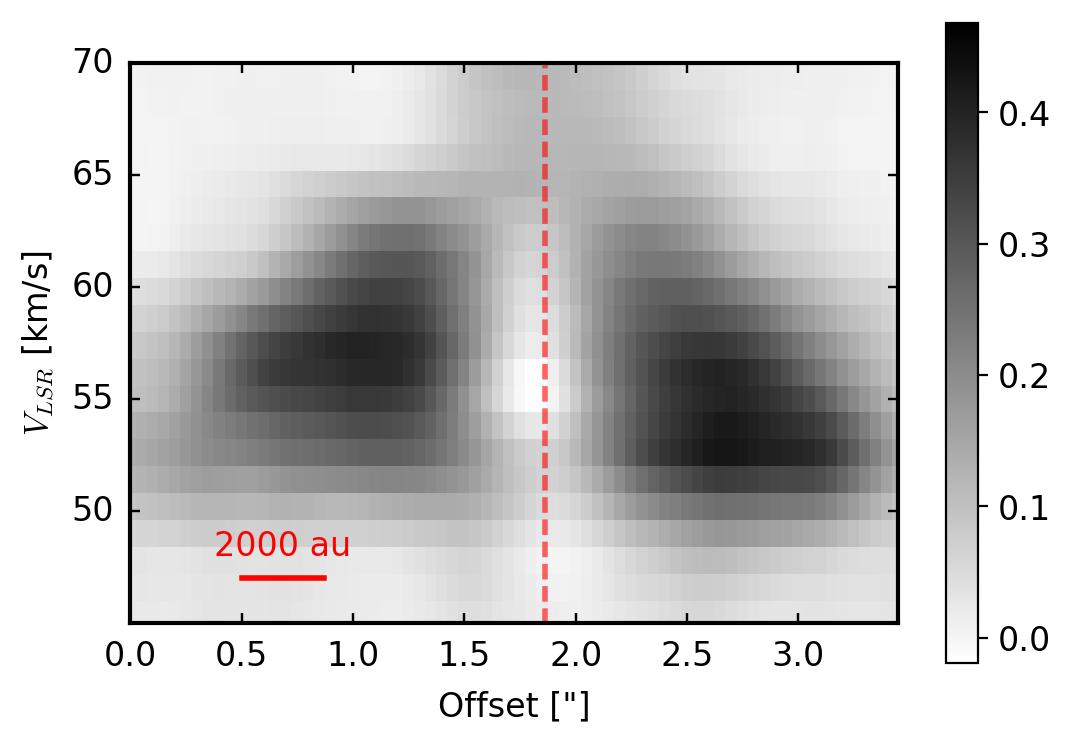}
{Position-velocity diagrams of the W51 e2e core taken at PA=$35\deg$,
perpendicular to the main outflow axis.  The vertical dashed line shows the
position of peak continuum emission. The lines are (a) \methylformate
$17_{3,14}-16_{3,13}$ 218.28083 GHz and (b) \methanol $8_{0,8}-7_{1,6}$
220.07849 GHz.  The spectral resolution is 0.5 \kms in (a) and 1.2 \kms in (b).
The data have been continuum subtracted, highlighting the low line-to-continuum
contrast near the source.  The \methylformate line was selected because the
molecule approximately traces the same material as \methanol, but the pair of
\methylformate J=17 lines were in our high spectral resolution window, so the
velocity substructure can be seen.
}
{fig:e2epvdiagrams}{1}{9cm}

We repeated this exercise for e8 and North, though their outflow directions
are more ambiguous, and we found similar features (i.e., a lack of any clear
rotation signature) at all plausible position angles.

While we failed to detect clear disk signatures toward these MYSOs, the
outflows driven from them suggests that disks are indeed present.  We suggest,
therefore, that the disks are either too small ($r<1000$ AU) or too optically
thick at 1 mm to be detected in our data.

\subsection{Ionizing vs non-ionizing radiation}
\label{sec:nonionizingradiation}
The formed and forming protostars are producing a total $\gtrsim10^7$ \lsun of
far infrared illumination \citep{Ginsburg2016b}.  This radiation heats the
cloud's molecular gas, affecting the initial conditions of future star
formation.

The ionizing radiation in W51 was discussed in detail in \citet{Ginsburg2016b}.
Ionizing radiation affects much of the cloud volume, but little of the
high-density prestellar material:  there is no evidence of increased molecular gas
temperatures in the vicinity of \hii regions.  While in Section
\ref{sec:chemistrymaps} we identify chemically enhanced regions as those where
radiative feedback has heated the dust and released ices into the gas phase, no
such regions are observed surrounding the compact \hii regions.

The chemical maps shown in Section \ref{sec:chemistrymaps} show the volumes of
gas clearly affected by newly forming high-luminosity stars.  The
\methanol-enhanced region around W51e2e extends 0.04 pc, or 8500 AU (see Section
\ref{sec:ch3ohtem}). Other locally enhanced species, especially the nitrogenic
molecules HNCO and \formamide, occupy a smaller and more asymmetric region
around e2e and e2w (Figure \ref{fig:e2methanolhnco}).  These chemically enhanced
regions are most prominent around the weakest radio sources or regions with
no radio detection; they are most likely heated by direct infrared radiation
from these sources.

The luminosities of the other \uchii and \hchii regions throughout the observed
area are high enough, $L\gtrsim10^4$ \lsun, to produce chemically enhanced
molecular envelopes if they were surrounded by dense ($n(\hh)\gtrsim10^4$
\percc) molecular gas.  Since few such regions are detected, we conclude that
these \hii regions are not surrounded by such high-density gas but instead are
traveling through a lower-density medium.

There are two counterexamples, e2w and d2, which are extremely compact \hchii regions
that  exhibit some enhanced molecular emission around them, though with a smaller
radial extent than the hot cores.  For e2w, it is difficult to estimate the extent
of the enhanced region, since e2w is embedded in a common core with e2e, but we can
set an upper limit of $1\arcsec\approx5400$ AU.  Around d2, the extent is
$0.6\arcsec\approx3000AU$.  Both of these objects likely turned on their ionizing
radiation (contracted onto the main sequence) only recently.  The enhanced
molecular emission is either from the remnant core that was heated during the
star's pre-ionizing phase, or it is presently being heated with photons that
have been absorbed and re-emitted as non-ionizing radiation.

\FigureTwo
{f24}
{f25}
{Image of \methanol $8_{0,8}-7_{1,6}$ (red), HCNO $10_{0,10}-9_{0,9}$ (green), and 225 GHz
continuum (blue) toward (a) W51e2 (b) W51e8.  The contours show Ku-band radio continuum
emission tracing the \hii regions (a) W51 e2w and (b) W51e1, e3, e4, e9, and
e10.  The \methanol emission is relatively symmetric around the high-mass
protostar W51 e2e and the weak radio source W51 e8, suggesting that these
forming stars are responsible for heating their surroundings.  By contrast, the
\hii regions do not exhibit any local molecular brightness enhancements (except
e8), indicating that the \hii regions are not heating their local dense
molecular gas.}
{fig:e2methanolhnco}{1}{8.5cm}

\FigureOneCol
{f27}
{Image of \methanol $8_{0,8}-7_{1,6}$ (red), HCNO $10_{0,10}-9_{0,9}$ (green),
and 225 GHz continuum (blue) toward  North, as in Figure
\ref{fig:e2methanolhnco}.  The contours show Ku-band radio continuum emission
tracing the diffuse IRS 2 \hii region.}
{fig:northmethanolhnco}{1}{8.5cm}

\subsection{Outflows}
\label{sec:mainoutflows}
While many outflows were detected, we defer their discussion to Appendix
\ref{sec:outflows} because the details of these flows is not relevant to the
main point of the paper.  However, we note that out of the dozen or so outflows
detected, \emph{none} come from radio continuum sources (\hii regions).  All
outflows that have a clear origin come from millimeter-detected,
centimeter-faint sources, suggesting that these sources are accreting molecular
material and are not emitting ionizing radiation.

\section{Discussion}
\label{sec:discussion}
\subsection{The scales and types of feedback}
\label{sec:feedbackscales}
The most prominent features of our observations are the warm, chemically
enhanced regions surrounding the highest dust concentrations, and the
corresponding \emph{lack} of such features around the ionized nebulae.  This
difference implies that the immediate star formation process - that of gas
collapse and fragmentation from a molecular cloud - is primarily affected by
feedback from stars that are presently accreting and therefore emitting most of
their radiation in the infrared, \emph{not} from previous generations of
now-exposed main-sequence stellar photospheres.

On the scales relevant to the fragmentation process, i.e., the $\sim0.1$ pc
scales of prestellar cores, this decoupling can be explained simply.  Stellar
light is produced mostly in the UV, optical, and near-infrared.  As soon as a
star is exposed, either by consuming or destroying its natal core, that light
is able to stream to relatively large ($\gtrsim1$ pc) scales before being
absorbed.  At that point, the stellar radiation is poorly coupled to the scales
of direct star formation.  By contrast, stars embedded in their natal cores
will have all of their light reprocessed from UV/optical/NIR to the far-IR
within a $<0.1$ pc sphere, providing far-infrared illumination capable of
heating its surroundings.

The different effects of ionizing vs thermal radiation can be seen directly in
the three main massive star forming regions, e2, e8, and North.  Figures
\ref{fig:e2methanolhnco} and \ref{fig:northmethanolhnco} show both the
highly excited warm molecular gas in color and the free-free emission from
ionized gas in contours.  As described in Section
\ref{sec:nonionizingradiation}, the spatial differences indicate that the
ionizing radiation sources - the exposed OB stars - have little effect on the
star-forming collapsing and fragmenting gas.

The low impact of short-wavelength photospheric radiation on collapsing gas
suggests that second-generation star formation is relatively unaffected by its
surroundings.  Instead, the stars of the same generation - those currently
embedded and accreting - have the dominant regulating effect on the gas
temperature.  To the extent that gas temperature governs the IMF, then, the
formation of the IMF \emph{within clusters} is therefore predominantly
self-regulated, with little external influence.

\subsubsection{Hot core chemical structure}
\label{sec:cheminterp}
In Section \ref{sec:chemistrymaps}, we showed regions with enhanced emission
in a variety of complex chemical species over a large volume.  While it is
not generally correct to conclude that enhanced emission indicates enhanced
abundance, the additional analysis of the \methanol abundance 
in Section \ref{sec:ch3ohtem} suggests that there is a genuine enhancement in
complex chemical abundances toward these hot cores.

We have not performed a detailed abundance analysis of multiple species, but we
nonetheless suggest that these sharp-edged bubbles around the hot cores
represent desorption (sublimation) zones in which substantial quantities of
grain-processed materials are released into the gas phase.  The relatively
sharp edges likely reflect the radius at which the temperature exceeds
the sublimation temperature for each species \citep{Garrod2006a,Green2009a},
though some species may appear at temperatures above or below their sublimation
temperature if they are mixed into ices that have a different sublimation
temperature.  Other species may also form in the high-density, high-temperature
gas at smaller radii, such as the nitrogenic (HNCO, \formamide) species we
detected, suggesting that the cores are dominated by sublimated ices from
$R\sim2000-5000$ AU and by species formed in the gas phase at $R\lesssim2000$
AU.

Most of the lines identified in the hot cores e2e, e8, and North are also
present in a lower-luminosity hot core, ALMAmm14.  However, their extent is
greater toward the more luminous sources.  This difference suggests that an
examination of the relationship between the luminosity of the protostars and
the extent of their chemically enhanced zones will be useful for identifying
 very massive protostars in other regions.

\subsubsection{Outflows}
\label{sec:outflowdiscussion}
While the outflows described in Appendix \ref{sec:outflows} are impressive and
plentiful, they are obviously not the dominant form of feedback, as their area
filling factor is small compared to that of the various forms of radiative
feedback.  A low area filling factor implies a substantially smaller volume
filling factor and therefore a lower overall effect on the cloud.  However,
these outflows likely do punch holes through protostellar envelopes and the
surrounding cloud material, allowing radiation to escape.

The detection of widespread high-J \methanol emission around the highest-mass
protostars suggests that the use of \methanol as a bulk outflow tracer as
suggested by \citet{Kristensen2015a} is not viable in regions with forming
high-mass stars.  While mid-J \methanol emission associated with
the outflow (e.g., the J=10-9 transition) is detected, it is completely
dominated by the general `extended hot core' emission described in Section
\ref{sec:chemistrymaps}.

None of the outflows originate in \uchii or \hchii regions.  While a clear
origin cannot be determined for all of the outflows, it is clear that no cm
continuum sources lay at the base of any.  The lack of molecular outflows
toward these sources implies that they are accreting at most weakly.

\subsection{The accreting phase of high-mass star formation in W51 is not ionized}
\label{sec:accretionandoutflows}
The strong outflows observed around the highest-mass forming stars, e2e, e8,
and North are clear indications of ongoing accretion onto these sources.
However, the bright \hii regions, including e2w, e1, and d2, all lack any sign
of an outflow or a surrounding rotating molecular structure.  Most of these
sources lack any surrounding molecular material at all.

Some models of high-mass star formation suggest that accretion continues
through the ionized (\hii region) phase \citep{Keto2002b,Keto2003a}.  The lack
of molecular material around the majority of the compact \hii regions in W51
suggests instead that most of the accretion is done by the time an \hii region
ignites.  Additionally, the W51e2 source, which was invoked as an example of an
ionized accretion flow in \citet{Keto2008a}, is  resolved into the e2e
hot core driving an outflow and the e2w \hchii region that is not, so the
evidence for ionized accretion onto e2w is diminished.

There is one clear example of a \hchii region surrounded by molecular gas in
our sample, the source d2.  However, it does not have an associated molecular
outflow, so there is no direct evidence of ongoing accretion.
Following the discussion of this source in Section \ref{sec:w51d2}, we suggest that
this core contains an early B ($M~8-20$ \msun) star that has just recently
reached the main sequence, making it older and less massive than the three main
MYSOs.

\subsection{The accreting protostars in the massive hot cores}
\label{sec:stellarproperties}
In Sections \ref{sec:W51e2e} and \ref{sec:w51e8andnorth}, we noted that the
\emph{lower limit} luminosities for the three most massive cores correspond to
early B-type photospheres.  Such stars should emit enough radiation to ignite
luminous compact \hii regions. 

If we assume a uniform, spherical \hii region (a Strömgren sphere), we can
obtain the Lyman continuum luminosity $Q_{\textrm{lyc}}$ required to produce our
$2-\sigma$ upper limit for e2e:
\begin{equation}
    Q_{\textrm{lyc}} = \frac{4}{3} \pi R^2 \alpha_B EM
\end{equation}
where $\alpha_B = 3\ee{-13}$ cm$^{3}$ s$^{-1}$, the emission measure $EM$ has
units cm$^{-6}$ pc, and $Q_{\textrm{lyc}}$ has units \pers.  We infer the emission measure
using Equation 4.60 of \citet{Condon2007a} and inverting their Equation 4.61 to
get

\begin{equation}
EM = -3.05\ee{6} 
\left(\frac{T_e}{10^4\mathrm{K}}\right)^{1.35}
\left(\frac{\nu}{\mathrm{GHz}}\right)^{2.1}
\ln\left(1-\frac{T_{B,\nu}}{T_e}\right)
\end{equation}
using numerical constants
from \citet{Mezger1967a}.  If we use the centimeter
continuum beam size (FWHM) as the radius, the resulting $Q_{\textrm{lyc}}<6\ee{45}$ \pers 
is well below the lowest tabulated $Q_{\textrm{lyc}}$ in \citet{Vacca1996a} and
\citet{Sternberg2003a},
$Q_{\textrm{lyc}}(\mathrm{B0.5V}) = 5\ee{47}$ \pers.  Using the stellar parameters from
\citet{Pecaut2013a} and assuming the stars are pure blackbodies, the upper
limit Lyman continuum implies a star later than about B2, or $M<7$ \msun.
If instead we assume the \hii region is optically thick, the implied radius is
about $R_{\scriptsize{\hii}}<100$ AU, which gives a density limit but not
a luminosity limit.  We favor the optically thin assumption because the outflow
should provide an escape outlet for the ionized gas to expand into.

The stellar luminosity inferred for e2e from the dust continuum is
$L>2.3\ee{4}$ \lsun (Section \ref{sec:W51e2e}), which corresponds to at least a
B1V star, or $M>11$ \msun.  The upper limits on the Lyman continuum luminosity
and lower limits on the bolometric luminosity are similar for e8 and North.
The contradiction between the two luminosity limits implies that the accreting
stars are not yet on the main sequence.

This result is surprising, since the Kelvin-Helmholtz timescale for a massive
star is extremely short, $\tau_{KH} < 5\ee{4}$ yr for a $M>20$ \msun star.  The
short contraction timescale suggests that these stars should reach the main
sequence and begin ionizing their surroundings while they are still accreting
\citep{Zinnecker2007a}.

While it is possible that we have caught three stars at a nearly simultaneous,
extremely short-lived phase in their evolution, it is also possible that they
are contracting more slowly than the Kelvin-Helmholtz timescale.  The large
observed mass reservoir suggests that high accretion rates are likely, and the
bright molecular outflows show that accretion is proceeding vigorously.  Rapid
accretion, and in particular rapid and \emph{variable} accretion, can change
the properties of the underlying star, bloating the star and reducing its
effective photospheric temperature \citep{Hosokawa2009a, Smith2012h,
Hosokawa2016a}.  Such stars can achieve radii $R\gtrsim200~\rsun \sim 1$ AU
while retaining photospheric temperatures $T\lesssim5000$ K.  Bloated central
stars would therefore explain why they do not produce \hii regions.

An alternative possibility is that the high accretion rates have created a
quenched \hii region \citep{Walmsley1995b,Osorio1999a,Keto2006a}.  In the
spherically symmetric version of this scenario, the accretion rate is faster
than the ionization rate, such that there is always fresh neutral material 
 at the surface of the star.  The critical rate for \hii region quenching
is small, $\dot{M} \sim 4\ee{-6}$ \msun \peryr for a B0 star, so it is likely
that, even if the central star has a hot photosphere, it is not capable of
driving an expanding \hii region.  The main reason to disregard this scenario
is the assumption of spherical symmetry: if the accretion is proceeding via a
disk, as evidenced by the presence of outflows, there ought to be a substantial
fraction of the stellar surface that is not directly accreting and therefore is
not quenched.  If there is a disk and an ionizing photosphere, there should be
an expanding bipolar \hii region \citep{Keto2006a}.  The lack of such a feature
suggests that the stellar photosphere is not emitting ionizing photons.

\subsubsection{Multiplicity}
High-mass stars preferentially form in multiples \citep{Zinnecker2007a}.  One
explanation for the low ionizing luminosity but high total luminosity observed
in the massive hot cores would be the presence of many moderate-mass
($M\lesssim8$ \msun) stars forming together.  We can rule out this situation,
since $\sim10-15$ main-sequence $M\sim6$ \msun stars would be required within a
tiny volume ($r<1000$ AU) to produce the observed total luminosity lower limit
while staying under the Lyman continuum upper limit.

We do not see any clear signs of multiple outflows toward any of the hot cores,
so if there are multiple stars forming, it is possible they are accreting from
a common disk.
Given the mass reservoirs available, however, there is little reason
to believe that multiplicity is the \emph{only} explanation for the
high-luminosity, low-ionizing luminosity sources: if multiples are forming, we
should expect at least one of them to reach O-star mass.  In short, on the
scales we can probe, multiplicity does not produce any obvious observational
effects.

\subsection{Fragmentation: Jeans analysis}
\label{sec:fragmentation}
Fragmentation is one of the critical problems in high-mass star formation.
Assuming typical initial conditions for molecular clouds, with temperatures of
order 10 K, gas is expected to fragment into sub-solar-mass cores, preventing gaseous
material from accreting onto single high-mass stars \citep{Krumholz2015d}.
Even after high-mass stars successfully form, further fragmentation could
halt the growth of these stars and limit their final mass
\citep{Peters2010a,Girichidis2012b}.

Turbulence provides another mechanism for gas to fragment.  In a supersonically
turbulent medium, intersecting shocks create local overdensities that can
greatly exceed the local Jeans mass even if gas temperatures are quite high.
In general, the properties of turbulence in the regime we are exploring
($T\gtrsim100$ K, $n\gtrsim10^6$ \percc) are not well explored.  If we take the
observed linewidths (\S \ref{sec:kinematics}) as purely turbulent motion, the
cores are extremely turbulent and barely bound, if at all.  However, as noted
in that section, some portion of the large linewidths can be resolved into
individual components, and the linewidths may therefore indicate that there are
many overlapping kinematically coherent flows along each line of sight.  For
the rest of this section, we assume that the gas flows are predominantly
coherent and that thermal support is therefore a relevant physical process.

Thermal Jeans fragmentation can be limited or suppressed entirely if the gas is
warm enough.  The high observed gas temperatures, $T\sim100-600$ K over
$\sim10^4$ AU, around the high mass protostars indicate that their radiative
feedback in the infrared has a dramatic effect on the gas.  The heated region
qualitatively matches that of \citet{Krumholz2006a}, who described
a core heated only by accretion luminosity down to $R=10$ AU and therefore
gave a lower limit on the total heating. 

We examined the temperature structure around the highest-mass cores in Section
\ref{sec:ch3ohtem} and the mass structure in Section \ref{sec:radialmass}.  We
put these together to measure the Jeans mass, $M_J = (\pi / 6) c_s^3
G^{-3/2} \rho^{-1/2}$
, and length, $\lambda_J = c_s G^{-1/2} \rho^{-1/2}$, 
in
Figure \ref{fig:mjeans}.  These plots show the azimuthally averaged $M_J$
and $R_J=\lambda_J/2$,
i.e., they show the Jeans mass if the medium
were of uniform density and temperature at the spherical-shell-average density
and azimuthal average temperature at each plotted radius.

The mass figure shows that the gas is stable on a beam size scale ($\sim1000$
AU), while the length figure shows that on larger scales, the gas could be
unstable to fragmentation.  On these larger scales, the Jeans length is about
the same as the hot core size, so we should not observe Jeans fragmentation on
the scale of individual beams.

Within this large reservoir, there are few detected fragments.  In our data,
within 6500 AU of W51 North, there is only one (ALMAmm35), around e2e there is
the HII region e2w and possibly two to three others between 5000 and 6500 AU,
and around e8 there are none.  Admittedly, our data are not as sensitive in the
areas immediately surrounding these cores because of dynamic range limitations.
Nonetheless, the lack of compact core detections around the most massive
sources is consistent with the interpretation that fragmentation is suppressed.

\FigureTwo
{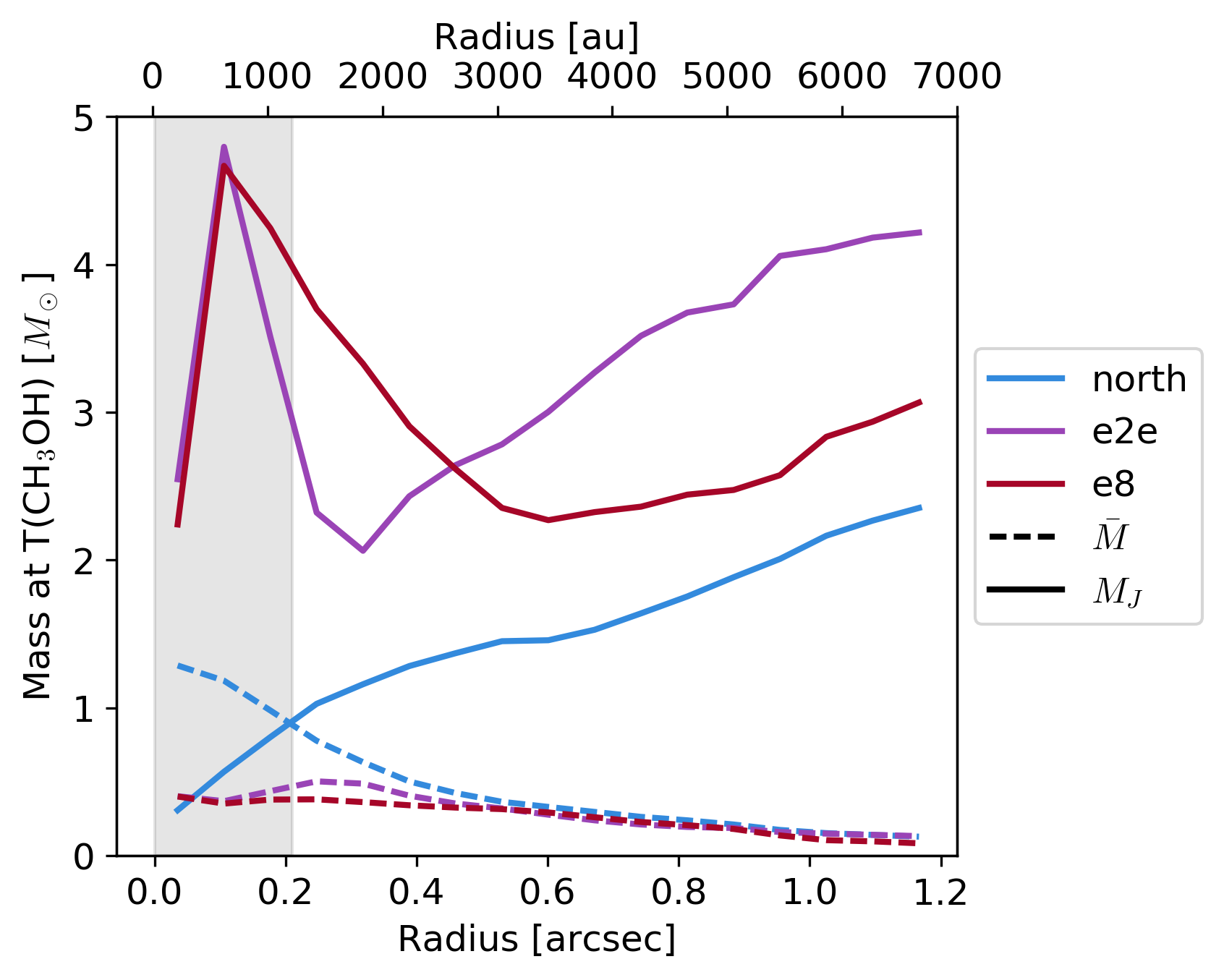}
{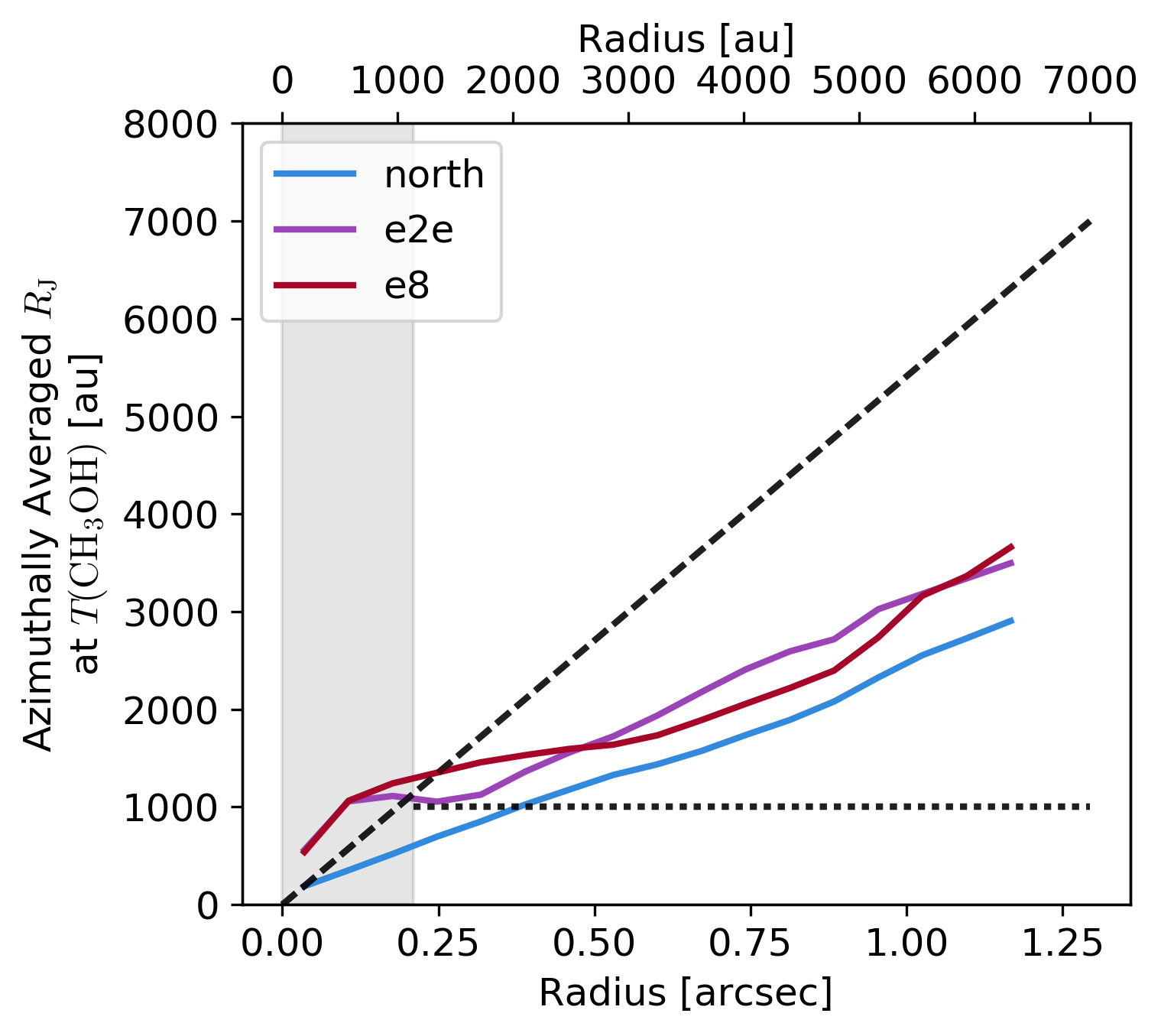}
{The azimuthally averaged Jeans mass surrounding the
three most massive cores.  We used the \methanol temperature from
\ref{sec:methanol}, Figure \ref{fig:hmradprof}b in both the Jeans mass
calculation and the dust-based mass determination.
The density used for the mass calculation is assumed to be distributed
over spherical shells.  The dashed lines show the measured mean mass per
$\sim1000$ AU beam at each radius.  Since these masses are lower than
the local Jeans mass, the gas is stable against fragmentation.
The high variation seen at small radii (below 0.2\arcsec, shaded area) is due
to sub-resolution noise.
In (b), the horizontal dotted line shows the beam scale. Above this line,
gas within a single beam is stable against Jeans fragmentation.
}
{fig:mjeans}{1}{8cm}

Given the current structure of the observed cores and their stability against
fragmentation on small scales, but susceptibility to fragmentation on larger
scales, it is unlikely that they could have existed at all without the presence
of a central heating source.  Should these $>200$ \msun cores have been present
before high-mass star formation had initiated, resting at $T\sim20$ K as in a
typical molecular cloud, they would have been subject to Jeans fragmentation on
a much smaller scale and would have formed a cluster of smaller stars
(\citet{Longmore2011a} reached the same conclusion that high-mass cores cannot
be formed with only low-mass stellar feedback as a heating source by examining
an earlier-stage high-mass star-forming region).  This prior instability
implies that the mass currently in the core had to be assembled from larger
scales while suppressing or slowing collapse on smaller scales, which is
essentially the opposite of inside-out collapse \citep{Naranjo-Romero2015a}.
In turn, such a core assembly implies that aspects of both the `competitive
accretion' and `core accretion' models may apply, with mass dumping onto a sink
source from large physical scales, yet assembling a quasi-stable core.

Our observations of warm cores with inhibited fragmentation suggests that,
during their formation process, these massive stars may be the only accreting
objects within a few thousand AU neighborhood.  Contrary to the
``fragmentation-induced starvation'' problem
\citep{Peters2010a,Peters2010c,Girichidis2012b}, in which surrounding gas
rapidly fragments and chokes off further accretion, these stars regulate their
own environment.  This `enforced isolation' is a way for massive
star formation to proceed similarly independent of the size of the parent
cloud: high-mass stars will form from similar size and shape cores whether in
an `isolated' or `clustered' region because they govern gas conditions in their
own surroundings.  The `enforced isolation' scenario says nothing about the
initial conditions that allow high-mass stars to form, but it suggests that all
high-mass protostellar cores will look approximately the same independent of
environment: a forming massive star can effectively create its own core by
heating the material that would otherwise form a small cluster.

\subsubsection{What about magnetic fields?}
A few authors \citep{Tang2009a,Tang2013a,Zhang2014a} have measured dust
polarization to get the field direction in W51e2.  \citet{Etoka2012a} used OH
masers to measure the field strength, obtaining 2-7 mG, which
likely come from high-density ($n\sim10^6-10^7$ \percc) gas \citep{Fish2003a}.
\citet{Koch2012a}
showed that even with this high field strength, the central core at
$r<1\arcsec$ is magnetically supercritical, i.e., dominated by self-gravity.
The presence of such strong magnetic fields throughout the hot cores may
suppress fragmentation and slow down accretion, though in simulations magnetic
fields have little effect on already-collapsing regions
\citep{Myers2013b,Krumholz2016a}.  At least, it is clear that magnetic fields
have not prevented the formation of the observed  massive cores.

\subsection{High-mass star formation within dense protoclusters:\\
Cooperative accretion or assembly line}
\label{sec:cooperative}
The conditions we currently observe in W51 are not ``initial conditions'', but
they are intermediate preconditions for high-mass (and maybe very-high-mass)
star formation.  We are observing a state in which high-mass stars have
recently reached the main sequence within the same volume of gas as
presently forming, rapidly accreting protostars.  These conditions are rare in
our galaxy, appearing only in a handful of high-mass star-forming regions,
specifically in the dense, dusty protoclusters (e.g., W49, Sgr B2, G333) that
are the likely precursors of $>10^4$ \msun young massive clusters (e.g., NGC
3606, Trumpler 14, the Arches).  In this section, we discuss the possible
interactions between individual high-mass protostars and the dense (in both gas
and stars) protocluster environment.

The simultaneous presence of hyper- and ultracompact \hii regions and
molecular cores indicates that there are at least two generations (likely
separated by at most a few thousand years) of massive stars forming within each
of the clusters.  How do these generations interact, and did the first
generation - the present-day \hchii and \uchii drivers, which we refer to here
as main-sequence stars - affect the second?

One possibility is that the first generation helped produce the second,
which we call ``cooperative accretion.''
If the main-sequence stars formed in the same way as the current generation of
forming stars, i.e., they heated their own surrounding cores as suggested in
the `enforced isolation' scenario above, they may have completely changed the
conditions of the parent cloud.  If we assume they reached the main sequence
before consuming all of the material they heated, and we assume that they
decoupled from the gas and stopped accreting soon after reaching the main
sequence, they must have left a substantial amount of much warmer gas behind.
Assuming that the thermal fragmentation scale is relevant for determining
the mass of new stars, the second generation would form from warmer material
and would therefore be higher mass than the first.

This toy model is analogous to the ``cooperative accretion'' mode suggested by
\citet{Zinnecker2007a}, but at a much earlier stage in the cluster development
when the gas is still molecular and dusty and therefore capable of efficient
cooling.  In the ionized cooperative accretion scenario, the most massive star
in a forming cluster will accrete the most material because its potential well
is deepest, and that star will continue to grow until it reaches a
pseudo-Eddington limit in which its own radiation produces a pressure that
reduces its effective potential, halting or reducing accretion.  At that point,
the second most massive star will dominate the accretion, and so on until the
gas is all gone.  Since we observe no direct evidence for ionized accretion in
W51 (Section \ref{sec:accretionandoutflows}), the ionized version of the
cooperative model is not likely to be significant in this particular region.

The molecular cooperative accretion model is also similar to the results of
\citet{Krumholz2011a}, in which radiative heating drove up the peak of the IMF.
In this case, though, we suggest that the affected region is smaller (not the
whole cloud).  Over the small heated region, the IMF is driven to be more
top-heavy than in the initial cooler cloud, permitting the formation of more
massive stars.

In this scenario, the highest mass stars (probably ``very massive stars'',
$M\gtrsim50$ \msun) would preferentially form within dense, clustered
environments, since suppressed fragmentation would allow the buildup of
more mass.  The first generation of stars forming from `primordial' gas
would come from a slightly different mass function than subsequent stars.  The
process would continue pushing the IMF higher until the gas is either exhausted
\citep{Kruijssen2012b,Ginsburg2016b} or expelled.

Our observations are consistent with this model given that the stars are able
to dynamically decouple from the gas.  If the previous generation were
responsible for substantial gas heating, we might expect to see warm gas
surrounding the \hchii regions.  Instead, we see these stars barely interacting
with the dense gas.  It is possible, though, that these stars are only
effective at dense gas heating \emph{before} they ignite Lyman continuum
emission and blow out cavities, and afterward they are merely uninteractive
witnesses to continued collapse \citep{Peters2010c}.  In this case, it is
critical that the main-sequence stars somehow became dynamically decoupled from
their dense cores.

A second, and contradictory, possibility is that the stars we see now formed
from a larger-scale cloud and have collapsed into their current configuration.
If the now-main-sequence stars formed somewhat earlier in a more distributed
manner and fell into a common central potential \citep[e.g., the `conveyor
belt' scenario for cluster
formation,][]{Ginsburg2012a,Longmore2014a,Walker2016a}, they would have had
little effect on the dense gas temperature.

This latter possibility may be possible to test by
measuring stellar kinematics in the forming clusters.  If the cluster of \hchii
regions in W51 e1 or the cluster of high-mass stars in W51 IRS2 originated in a
more widespread distribution and collapsed into their current configuration,
the individual stars must have formed in  effectively isolated regions.  If
instead the stars are very close to their birthplace, they would have heated the
dense gas before they ignited \hii regions and thereby established the initial
conditions for the current generation of high-mass star formation.  Comparing
the velocities of the \hii regions and the dense gas can distinguish
between these scenarios.  Current data \citep[i.e., H77$\alpha$ measurements
in][]{Ginsburg2016a} are consistent
with the more interactive cooperative accretion scenario, but the errors on the
ionized gas velocities are too large to perform a definitive test, and it is not
clear whether the recombination lines trace the stellar velocities.

\subsection{The population of faint, compact sources}
\label{sec:faintsrcs_discussion}
In Section \ref{sec:sourceid}, we reported the identification of 75 or 113
sources, depending on identification method, but we have not discussed them since.
In this section, we briefly discuss why we have chosen to ignore these sources
and offer some cautions and prospects for the interpretation of millimeter sources
in crowded star-forming regions. 

First, observational considerations have hindered the detection of `cores' down
to our theoretical mass limit $M\sim1\msun$: the  extended millimeter continuum
structure resulted in imaging artifacts that limited the dynamic range in the
vicinity of bright sources.  This effect means that we have a wildly varying
completeness to compact sources across our image, and the completeness is
difficult or perhaps impossible to measure.

Second, the nature of the sources makes determination of their masses quite
difficult.  In Appendix \ref{sec:contsrcs}, we discuss our attempts to
determine the temperature and structure of the individual sources and
characterize them based on their morphology.  Most of these sources are
centrally concentrated with high central brightness temperatures, suggesting
that they are protostellar, which in turn means that any dust-derived mass is
subject to dramatic uncertainties in the unresolved temperature structure.
More diffuse and cool sources, those that appear similar to local prestellar
cores, are theoretically detectable in our data, and some even are detected
(those with classification \texttt{-C-} in Table \ref{tab:photometry1}), but in
practice can only be seen in regions with no other substructure, especially
free-free emission, that are rare in the observed area.

These fainter sources are the most prominent members of the next generation of
stars forming in this cloud, but we do not yet have the tools needed to
understand them.  In order to probe the formation of the IMF, we need to be
able to determine the masses of stars forming in the millimeter sources, which
likely requires multiwavelength data.  

These limitations prevent us from drawing any robust conclusions from these
data.  However, they are presented in full detail in Appendix
\ref{sec:contsrcs}, including a description of the spatial structure,
characterization technique, and flux distribution.

\section{Conclusions}
\label{sec:conclusion}

We have presented ALMA 1.3 mm (227 GHz) continuum and line observations of the
high-mass protocluster W51 at $\sim0.2\arcsec$ ($\sim1000$ AU) resolution.  We
examined the three most massive forming stars and the surrounding population of
forming stars.

The key observational results include:
\begin{itemize}
    \item We identified chemically enhanced regions around high-mass protostars
        and suggest that they are radiatively heated zones in which previously
        frozen-out chemical species have desorbed into the gas phase.
    \item We measured the temperature and mass structure surrounding the three
        highest-mass cores, W51 e2e, e8, and North.  All three have masses
        $M>200$ \msun within $R\lesssim5000$ AU.  The core temperatures
        inferred from LTE modeling of \methanol are $100 \mathrm{K} < T < 600
        \mathrm{K}$, which brightness-based lower limits $T>100$ K confirm.
        Their centers are likely to be optically thick in the dust continuum.
    \item We searched for signs of disks around the central protostars in these
        hot cores, but found no clear signatures of disks or rotating
        structures perpendicular to the observed outflows on scales of
        10$^3$-10$^4$ AU.
    \item The flux density recovered in the ALMA map above a threshold of 10
        mJy \perbeam is $\sim30\%$ of that seen in single-dish data.  At this
        brightness, cold (20 K) dust would correspond to a column density
        $N(\hh)>10^{25}$ \persc, leading us to conclude that a large fraction
        of the cloud - likely all of the gas emitting at over 10 mJy \perbeam -
        is warmer than 20 K.
    \item We used three techniques for estimating the temperature of
        dust continuum sources with ALMA: limits from dust brightness
        temperature, limits from line brightness temperature, and LTE modeling
        of \methanol lines.  While these methods do not necessarily agree
        or provide direct and accurate measurements of the temperature, they
        provide strong enough constraints to draw substantial physical conclusions.
    \item We cataloged and classified 75 continuum sources on the basis of their
        dust emission, line emission, and morphology.  These sources are
        prestellar cores, dust-enshrouded protostars, and hypercompact \hii
        regions.
\end{itemize}

From these observations, we have inferred the following: 
\begin{itemize}
    \item During the earliest stages of their formation, before they have 
        ignited \hii regions, high-mass protostars heat a large volume
        ($R\sim5000 AU$), and correspondingly large mass, of gas around them.
    \item Older massive stars, those with surrounding \hii regions, appear
        to have little effect on the temperature of dense gas around them.
        Instead, their feedback primarily affects larger ($\gtrsim 0.1$ pc)
        scales.
    \item Because older generation stars have little impact on the star-forming
        molecular gas, it appears that the star formation process and the IMF
        are \emph{self-regulated} within clusters; they are unaffected by
        external forces like main-sequence OB-star feedback.
    \item Heated massive cores surrounding the highest-mass protostars in
        W51 show no evidence of ongoing fragmentation and are warm enough
        to significantly inhibit Jeans fragmentation.  Hot massive cores
        therefore serve as a mass reservoir for accretion onto possibly single
        massive stars.
    \item Hot massive cores are heated by the accreting protostars within them,
        implying that young massive stars self-regulate their core structures.
    \item Because the hot cores surround sources with tight upper limits
        on their free-free emission, yet the central stellar luminosities
        are consistent with main-sequence stars that would produce ionizing
        radiation, we suggest that the central stars are rapidly accreting and
        bloated.  The high accretion rate and bloating suppresses the
        photospheric temperature, preventing the production of ionizing
        radiation.
\end{itemize}

Our work suggests that thermal feedback from high-mass stars has dramatic
effects on their environments before they reach the main sequence.  Radiative
feedback effects on dense gas therefore need to be included in theories of
high-mass star and cluster formation.

\textbf{Acknowledgments}
The National Radio Astronomy Observatory is a facility of the National Science
Foundation operated under cooperative agreement by Associated Universities,
Inc.
This paper makes use of the ALMA data set 2013.1.00308.S and VLA data sets
12B-365, 13A-064, and 12A-074.
ALMA is a partnership of ESO (representing its member states), NSF (USA) and
NINS (Japan), together with NRC (Canada), NSC and ASIAA (Taiwan), and KASI
(Republic of Korea), in cooperation with the Republic of Chile. The Joint ALMA
Observatory is operated by ESO, AUI/NRAO and NAOJ.
JMDK gratefully acknowledges funding in the form of an
Emmy Noether Research Group from the Deutsche Forschungsgemeinschaft (DFG),
grant number KR4801/1-1.
K.W. is supported by grant WA3628-1/1 of the German Research Foundation (DFG)
through the priority program 1573 (``Physics of the Interstellar Medium'').

\facilities{ALMA, JVLA}

\software{
    astropy \citep{Astropy-Collaboration2013a}
    ipython \citep{Perez2007a}
    CASA \url{https://casa.nrao.edu/}
    pyspeckit \url{http://pyspeckit.bitbucket.org} \citet{Ginsburg2011c}
    aplpy \url{https://aplpy.github.io/}
    wcsaxes \url{http://wcsaxes.readthedocs.org}
    pvextractor \url{http://pvextractor.readthedocs.org}
    spectral cube \url{http://spectral-cube.readthedocs.org}
    ds9 \url{http://ds9.si.edu}
    \texttt{dust\_emissivity} \url{http://dust\_emissivity.readthedocs.org}
    }

\appendix

\section{Additional observational details}
In this Appendix, we present the complete list of imaged lines.

\begin{table*}[htp]
\centering
\caption{Spectral Lines in SPW 0}

\begin{tabular}{ll}
\label{tab:linesspw0}
Line Name & Frequency \\
 & $\mathrm{GHz}$ \\
\hline
H$_2$CO $3_{0,3}-2_{0,2}$ & 218.22219 \\
H$_2$CO $3_{2,2}-2_{2,1}$ & 218.47564 \\
E-CH$_3$OH $4_{2,2}-3_{1,2}$ & 218.44005 \\
CH$_3$OCHO $17_{3,14}-16_{3,13}$E & 218.28083 \\
CH$_3$OCHO $17_{3,14}-16_{3,13}$A & 218.29787 \\
CH$_3$CH$_2$CN $24_{3,21}-23_{3,20}$ & 218.39002 \\
Acetone $8_{7,1}-7_{4,4}$AE & 218.24017 \\
O$^{13}$CS 18-17 & 218.19898 \\
CH$_3$OCH$_3$ $23_{3,21}-23_{2,22}$AA & 218.49441 \\
CH$_3$OCH$_3$ $23_{3,21}-23_{2,22}$EE & 218.49192 \\
CH$_3$NCO $25_{1,24} - 24_{1,23}$ & 218.5418 \\
CH$_3$SH $23_2-23_1$ & 218.18612 \\
\hline
\end{tabular}
\end{table*}

\begin{table*}[htp]
\centering
\caption{Spectral Lines in SPW 1}

\begin{tabular}{ll}
\label{tab:linesspw1}
Line Name & Frequency \\
 & $\mathrm{GHz}$ \\
\hline
H$_2$CO $3_{2,1}-2_{2,0}$ & 218.76007 \\
HC$_3$N 24-23 & 218.32471 \\
HC$_3$Nv$_7$=1 24-23a & 219.17358 \\
HC$_3$Nv$_7$=1 24-23a & 218.86063 \\
HC$_3$Nv$_7$=2 24-23 & 219.67465 \\
OCS 18-17 & 218.90336 \\
SO $6_5-5_4$ & 219.94944 \\
HNCO $10_{1,10}-9_{1,9}$ & 218.98102 \\
HNCO $10_{2,8}-9_{2,7}$ & 219.73719 \\
HNCO $10_{0,10}-9_{0,9}$ & 219.79828 \\
HNCO $10_{5,5}-9_{5,4}$ & 219.39241 \\
HNCO $10_{4,6}-9_{4,5}$ & 219.54708 \\
HNCO $10_{3,8}-9_{3,7}$ & 219.65677 \\
E-CH$_3$OH $8_{0,8}-7_{1,6}$ & 220.07849 \\
E-CH$_3$OH $25_{3,22}-24_{4,20}$ & 219.98399 \\
E-CH$_3$OH $23_{5,19}-22_{6,17}$ & 219.99394 \\
C$^{18}$O 2-1 & 219.56036 \\
H$_2$CCO 11-10 & 220.17742 \\
HCOOH $4_{3,1}-5_{2,4}$ & 219.09858 \\
CH$_3$OCHO $17_{4,13}-16_{4,12}$A & 220.19027 \\
CH$_3$CH$_2$CN $24_{2,22}-23_{2,21}$ & 219.50559 \\
Acetone $21_{1,20}-20_{2,19}$AE & 219.21993 \\
Acetone $21_{1,20}-20_{1,19}$EE & 219.24214 \\
Acetone $12_{9,4}-11_{8,3}$EE & 218.63385 \\
H$_2^{13}$CO $3_{1,2}-2_{1,1}$ & 219.90849 \\
SO$_2$ $22_{7,15}-23_{6,18}$ & 219.27594 \\
SO$_2$ $v_2=1$ $20_{2,18}-19_{3,17}$ & 218.99583 \\
SO$_2$ $v_2=1$ $22_{2,20}-22_{1,21}$ & 219.46555 \\
SO$_2$ $v_2=1$ $16_{3,13}-16_{2,14}$ & 220.16524 \\
\hline
\end{tabular}
\end{table*}

\begin{table*}[htp]
\centering
\caption{Spectral Lines in SPW 2}

\begin{tabular}{ll}
\label{tab:linesspw2}
Line Name & Frequency \\
 & $\mathrm{GHz}$ \\
\hline
$^{12}$CO $2-1$ & 230.538 \\
OCS 19-18 & 231.06099 \\
HNCO $28_{1,28}-29_{0,29}$ & 231.873255 \\
A-CH$_3$OH $10_{2,9}-9_{3,6}$ & 231.28115 \\
$^{13}$CS 5-4 & 231.22069 \\
NH$_2$CHO $11_{2,10}-10_{2,9}$ & 232.27363 \\
H30$\alpha$ & 231.90093 \\
CH$_3$OCHO $12_{4,9}-11_{3,8}$E & 231.01908 \\
CH$_3$CH$_2$OH $5_{5,0}-5_{4,1}$ & 231.02517 \\
CH$_3$OCH$_3$ $13_{0,13}-12_{1,12}$AA & 231.98772 \\
N$_2$D$^+$ 3-2 & 231.32183 \\
g-CH$_3$CH$_2$OH $13_{2,11}-12_{2,10}$ & 230.67255 \\
g-CH$_3$CH$_2$OH $6_{5,1}-5_{4,1}$ & 230.79351 \\
g-CH$_3$CH$_2$OH $16_{5,11}-16_{4,12}$ & 230.95379 \\
g-CH$_3$CH$_2$OH $14_{0,14}-13_{1,13}$ & 230.99138 \\
SO$_2$ $v_2=1$ $6_{4,2}-7_{3,5}$ & 232.21031 \\
CH$_3$SH $16_2-16_1$ & 231.75891 \\
CH$_3$SH $7_3-8_2$ & 230.64608 \\
\hline
\end{tabular}
\end{table*}

\begin{table*}[htp]
\centering
\caption{Spectral Lines in SPW 3}

\begin{tabular}{ll}
\label{tab:linesspw3}
Line Name & Frequency \\
 & $\mathrm{GHz}$ \\
\hline
A-CH$_3$OH $4_{2,3}-5_{1,4}$ & 234.68345 \\
E-CH$_3$OH $5_{-4,2}-6_{-3,4}$ & 234.69847 \\
A-CH$_3$OH $18_{3,15}-17_{4,14}$ & 233.7958 \\
$^{13}$CH$_3$OH $5_{1,5}-4_{1,4}$ & 234.01158 \\
PN $5-4$ & 234.93569 \\
NH$_2$CHO $11_{5,6}-10_{5,5}$ & 233.59451 \\
Acetone $12_{11,2}-11_{10,1}$AE & 234.86136 \\
SO$_2$ $16_{6,10}-17_{5,13}$ & 234.42159 \\
CH$_3$NCO $27_{2,26} - 26_{2,25}$ & 234.08812 \\
CH$_3$SH $15_2-15_1$ & 234.19145 \\
\hline
\end{tabular}

\end{table*}

\section{Outflows}
\label{sec:outflows}
We detected many outflows, primarily in CO 2-1 and SO $6_5-5_4$.  The flows are
weakly detected in some other lines, e.g. \formaldehyde, but we defer
discussion of outflow chemistry to a future work.

In this section, we discuss some of the unique outflows and unique features of
outflows in the W51 region.  We show the most readily identified outflows in
Figures \ref{fig:outflowscontinuumnorth} - \ref{fig:e8cooutflow}.

\subsection{The Lacy jet}
\label{sec:lacyjet}
A high-velocity outflow was discovered within the W51 IRS2 region by
\citet{Lacy2007a} in the 12.8 \um [Ne II] line and subsequently detected in
H77$\alpha$ by \citet{Ginsburg2016b}.  We have discovered the CO counterpart to
this
outflow, which comes from near the continuum source ALMAmm31 (Figure
\ref{fig:lacyjet}).
The outflow shows red- and blue-shifted lobes
that form the base of the ionized outflow reported by \citet[][Figure
\ref{fig:outflowscontinuumnorth}]{Lacy2007a},
confirming the Lacy et al. hypothesis that the jet came
from a molecular outflow punching out of a molecular cloud into an \hii region.
Strangely, the outflow is not directly centered on the millimeter continuum
source, but is slightly offset.  

The presence of the Lacy jet is important for ruling out outflows from \hii
regions.  It provides clear evidence that a molecular outflow that is
subsequently ionized can be easily detected in existing radio recombination
line data.  If outflows of comparable mass were being launched from the stars
at the centers of \hchii regions (e.g., e2w), we would detect these flows.
Their absence provides an upper limit on the outflow rate - and presumably the
accretion rate - onto these sources.  While we cannot yet make that limit
quantitative, it is clear that the \hchii region sources are accreting
substantially less than the dust continuum sources.

\FigureTwo
{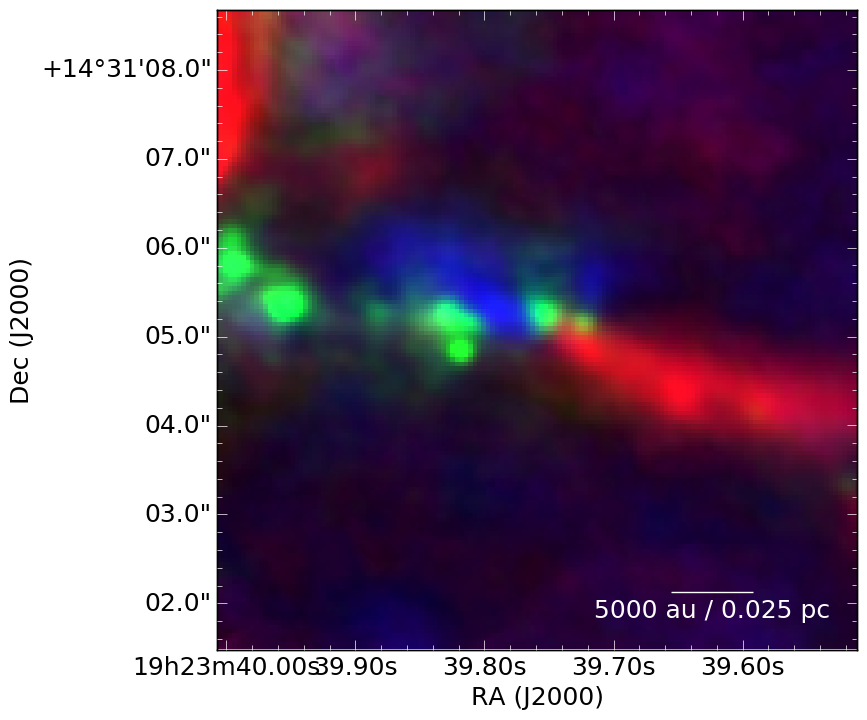}
{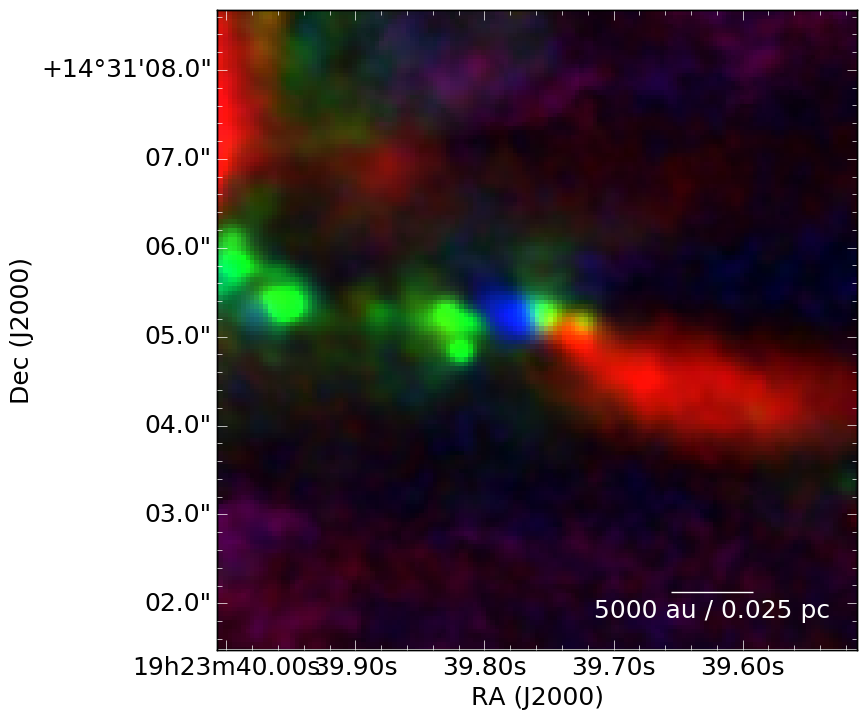}
{Outflows shown in red and blue for (a) CO 2-1 and (b) SO $6_5-5_4$ with
continuum in green.  This symmetric molecular outflow forms the base of the
\citet{Lacy2007a} ionized outflow detected further to the east.
The continuum source is offset from the line joining the red and blue outflow lobes.}
{fig:lacyjet}{1}{8cm}

\Figure{{f33}.png}
{Outflows in the W51 IRS2 region.  The green emission is NACO K-band continuum
\citep{Figueredo2008a,Barbosa2008a}, with ALMA 1.4 mm continuum contours in white and
H77$\alpha$ contours in blue.  The \citet{Lacy2007a} jet is prominent in
H77$\alpha$.}
{fig:outflowscontinuumnorth}{1}{0.8\textwidth}

\subsection{north}
The outflow from W51 North is extended and complex.  A jet-like high-velocity
feature appears directly to the north of W51 North in both CO and SO (Figure
\ref{fig:outflowscontinuumnorth}).  However, in SO, this feature begins to emit
at $\sim47$ \kms and continues to $\sim 100$ \kms.  The CO emission below $<70$
\kms is completely absent, presumably obscured by foreground material.  The
blueshifted component, by contrast with the red, points to the southeast and is
barely detected in CO, but again cleanly in SO.  It is sharply truncated,
extending only $\sim1 \arcsec$ ($\sim5000$ AU).  Unlike the Lacy jet, there is
no evidence that this outflow transitions into an externally ionized state.

The northernmost point of the W51 North outflow may coincide with
the \citet{Hodapp2002a} \hh and [Fe II] outflow.  There is some CO 2-1
emission coincident with the southernmost point of the \hh features,
and these all lie approximately along the W51 North outflow vector.
However, the association is only circumstantial.

\subsection{The e2e outflow}
The most prominent bipolar outflow in W51, which was previously detected by the
SMA \citep{Shi2010b,Shi2010a}, comes from the source e2e.  This outflow is
remarkable for its high velocity, extending nearly to the limit of our spectral
coverage in $^{12}$CO.  The ends of the flow cover at least $-50 < v_{LSR} <
160$ \kms, or a velocity $v\pm100$ \kms.  

The morphology is also notable.  Both ends of the outflow are sharply truncated
at $\sim2.5\arcsec$ (0.07 pc) from e2e (Figure \ref{fig:outflowscontinuume2}).
To the southeast, the high-velocity flow lies along a line that is consistent
with the extrapolation from the northwest flow, but at lower velocities ($10 <
v_{LSR} < 45$ \kms), it jogs toward a more north-south direction (Figure
\ref{fig:e2ecooutflow}).  In the
northwest, the redshifted part of this flow ($70 < v_{LSR} < 120$ \kms)
apparently collides with a \emph{blue}shifted flow from another source ($22 <
v_{LSR} < 45$ \kms), suggesting that these outflows intersect, though such a
scenario seems  implausible given their small volume filling factor.

The extreme velocity and morphology carry a few implications for the
accretion process in W51.  The sharp symmetric truncation at the outflow ends,
combined with the extraordinary velocity, suggests that the outflow is freshly
carving a cavity in the surrounding dense gas.  The observed velocities are
high enough that their bow shocks likely dissociated all molecules, so some
ionized gas is likely present at the endpoints; this ionized gas has not been
detected in radio images because of the nearby 100 mJy HCHII region e2w.  The
dynamical age of the outflow is $\sim600$ years at the peak observed velocity,
which is a lower limit on the true age of the outflow.

\Figure
{{f34}.png}
{Outflows in red and blue overlaid on mm continuum in green with cm continuum
contours in white.  The northern source is e2, the southern source at the tip
of the long continuum filament is e8.}
{fig:outflowscontinuume2}{1}{0.8\textwidth}

\Figure{{f35}.png}
{Channel maps of the e2e outflow in CO 2-1.  The dashed line approximately
connects the northwest and southeast extrema of the flow.}
{fig:e2ecooutflow}{1}{18cm}

Figure \ref{fig:outflowonmethanol} shows the e2e outflow overlaid on the
emission from a \methanol line.  It illustrates that the \methanol enhancement
is not produced by the outflow.

\FigureTwo
{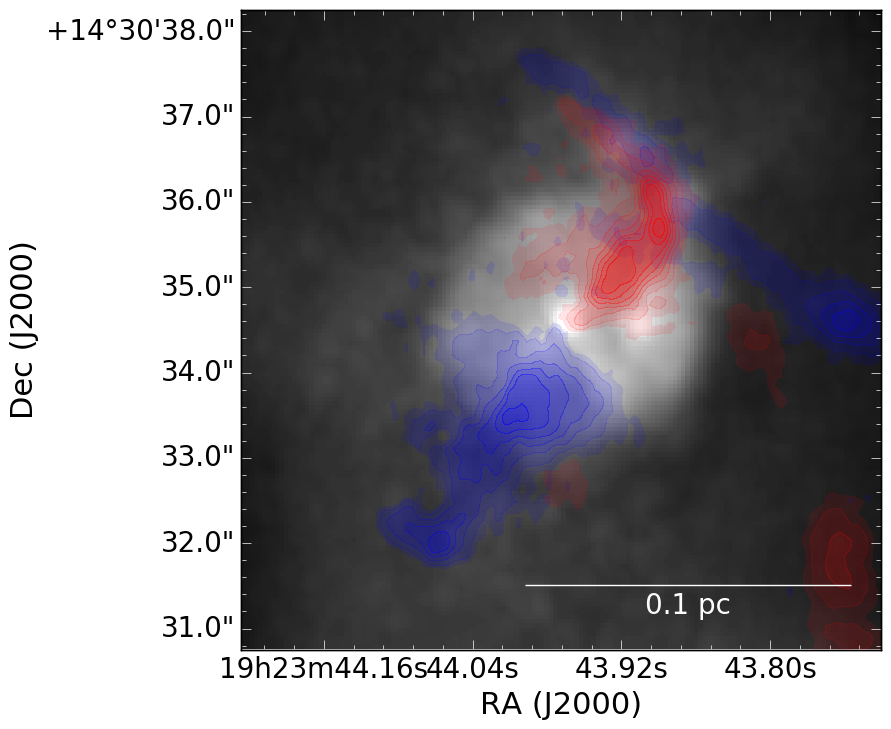}
{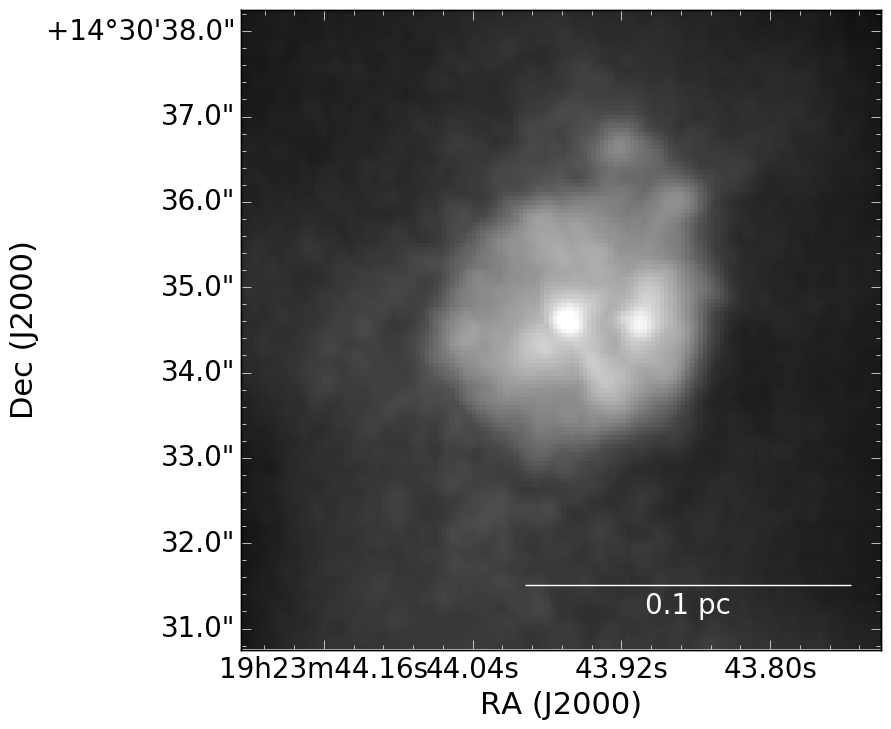}
{The e2e core as seen in the peak intensity of the \methanol $8_{0,8}-7_{1,6}$
line, with continuum included, is shown in greyscale.
(a) The integrated \twelveco 2-1 outflow is overlaid in red (73 to 180 \kms)
and blue (0 to 45 \kms).  The `core' is circularly symmetric, while the outflow
is clearly bipolar. Panel (b) is included to make the comparison clearer.}
{fig:outflowonmethanol}{1}{8cm}

\subsection{e8}
There are at least four distinct outflows coming from the e8 filament.
The e8 core is launching a redshifted outflow to the northwest.  A blueshifted
outflow is coming from somewhere south of the e8 peak and pointing straight
east.  While these originate quite near each other, they seem not to have
a common source, since the red and blue streams are not parallel (Figures
\ref{fig:outflowscontinuume2} and \ref{fig:e8cooutflow}).  The e8 outflows are too
confused and asymmetric for simple interpretation.

\Figure{{f39}.png}
{Channel maps of the e8 outflow in \twelveco 2-1.  The outflows here are more
erratic, with fewer clearly-connected red and blue lobes.}
{fig:e8cooutflow}{1}{18cm}

\section{Details of the extracted sources}
\label{sec:contsrcs}
We provide additional information and details about the continuum
source extraction, along with complete catalogs, in this Appendix.

\subsection{The spatial distribution of continuum sources}
\label{sec:corespatialdistribution}
The detected continuum sources are not uniformly distributed across the
observed region.  The most notable feature in the spatial distribution is their
alignment: most continuum sources collect along approximately linear features.
This is especially evident in W51 IRS2, where the core density is very high and
there is virtually no deviation from the line.  The e8 filament is also notably
linear, though there are a few sources detected just off the filament. 

On a larger scale, the e8 filament points toward e2, apparently tracing a
slightly longer filamentary structure that is either lower-column or resolved
out by our data.  With some imagination, this might be extended along the
entire northeast ridge to eventually connect in a broad half-circle with the
IRS2 filament (Figure \ref{fig:corepositions}).  This morphology hints at a
possible sequential star formation event, where some central bubble has swept
gas into these filaments.  However, there is reason to be skeptical of this
interpretation: this ring has no counterpart in ionized gas as would be
expected if it were driven as part of an expanding \hii region or a wind
bubble, and there is little reason to expect such circular symmetry from an
isolated molecular cloud, so the star forming circle may be merely a
coincidental alignment.

Whether it is physical or not, there is a relative lack of millimeter continuum
sources within the circle.  There is no lack of molecular gas, however, as both
CO and \formaldehyde emission fill the full field of view.

\Figure{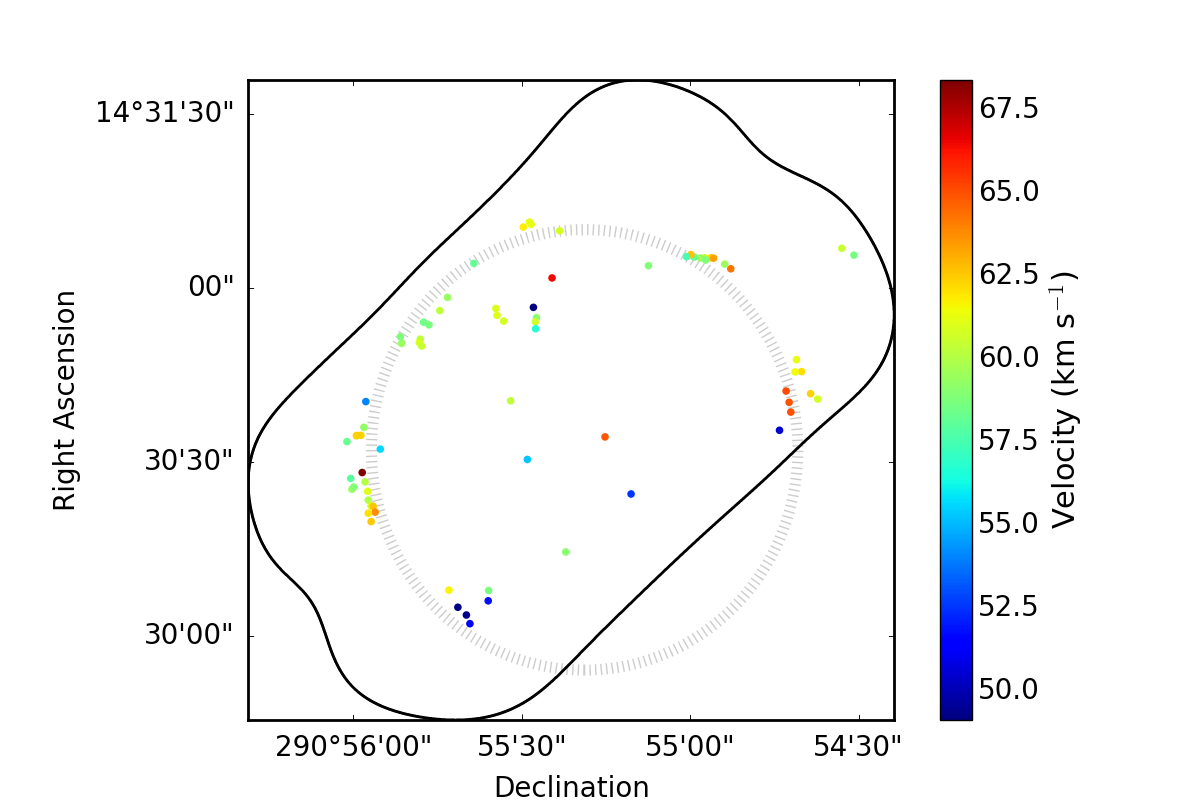}
{The spatial distribution of the hand-identified core sample.
The black outer contour shows the observed field of view.  The dashed circle
(with $r=1$ pc) shows a hypothetical ring of star formation.
The velocities shown are the mean of the velocity of peak intensity for many
lines.
}{fig:corepositions}{1}{16cm}

\subsection{Photometry}
\label{sec:photometry}
We created a catalog of the hand-extracted sources including their peak and mean
intensity, their centroid, and their geometric properties.  For each source,
we further extracted aperture photometry around the centroid in 6 apertures:
0.2, 0.4, 0.6, 0.8, 1.0, and 1.5\arcsec.  We performed the same aperture
photometry on the W51 Ku-band images from \citet{Ginsburg2016a} to estimate the
free-free contribution to the observed intensity measurements.  The
free-free contribution at $\sim227$ GHz will fall in a range between optically
thick, spectral index $\alpha_\nu=2$, and optically thin, $\alpha_\nu=0.1$,
which correspond to factors of $S_{227 \mathrm{GHz}} = 227 S_{15 \mathrm{GHz}}$
and $S_{227 \mathrm{GHz}} = 1.3 S_{15 \mathrm{GHz}}$, respectively.  These
measurements are reported in Table
\ref{tab:photometry1}.

The source flux density and intensity distribution are shown in Figure
\ref{fig:fluxhistograms}.  The most common nearest-neighbor separation between
cataloged sources is $\sim0.3\arcsec$, which implies that the larger apertures
double-count some pixels.  The smallest separation is 0.26\arcsec, so the
0.2\arcsec\ aperture contains almost only unique pixels.  The corresponding
masses are shown in Figure \ref{fig:masshistograms} assuming the dust
temperature is equal to the source's peak line brightness temperature (Section
\ref{sec:temperature}).

Except where noted below, the hand-selected sources are used for further
analysis because they are more reliable.

\begin{table*}[htp]
\caption{Continuum Source IDs and photometry Part 1}
\begin{tabular}{lllllllllllllllllllllllllllllllllllllllllllllllllllllllllllllllllllll}
\label{tab:photometry1}
Source ID & RA & Dec & $S_{\nu}(0.2\arcsec)$ & $S_{\nu}(0.4\arcsec)$ & $T_{B,max}$ & M$(T_B, 0.2\arcsec)$ & M$(T_B, \mathrm{peak})$ & Categories \\
 &  &  & $\mathrm{mJy}$ & $\mathrm{mJy}$ & $\mathrm{K}$ & $\mathrm{M_{\odot}}$ & $\mathrm{M_{\odot}}$ &  \\
\hline
ALMAmm1 & 19:23:42.864 & 14:30:07.92 & 3.7 & 6.9 & 11 & 2.6 & 2.6 & fCc \\
ALMAmm2 & 19:23:42.394 & 14:30:07.86 & 4.2 & 7.3 & 4 & 3 & 12 & fCc \\
ALMAmm3 & 19:23:42.398 & 14:30:06.08 & 4.2 & 11 & nan & 3 & 2.9 & f-- \\
ALMAmm4 & 19:23:42.614 & 14:30:02.14 & 7.7 & 16 & 11 & 5.4 & 6.1 & -Cc \\
ALMAmm5 & 19:23:42.658 & 14:30:03.63 & 1 & 19 & 5.9 & 7.3 & 8.8 & -Cc \\
ALMAmm6 & 19:23:42.758 & 14:30:04.97 & 2.8 & 9.2 & 3.8 & 2 & 1.1 & fC- \\
ALMAmm7 & 19:23:40.702 & 14:30:24.5 & 3.5 & 7.4 & 1.4 & 2.5 & 7.1 & -Cc \\
ALMAmm9 & 19:23:41.481 & 14:30:14.6 & 21 & 46 & 5.8 & 15 & 9 & -Cc \\
ALMAmm10 & 19:23:38.738 & 14:30:47.66 & 3.6 & 7.8 & 5.3 & 2.6 & 1 & -Cc \\
ALMAmm11 & 19:23:38.684 & 14:30:45.57 & 19 & 36 & 12 & 14 & 12 & -Cc \\
ALMAmm12 & 19:23:38.755 & 14:30:45.54 & 5.2 & 2 & 11 & 3.7 & 11 & -C- \\
ALMAmm13 & 19:23:38.825 & 14:30:40.31 & 7 & 12 & 11 & 5 & 8.6 & -Cc \\
ALMAmm14 & 19:23:38.57 & 14:30:41.79 & 67 & 14 & 36 & 23 & 23 & --c \\
ALMAmm15 & 19:23:38.486 & 14:30:40.86 & 14 & 31 & 35 & 4.9 & 8.1 & --c \\
ALMAmm16 & 19:23:38.2 & 14:31:06.85 & 23 & 45 & 5.6 & 16 & 32 & -Cc \\
ALMAmm17 & 19:23:42.214 & 14:30:54.31 & 15 & 26 & 12 & 11 & 5 & fCc \\
ALMAmm18 & 19:23:42.293 & 14:30:55.29 & 15 & 31 & 5.6 & 11 & 4.1 & -Cc \\
ALMAmm19 & 19:23:42.307 & 14:30:56.49 & 4.9 & 13 & 21 & 3.2 & 1.4 & --- \\
ALMAmm20 & 19:23:41.64 & 14:31:01.75 & 8 & 21 & 6.6 & 5.7 & 4 & -C- \\
ALMAmm21 & 19:23:41.981 & 14:31:10.52 & 8.9 & 15 & nan & 6.3 & 31 & --c \\
ALMAmm22 & 19:23:41.909 & 14:31:11.38 & 8.6 & 23 & nan & 6.1 & 18 & --- \\
ALMAmm23 & 19:23:40.496 & 14:31:03.94 & 22 & 65 & 26 & 11 & 5.2 & --- \\
ALMAmm24 & 19:23:39.953 & 14:31:05.35 & 29 & 6 & 72 & 46 & 32 & -Hc \\
ALMAmm25 & 19:23:42.132 & 14:30:40.57 & 18 & 38 & 3.8 & 13 & 6 & fCc \\
ALMAmm26 & 19:23:43.102 & 14:30:53.66 & 13 & 32 & 8.9 & 9.5 & 14 & -Cc \\
ALMAmm27 & 19:23:42.967 & 14:30:56.18 & 1 & 27 & 5 & 7.2 & 8.2 & fC- \\
ALMAmm28 & 19:23:43.68 & 14:30:32.24 & 16 & 33 & 19 & 11 & 6.5 & -Cc \\
ALMAmm29 & 19:23:41.933 & 14:30:30.45 & 9 & 16 & nan & 6.4 & 11 & f-c \\
ALMAmm30 & 19:23:43.164 & 14:30:54.12 & 8.3 & 2 & 4.3 & 5.9 & 2 & fCc \\
ALMAmm31 & 19:23:39.754 & 14:31:05.24 & 17 & 36 & 46 & 43 & 16 & --c \\
ALMAmm32 & 19:23:39.724 & 14:31:05.15 & 87 & 23 & 94 & 1 & 13 & -H- \\
ALMAmm33 & 19:23:39.828 & 14:31:05.23 & 19 & 51 & 48 & 48 & 18 & --- \\
ALMAmm34 & 19:23:39.878 & 14:31:05.19 & 8 & 24 & 9 & 1 & 1 & -H- \\
ALMAmm35 & 19:23:39.991 & 14:31:05.77 & 2 & 58 & 34 & 76 & 18 & --- \\
ALMAmm36 & 19:23:39.518 & 14:31:03.33 & 22 & 47 & 11 & 16 & 6.5 & -Cc \\
\hline
\end{tabular}

\end{table*}
\begin{table*}[htp]
\caption{Continuum Source IDs and photometry Part 2}
\begin{tabular}{lllllllllllllllllllllllllllllllllllllllllllllllllllllllllllllllllllll}
\label{tab:photometry2}
Source ID & RA & Dec & $S_{\nu}(0.2\arcsec)$ & $S_{\nu}(0.4\arcsec)$ & $T_{B,max}$ & M$(T_B, 0.2\arcsec)$ & M$(T_B, \mathrm{peak})$ & Categories \\
 &  &  & $\mathrm{mJy}$ & $\mathrm{mJy}$ & $\mathrm{K}$ & $\mathrm{M_{\odot}}$ & $\mathrm{M_{\odot}}$ &  \\
\hline
ALMAmm37 & 19:23:41.825 & 14:30:54.9 & 17 & 36 & 34 & 6.4 & 3.1 & --c \\
ALMAmm38 & 19:23:41.011 & 14:30:34.34 & 2.4 & 4.8 & 2 & 1.7 & 15 & -Cc \\
ALMAmm39 & 19:23:41.887 & 14:31:11 & 7.7 & 22 & 3.4 & 5.5 & 25 & -C- \\
ALMAmm40 & 19:23:41.549 & 14:31:09.87 & 7.5 & 18 & 9.2 & 5.3 & 5.5 & -Cc \\
ALMAmm41 & 19:23:43.85 & 14:30:40.43 & 2 & 29 & 3 & 8.6 & 13 & --c \\
ALMAmm43 & 19:23:39.59 & 14:31:04.13 & 19 & 53 & 17 & 14 & 4.8 & fC- \\
ALMAmm44 & 19:23:38.054 & 14:31:05.68 & 8.4 & 19 & 5.7 & 6 & 28 & -Cc \\
ALMAmm45 & 19:23:38.76 & 14:31:07.22 & 7.3 & 2 & 4.4 & 5.2 & 1.8 & -C- \\
ALMAmm46 & 19:23:41.834 & 14:30:52.99 & 15 & 37 & 24 & 8.5 & 3 & --- \\
ALMAmm47 & 19:23:42.569 & 14:31:04.27 & 6.3 & 14 & 4.9 & 4.5 & 9.2 & -Cc \\
ALMAmm48 & 19:23:42.881 & 14:30:58.4 & 9.4 & 19 & 8.1 & 6.7 & 13 & -Cc \\
ALMAmm49 & 19:23:43.205 & 14:30:51.2 & 15 & 46 & 21 & 9.8 & 7.6 & --- \\
ALMAmm50 & 19:23:43.217 & 14:30:50.6 & 2 & 53 & 14 & 14 & 6.3 & -C- \\
ALMAmm51 & 19:23:43.188 & 14:30:50.01 & 19 & 46 & 16 & 13 & 2.7 & -C- \\
ALMAmm52 & 19:23:38.806 & 14:30:38.62 & 4.9 & 9.7 & 7.9 & 3.5 & 1 & -Cc \\
ALMAmm53 & 19:23:38.861 & 14:30:42.25 & 8 & 18 & 13 & 5.7 & 5.6 & -Cc \\
ALMAmm54 & 19:23:38.94 & 14:30:35.48 & 5.1 & 9.7 & 2.5 & 3.6 & 9.7 & -Cc \\
ALMAmm55 & 19:23:43.426 & 14:30:50.46 & 11 & 29 & 4.5 & 7.7 & 6.5 & -C- \\
ALMAmm56 & 19:23:43.44 & 14:30:51.61 & 9 & 25 & 5.3 & 6.4 & 5.4 & -C- \\
ALMAmm57 & 19:23:41.731 & 14:30:52.99 & 5.6 & 1 & 2 & 4 & 1.2 & fCc \\
d2 & 19:23:39.818 & 14:31:04.83 & 16 & 43 & 99 & 18 & 15 & -H- \\
e1mm1 & 19:23:43.86 & 14:30:26.58 & 16 & 42 & 23 & 95 & 31 & --- \\
e2e & 19:23:43.956 & 14:30:34.57 & 69 & 19 & 84 & 94 & 61 & -H- \\
e2e peak & 19:23:43.963 & 14:30:34.56 & 74 & 18 & 1 & 8 & 68 & -Hc \\
e2nw & 19:23:43.874 & 14:30:35.99 & 22 & 51 & 4 & 69 & 33 & --c \\
e2se & 19:23:44.076 & 14:30:33.53 & 36 & 98 & 93 & 4.4 & 7.3 & -H- \\
e2w & 19:23:43.91 & 14:30:34.61 & 54 & 12 & 85 & 72 & 6 & fHc \\
e3mm1 & 19:23:43.829 & 14:30:24.95 & 5 & 18 & 23 & 29 & 9.4 & --- \\
e5 & 19:23:41.862 & 14:30:56.69 & 25 & 31 & nan & 18 & 34 & F-c \\
e8mm & 19:23:43.894 & 14:30:28.2 & 68 & 17 & 18 & 43 & 16 & -H- \\
eEmm1 & 19:23:44.016 & 14:30:25.32 & 38 & 1 & 39 & 12 & 7 & --- \\
eEmm2 & 19:23:43.994 & 14:30:25.7 & 37 & 1 & 22 & 24 & 6.5 & --- \\
eEmm3 & 19:23:44.03 & 14:30:27.17 & 43 & 95 & 22 & 27 & 8.1 & --c \\
eSmm1 & 19:23:43.822 & 14:30:23.44 & 69 & 18 & 34 & 26 & 2 & --- \\
eSmm2 & 19:23:43.788 & 14:30:22.42 & 68 & 17 & 29 & 31 & 23 & --c \\
eSmm2a & 19:23:43.764 & 14:30:22.38 & 5 & 13 & 24 & 28 & 18 & --- \\
eSmm3 & 19:23:43.74 & 14:30:21.37 & 52 & 99 & 29 & 23 & 16 & --c \\
eSmm4 & 19:23:43.822 & 14:30:21.18 & 36 & 96 & 27 & 17 & 1 & --- \\
eSmm6 & 19:23:43.788 & 14:30:19.74 & 5 & 11 & 23 & 29 & 15 & --c \\
north & 19:23:40.044 & 14:31:05.42 & 72 & 17 & 69 & 12 & 59 & -Hc \\
\hline
\end{tabular}
\par
The Categories column consists of three letter codes as described in Section \ref{sec:contsourcenature}.  In column 1, \texttt{F} indicates a free-free dominated source, \texttt{f} indicates significant free-free contribution, and \texttt{-} means there is no detected cm continuum.  In column 2, the peak brightness temperature is used to classify the temperature category.  \texttt{H} is `hot' ($T>50$ K), \texttt{C} is `cold' ($T<20$ K), and \texttt{-} is indeterminate (either $20<T<50$K or no measurement).  In column 3, \texttt{c} indicates compact sources, and \texttt{-} indicates a diffuse source.
\end{table*}

\Figure{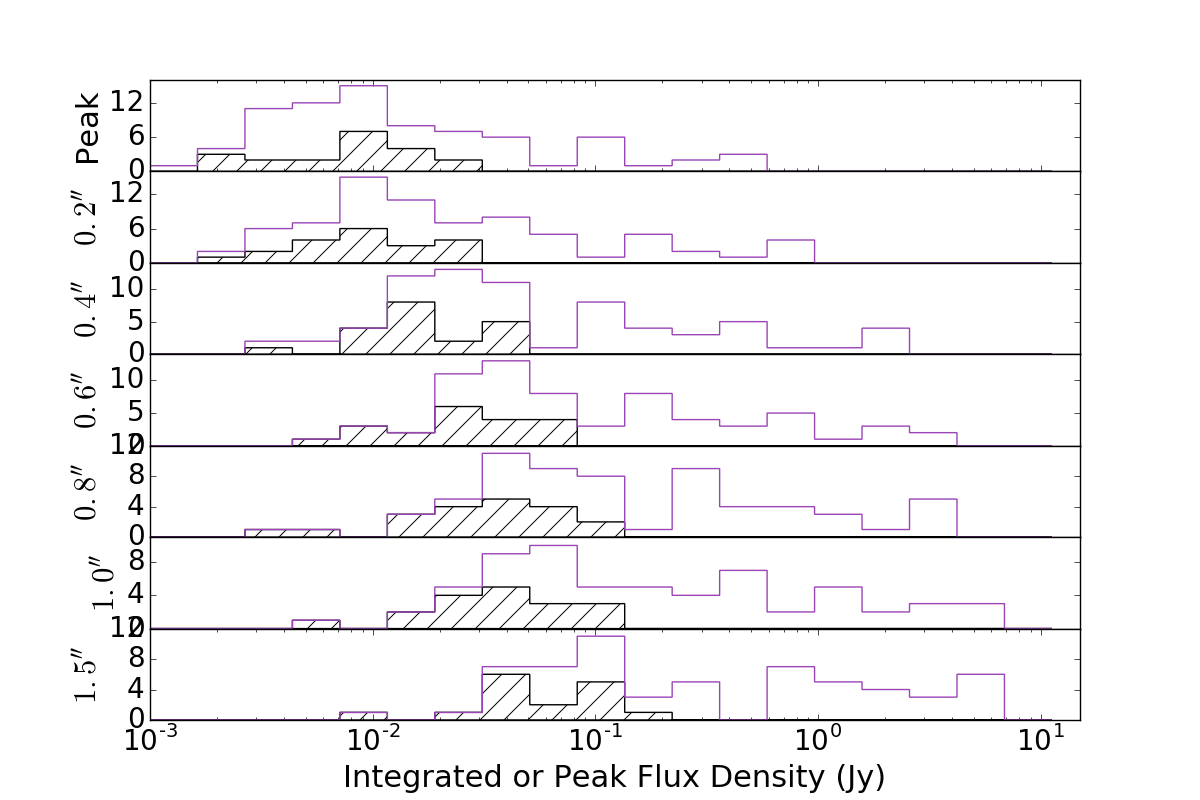}
{Histograms of the core flux densities measured with circular apertures centered
on the hand-extracted core positions.  The aperture size is listed 
in the y-axis label.  For the top plot, labeled `Peak', this is the peak
intensity in Jy/beam.  For the rest, it is the integrated flux density
in the specified aperture.  The unfilled data show all sources and the hashed
data are for starless core candidates (Section \ref{sec:contsourcenature}).
See Figure \ref{fig:masshistograms} for the corresponding masses.}
{fig:fluxhistograms}
{1}{16cm}

\subsection{Temperature estimation of the continuum sources}
\label{sec:temperature}
The temperature is a critical ingredient for determining the total mass of each
continuum source or region. Since we do not have any means of directly
determining the dust temperature, as the SED peak is well into the THz regime
and inaccessible to any existing instruments at the requisite resolution, we
employ alternative indicators.  Above a density $n\gtrsim10^5-10^6$ \percc,
the gas and dust become strongly collisionally coupled, meaning the gas
temperature should accurately reflect the dust temperature.  Below this density,
the two may be decoupled.

The average dust temperature, as estimated from Herschel Hi-Gal SED fits
\citep{Molinari2016a,Wang2015a}, is 38 K when including the 70 \um data or 26 K
when excluding it.  This average is obtained over a $\sim45\arcsec$ ($\sim 1$
pc) beam and therefore is likely to be strongly biased toward the hottest dust
in the \hii regions.  Despite these
uncertainties, this bulk measurement provides us with a reasonable range to
assume for the uncoupled, low-density dust, which (weakly) dominates the mass
(see Section \ref{sec:massbudget}).

One constraint on the dust temperature we can employ is the absolute surface
brightness.  For some regions, especially the e8 filament and the hot cores,
the surface brightness is substantially
brighter than is possible for a beam-filling, optically thick blackbody at 20
K, providing a lower limit on the dust temperature ranging from 20 K (35
mJy/beam) to $\sim300$ K (0.5 Jy/beam).  Toward most of this emission, optically-thick
free-free emission can be strongly ruled out as the driving mechanism: 
existing data limits the free-free contribution to be $<50\%$ if it is
optically thick and negligible ($\ll1\%$) if it is optically thin at radio
wavelengths \citep{Ginsburg2016b, Goddi2016a}.

To gain a more detailed measurement of the dust temperature in regions where it
is likely to be coupled to the gas, we use the peak brightness temperature
$T_{B,max}$ of spectral lines along the line of sight.  If the observed
molecule is in local thermal equilibrium, as is expected if the density is high
enough to be collisionally coupled to the dust, and it is optically thick, the
brightness temperature provides an approximate measurement of the local
temperature near the $\tau=1$ surface.  If any of these assumptions do not
hold, $T_{B,max}$ will set a lower limit on the true gas temperature.  Only 
two mechanisms can push $T_{B,max} > T_{dust}$: nonthermal (maser) emission, which
is not known for any of the observed lines nor expected given the reasonable
$T_B$ observed, or a dust-emitting region that has a smaller beam filling
factor than the gas-emitting region, which is unlikely when the dust emission
structure is resolved, as is the case toward most sources.

One potential problem with this approach is whether the gas becomes optically thick
before probing most of the dust, in which case spectral line self-absorption
will occur.  Some transitions of more abundant molecules, e.g., CO and
\formaldehyde, are likely to be affected by this issue.  However, many of the
molecules included in the observations (Tables
\ref{tab:linesspw0}-\ref{tab:linesspw3}) have lower abundances, especially in
lower-density gas, and are likely to be optically thin along most of the lines
of sight.

Some sources have no detected line emission aside from the molecular cloud
species CO and \formaldehyde, which are also associated with more diffuse gas
and not isolated to the compact continuum sources but in these cases peak
locally on the compact source.  The minimum density requirement imposed by a
continuum detection at our limit of 1.6 mJy is $n>10^{7.5}$ \percc for a
spherical source.  At such high density, it is unlikely that the species are
undetected because they are subthermally excited.  More likely, the
line-nondetection sources have an underyling emission source that is very
compact, optically thick, and/or cold.

Figure \ref{fig:peaktbhist} shows the distribution of peak line brightnesses
for the continuum sources.  The spectra used to determine this brightness are
the spectra obtained from the brightest continuum pixel within the source
aperture.  To obtain the peak line brightness, we fit Gaussian profiles to each
identified line listed in Tables \ref{tab:linesspw0}-\ref{tab:linesspw3},
rejecting those with poor fits.  The line brightnesses reported in the figure
are the sum of the continuum subtracted peak line brightness and the continuum
brightness (i.e., they are the raw observed peak brightness).  Excepting CO and
\formaldehyde, which are excluded from the plot, \methanol is the brightest
line toward most sources. 

\Figure{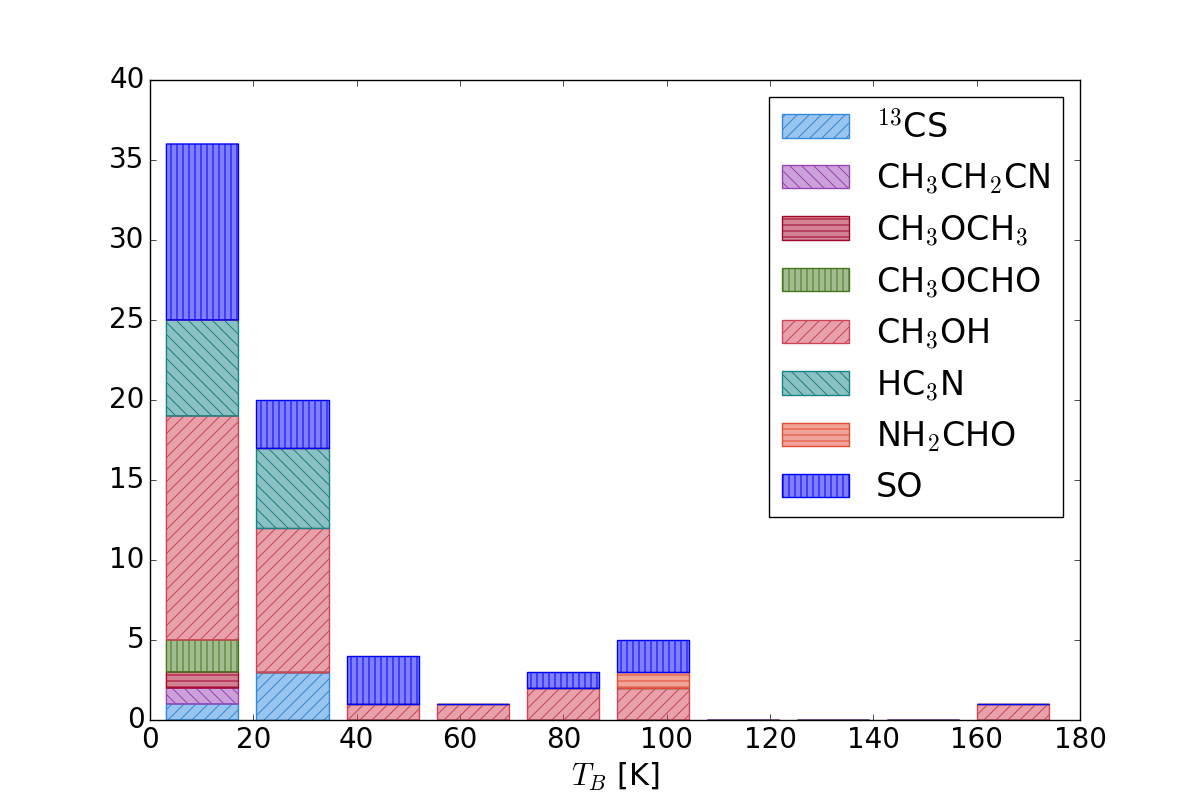}
{Histogram of the brightest line toward each continuum source.
The bars are colored by the molecular species associated with the brightest
line that is not associated with extended molecular cloud emission,
i.e., CO and its isotopologues and \formaldehyde are excluded.}
{fig:peaktbhist}{1}{0.9\textwidth}

We use these peak line brightness temperatures to compute the masses of the
continuum sources.  For sources with $T_{B,max} < 20$ K, we assume $T_{dust} =
20$ K to avoid producing unreasonably high masses; in such sources the lines
are likely to be optically thin and/or subthermally excited.  This correction is
illustrated in Figure \ref{fig:moftbvsm20k}.

\Figure{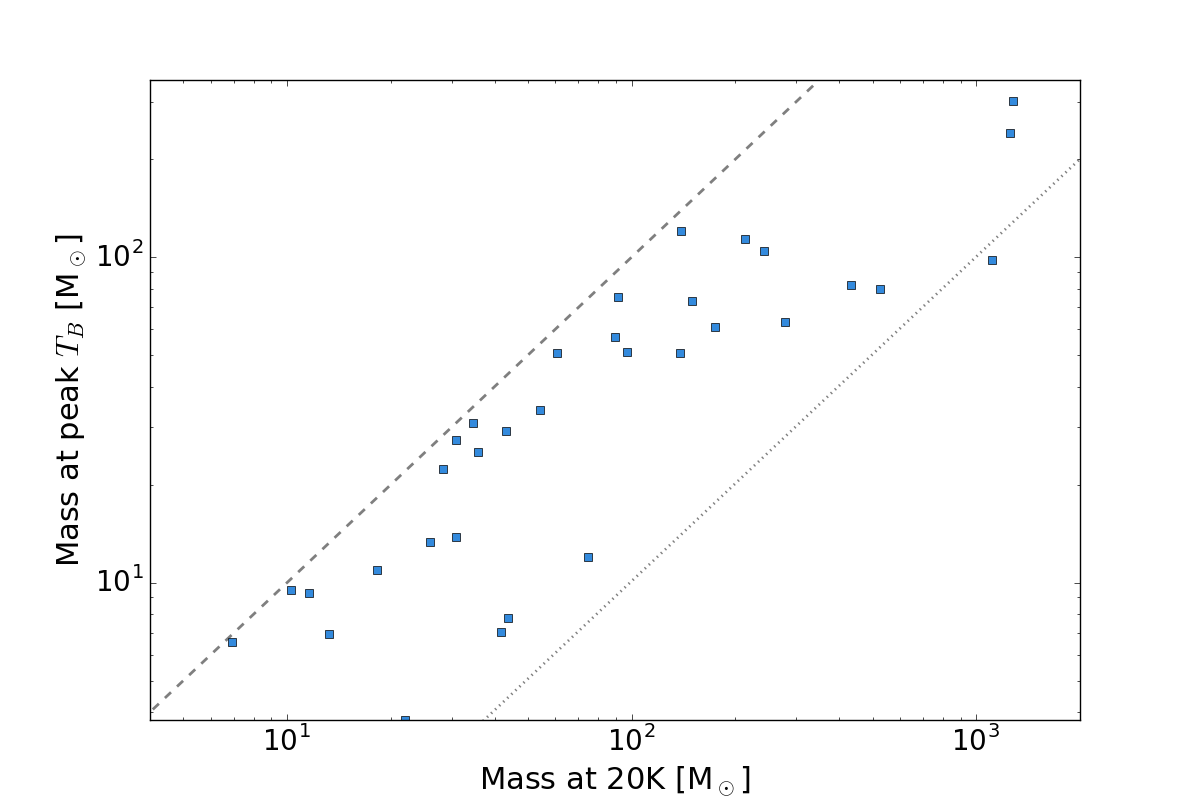}
{The mass computed assuming the dust temperature is the peak brightness
temperature vs. that computed assuming $T_{dust}=20$ K  for the aperture extracted
continuum sources.
The dashed line shows $M(T_{B,max}) =
M(20\textrm{K})$ and the dotted line shows $M(T_{B,max}) = 0.1 
M(20\textrm{K})$ 
}{fig:moftbvsm20k}{1}{16cm}

This section has provided some simple temperature estimates across all of the
detected continuum sources.  In Section \ref{sec:ch3ohtem}, we  examine
the thermal structure of the hot cores in more detail.

\Figure{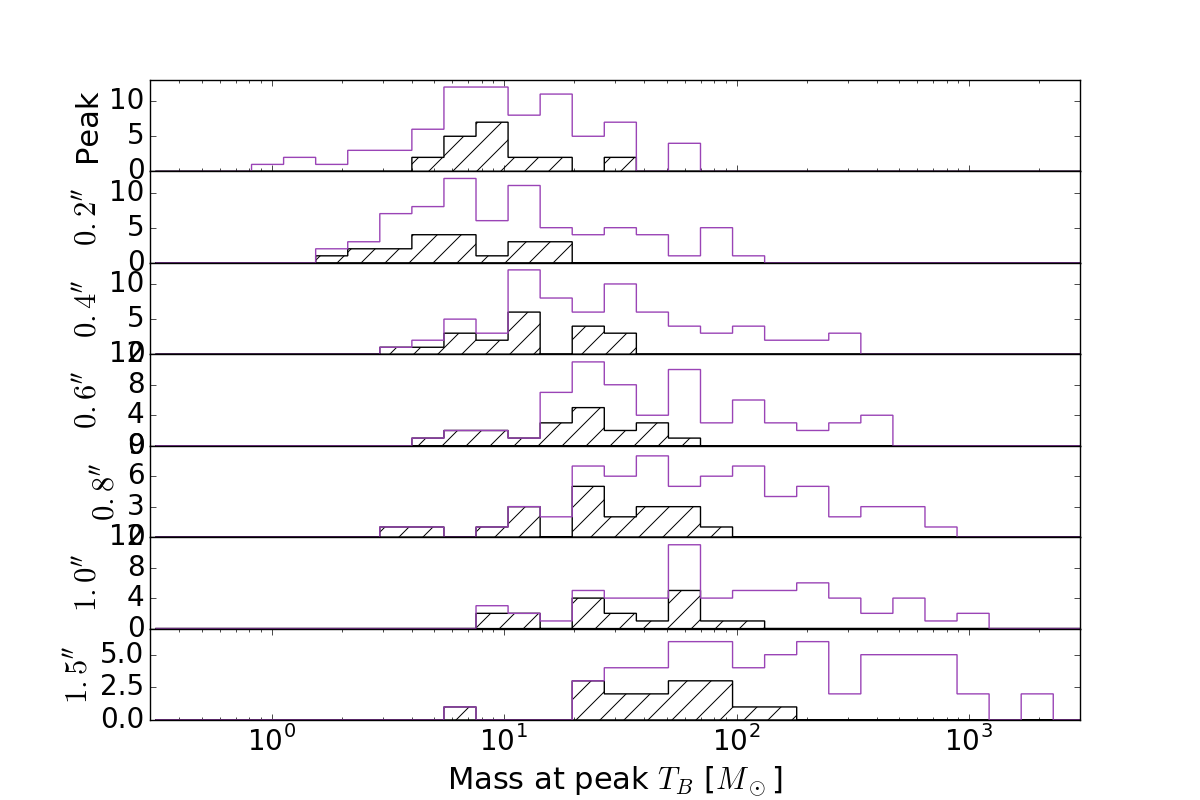}
{Histograms of the core masses computed from the flux density measurements
shown in Figure \ref{fig:fluxhistograms} using the peak brightness temperature 
toward the center of that source as the dust temperature.
The aperture size is listed in the y-axis label.  For the top plot, labeled
`Peak', the mass is computed from peak
intensity in Jy/beam.  For the rest, it is the integrated flux density in
the specified aperture in Jy.  The unfilled data show all sources and the
hashed
data are for starless core candidates (Section \ref{sec:contsourcenature}).}
{fig:masshistograms}
{1}{16cm}

\subsection{The nature of the continuum sources}
\label{sec:contsourcenature}
Millimeter continuum sources in star-forming regions are usually assumed to be
either protostars or starless cores.  However, in this high-mass star-forming
region, we have to consider not only those possibilities but also potential
free-free sources and high-luminosity main-sequence stars embedded in dust.

To distinguish these possibilities, we measure both the spectral lines and
features of the continuum emission toward the compact continuum sources.  Main
sequence OB stars and their illuminated ionized nebulae are in principle easily
identified by their free-free emission, so we use centimeter continuum and
radio recombination line emission to identify these sources.  Starless cores,
protostellar cores, and their variants are more difficult to identify, so we
used a combination of gas temperature and continuum concentration parameter to
classify them.

To estimate the gas temperature toward the compact sources, we fit each of up
to $\sim50$ lines (see Tables \ref{tab:linesspw0}-\ref{tab:linesspw3}) with
Gaussian profiles to attempt to determine the relative line strengths toward
each source.  Most sources were detected in at least $\sim5-10$ lines, though
some of these are associated with interstellar rather than circumstellar
material, i.e., \formaldehyde, CO, $^{13}$CS.  For sources with detections in
non-interstellar lines, we used the peak brightness temperature of the line as
an estimated lower limit on the core temperature.

In the continuum, we measured a `concentration parameter', which is the ratio
of the flux density in a 0.2\arcsec aperture to that in a 0.2\arcsec-0.4\arcsec
annulus divided by three to account for the annulus' larger area.  A uniform
source with $r>0.4\arcsec$ would have a concentration $C=1$ by this
definition, while an unresolved point source would have a Gaussian profile
resulting in $C=14$.  Only one source approaches this extreme, the HII region
e5, while the rest have $C\leq7$.  We set the threshold for a `concentrated'
source to be $C>2$, which is arbitrary, but does a reasonable job of
distinguishing the sources with a clear central concentration from those that
have none.

We classified each of the 75 hand-selected sources on the following parameters:
\begin{enumerate}
    \item Free-free dominated sources ($S_{15 GHz} > 0.5 S_{227 GHz}$) are \hii
        regions
    \item Free-free contaminated sources ($S_{15 GHz} > 0.1 S_{227 GHz}$) are
        likely to be dust-dominated but with \hii region contamination; these
        are either dusty sources superposed on or embedded in a large \hii
        region or they are compact, dusty \hii regions
    \item Starless core candidates were identified as those with cold peak
        brightness temperatures $T_B < 20$ K and with a high concentration
        parameter ($C>2$)
    \item Hot core candidates are those with peak $T_B>50$ K and $C>2$
    \item Extended cold core and hot core candidates are those with $T_B<20$ K
        and $T_B>50$ K and $C<2$, respectively.
    \item The remaining sources with $S_{15 GHz} < 0.1 S_{227 GHz}$ and $50 >
        T_B > 20$ K were classed as uncertain compact ($C>2$) or uncertain extended
        ($C<2$)
\end{enumerate}

These classifications are set in the `Categories' column of Table
\ref{tab:photometry1}.  They serve as a broad guideline for further
analysis.   Table \ref{tab:photometry2} lists the measured source properties and classifications.

\section{Additional kinematic plots}
Kinematics figures for the e8 and North cores are shown in this Appendix to
minimize clutter in the main text.

Figures \ref{fig:kinematicse8} and \ref{fig:kinematicsnorth} show the moment 1
and moment 2 maps of two \methanol lines as described in Section
\ref{sec:kinematics}.  W51 e8 has notably narrower lines in some parts of 
the core than either e2 or North.

\Figure{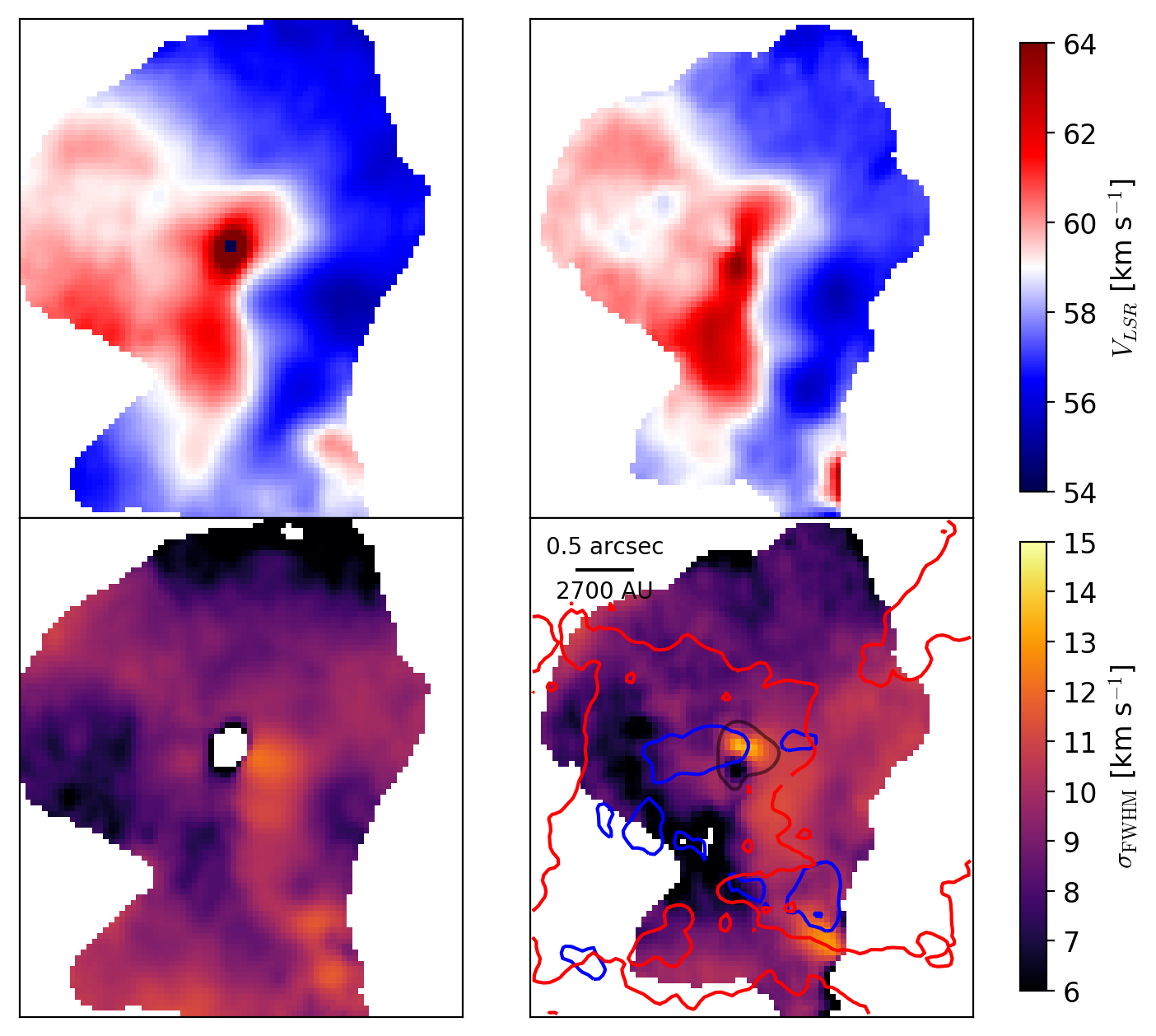}
{Moment 1 and 2 maps of the W51 e8 core over the velocity
range 48-68 \kms.  While there is  outflowing $^{12}$CO, shown in the
lower-right panel, there is not a clear bipolar outflow. See Figure
\ref{fig:kinematicse2}.}
{fig:kinematicse8}{1}{0.9\textwidth}

\Figure{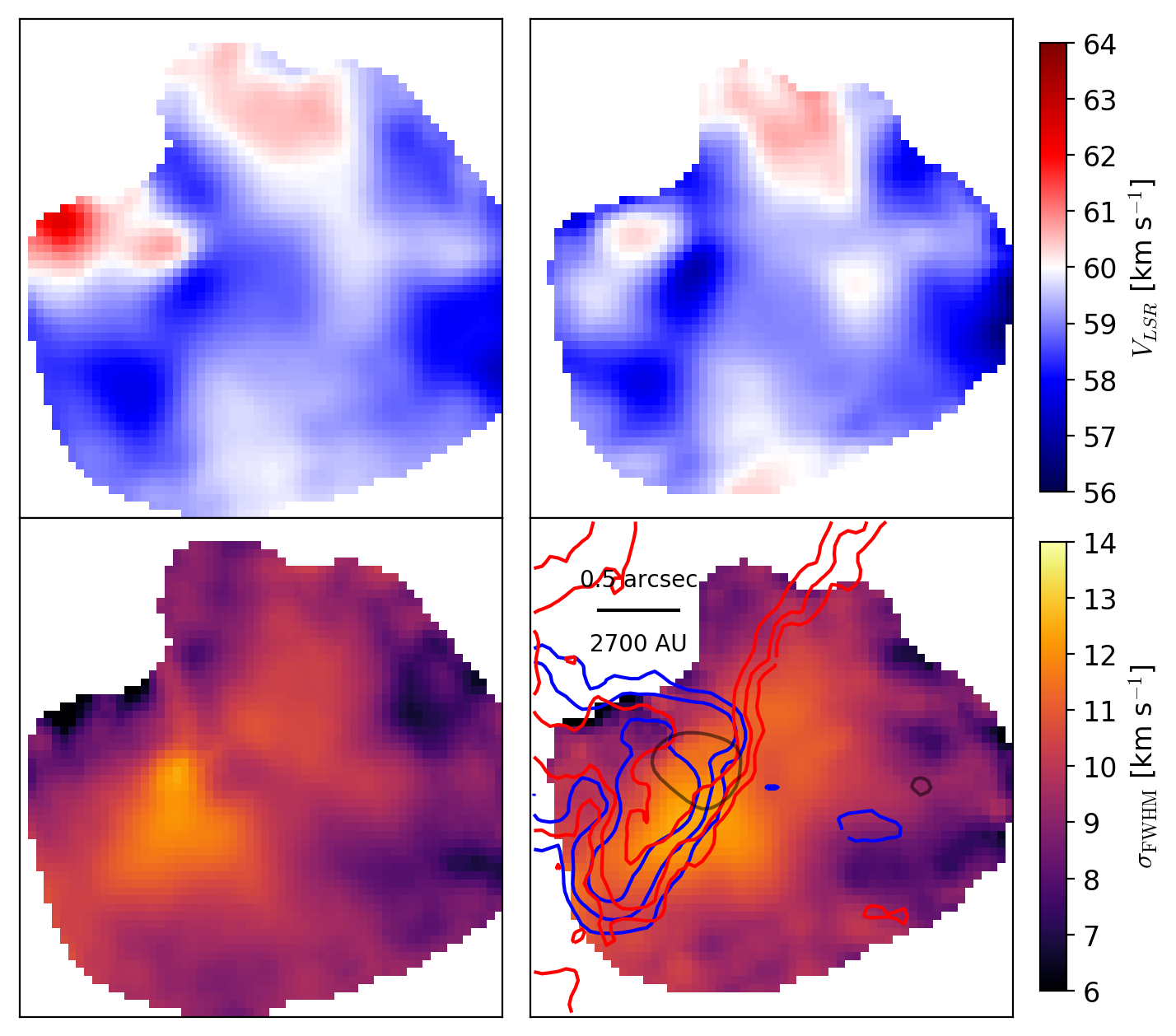}
{Moment 1 and 2 maps of the W51 North core over the velocity
range 48-70 \kms.  In the lower-right panel, the central
continuum source is offset from the center of the molecular core
because the northeast component of the core is interacting
with the IRS2 \hii region. See Figure \ref{fig:kinematicse2}.}
{fig:kinematicsnorth}{1}{0.9\textwidth}

Figures \ref{fig:e8pvdiagrams} and \ref{fig:northpvdiagrams} show
position-velocity diagrams of a \methanol line and a \methylformate line as
described in Section \ref{sec:kinematics}.  They illustrate that there is no
clear rotation signature indicating the presence of a large ($>1000$ AU) disk.

\FigureTwo
{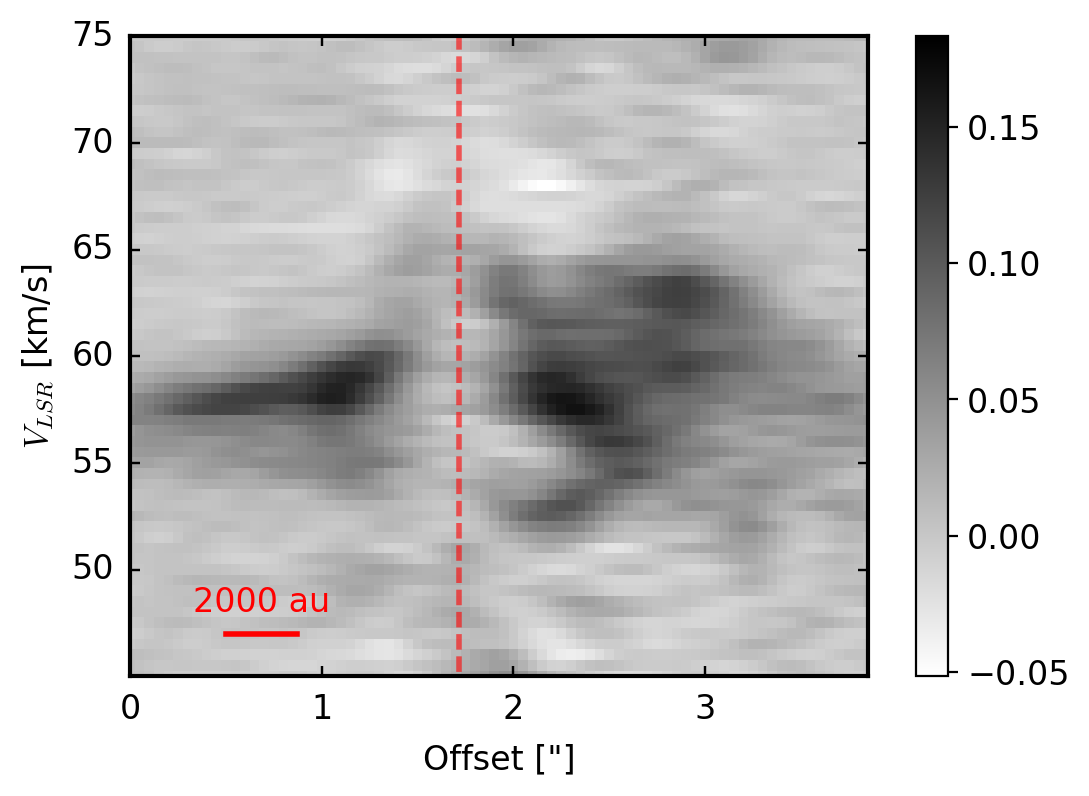}
{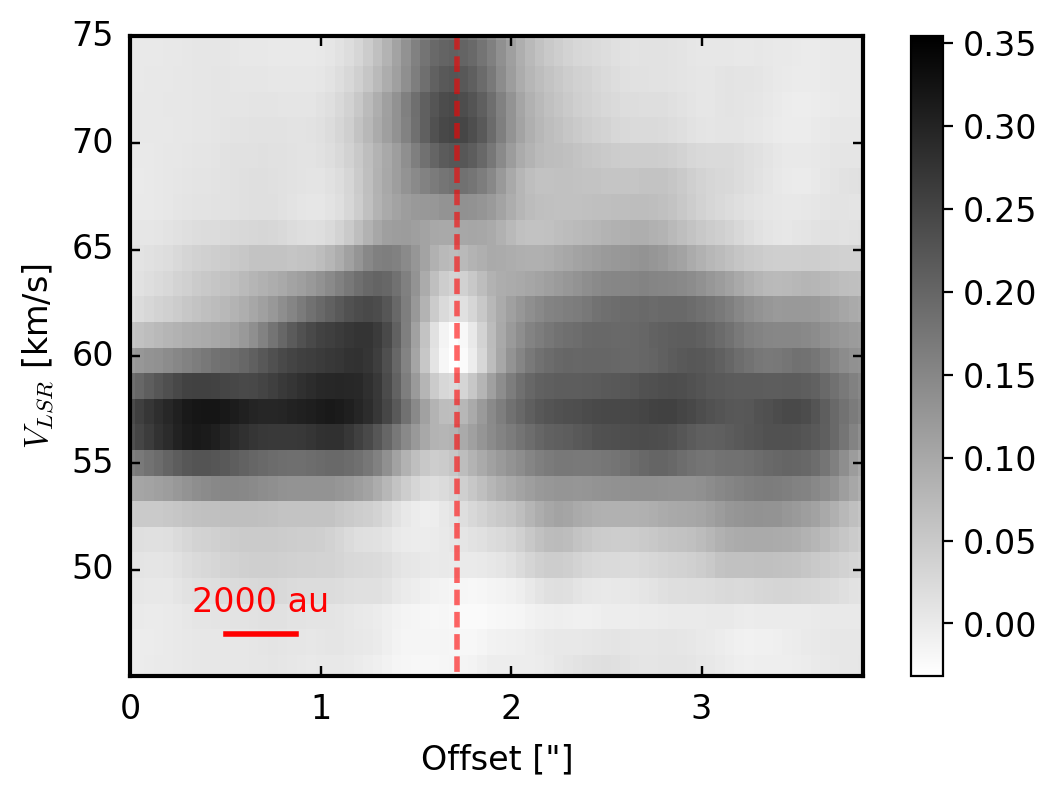}
{Position-velocity diagrams of the W51 e8 core taken at PA=$8\deg$,
perpendicular to the most clearly linear outflow axis; e8 does not drive an
unambiguous bipolar outflow on small scales.  The vertical dashed line shows the
position of peak continuum emission. The lines are (a) \methylformate
$17_{3,14}-16_{3,13}$ 218.28083 GHz and (b) \methanol $8_{0,8}-7_{1,6}$
220.07849 GHz.  The spectral resolution is 0.5 \kms in (a) and 1.2 \kms in (b).
The data have been continuum subtracted, highlighting the low line-to-continuum
contrast near the source.  The \methylformate line was selected because the
molecule approximately traces the same material as \methanol, but the pair of
\methylformate J=17 lines were in our high spectral resolution window, so the
velocity substructure can be seen.
}
{fig:e8pvdiagrams}{1}{8cm}

\FigureTwo
{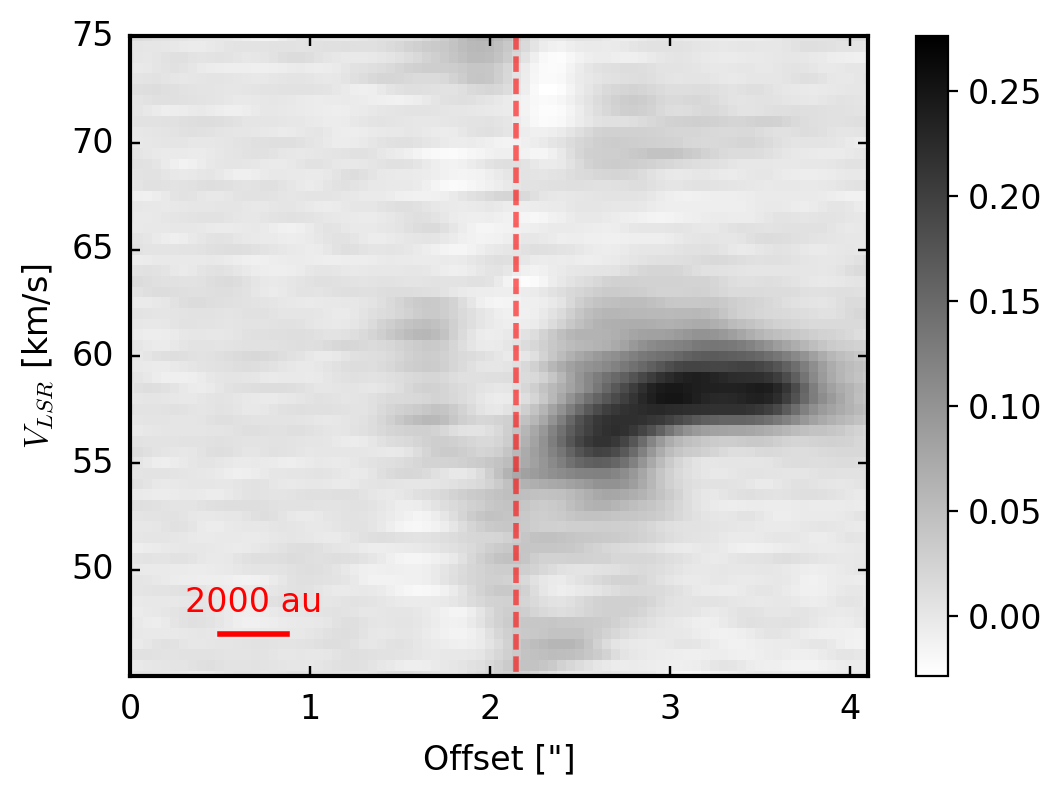}
{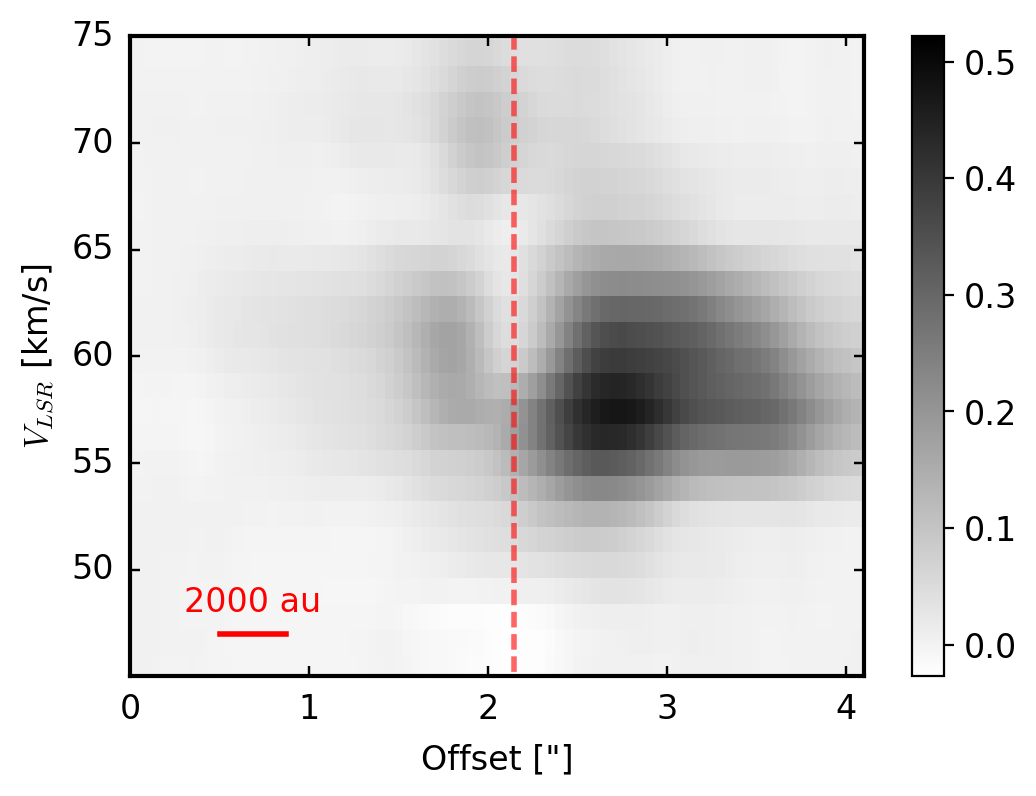}
{Position-velocity diagrams of the W51 north core taken at PA=$58\deg$,
perpendicular to the large-scale CO outflow.  The vertical dashed line shows the
position of peak continuum emission. The lines are (a) \methylformate
$17_{3,14}-16_{3,13}$ 218.28083 GHz and (b) \methanol $8_{0,8}-7_{1,6}$
220.07849 GHz.  The spectral resolution is 0.5 \kms in (a) and 1.2 \kms in (b).
The data have been continuum subtracted, highlighting the low line-to-continuum
contrast near the source.  The \methylformate line was selected because the
molecule approximately traces the same material as \methanol, but the pair of
\methylformate J=17 lines were in our high spectral resolution window, so the
velocity substructure can be seen.
The left half of the core is missing because it intersects with the \hii
region, as can be seen in Figure \ref{fig:northmethanolhnco}.
}
{fig:northpvdiagrams}{1}{8cm}

\section{A bubble around e5}
\label{sec:e5bubble}
There is evidence of a bubble in the continuum around e5 with a radius of
6.2\arcsec (0.16 pc; Figure \ref{fig:e5bubble}).  The bubble is completely
absent in the centimeter continuum, so the observed emission is from dust.  The
bubble edge can be seen from 58 \kms to 63 \kms in \ceighteeno and
\formaldehyde, though it is not contiguous in any single velocity channel.
There is a collection of compact sources (protostars or cores) along the
southeast edge of the bubble.

The presence of such a bubble in dense gas, but its absence in ionizing gas, is
surprising.  The most likely mechanism for blowing such a bubble is ionizing
radiative feedback, especially around a source that is currently a hypercompact
HII region, but since no free-free emission is evident within or on the edge of
the bubble, it is at least not presently driving the bubble.  A plausible
explanation for this discrepancy is that e5 was an exposed O-star within the
past Myr, but has since begun accreting heavily (or has traveled into a region
of high density) and therefore had its HII
region shrunk.  This model is marginally supported by the presence of a `pillar'
of dense material pointing from e5 toward the south.

The total flux in the north half of the `bubble', which shows no signs of
free-free contamination, is about 1.5 Jy.  The implied mass in just this
fragment of the bubble is about $M\sim350$ \msun for a relatively high assumed
temperature $T=50$ K.  The total mass of the bubble is closer to $M\sim1000$
\msun, though it may be lower ($\sim500$ \msun) if the southern half is
dominated by free-free emission.

With such a large mass, the implied density of the original cloud, assuming it
was uniformly distributed over a 0.2 pc sphere, is $n(\hh) \approx 2-5\ee{5}$
\percc.

To evaluate the plausibility of the \hii-region origin of the bubble, we compare
to classical equations for \hii regions.
The Str\"omgren radius is \\
\begin{eqnarray}
R_{\rm s}=\left(\frac{3Q_{\rm H}}{4\pi\alpha_{\rm B} n^{2}}\right)^{\frac{1}{3}}.
\end{eqnarray} 
For $Q_{\rm H}\sim10^{49}$ \pers (corresponding to an $M\approx40$\msun
main-sequence star), $\alpha_{\rm B}=3\times10^{-13}$\,cm$^{3}$\,s$^{-1}$,
$R_{\rm s}\approx0.01$\,pc.\\
\\
The Spitzer solution for HII region expansion gives\\
\begin{eqnarray}
R_{\rm HII}(t)=R_{\rm s}\left(1+\frac{7}{4}\frac{c_{\rm II}t}{R_{\rm s}}\right)^{\frac{4}{7}}.
\end{eqnarray} 
With $c_{\rm II}=7.5$\,km\,s$^{-1}$ and $t=10^{4}$\,yr,
$R_{\rm HII}(t)\approx0.04$\,pc, while at $t=10^5$\,yr, it is $R_{\rm
HII}\approx0.16$\,pc, which is comparable to the observed radius
($r_{obs} \sim 0.13-0.19$ pc)\\
\\
\citet{Whitworth1994a} give the fragmentation timescale as
\begin{eqnarray}
t_{\rm frag}\sim1.56\left(\frac{c_{\rm s}}{0.2{\rm km\,s}^{-1}}\right)^{\frac{7}{11}}\left(\frac{Q_{\rm H}}{10^{49}{\rm s}^{-1}}\right)^{-\frac{1}{11}}\left(\frac{n}{10^{3}{\rm cm}^{-3}}\right)^{-\frac{5}{11}}{\rm Myr}.
\end{eqnarray} 
Plugging in our numbers gives $t_{\rm frag}\approx1.0\times10^{5}$\,yr, or
$10\times$ longer than the expansion time.\\
\\

These values are consistent with a late O-type star having been exposed,
driving an \hii region, for $\sim10^4-10^5$ years, after which a substantial
increase in the local density quenched the ionizing radiation from the star,
trapping it into a hypercompact ($r<0.005$ pc) configuration.  The
recombination timescale is short enough that the ionized gas would disappear
almost immediately after the continuous ionizing radiation source was hidden.
This is essentially the scenario laid out in \citet{de-Pree2014a} as an
explanation for the compact \hii region lifetime problem.  In this case,
however, it also seems that the \hii region has effectively driven the
``collect'' phase of what will presumably end in a collect-and-collapse style
triggering event.

Technically, it is possible that e5 actually represents an optically thick
high-mass-loss-rate wind rather than an ultracompact HII region. 
For example, $\eta$ Car would have a flux of $\sim0.5$ Jy at 2 cm
and $\sim5$ Jy at 1 mm at the distance of W51.  While we cannot rule out
this possibility, it would render the association of e5 with the `bubble'
purely coincidental.

\FigureTwo{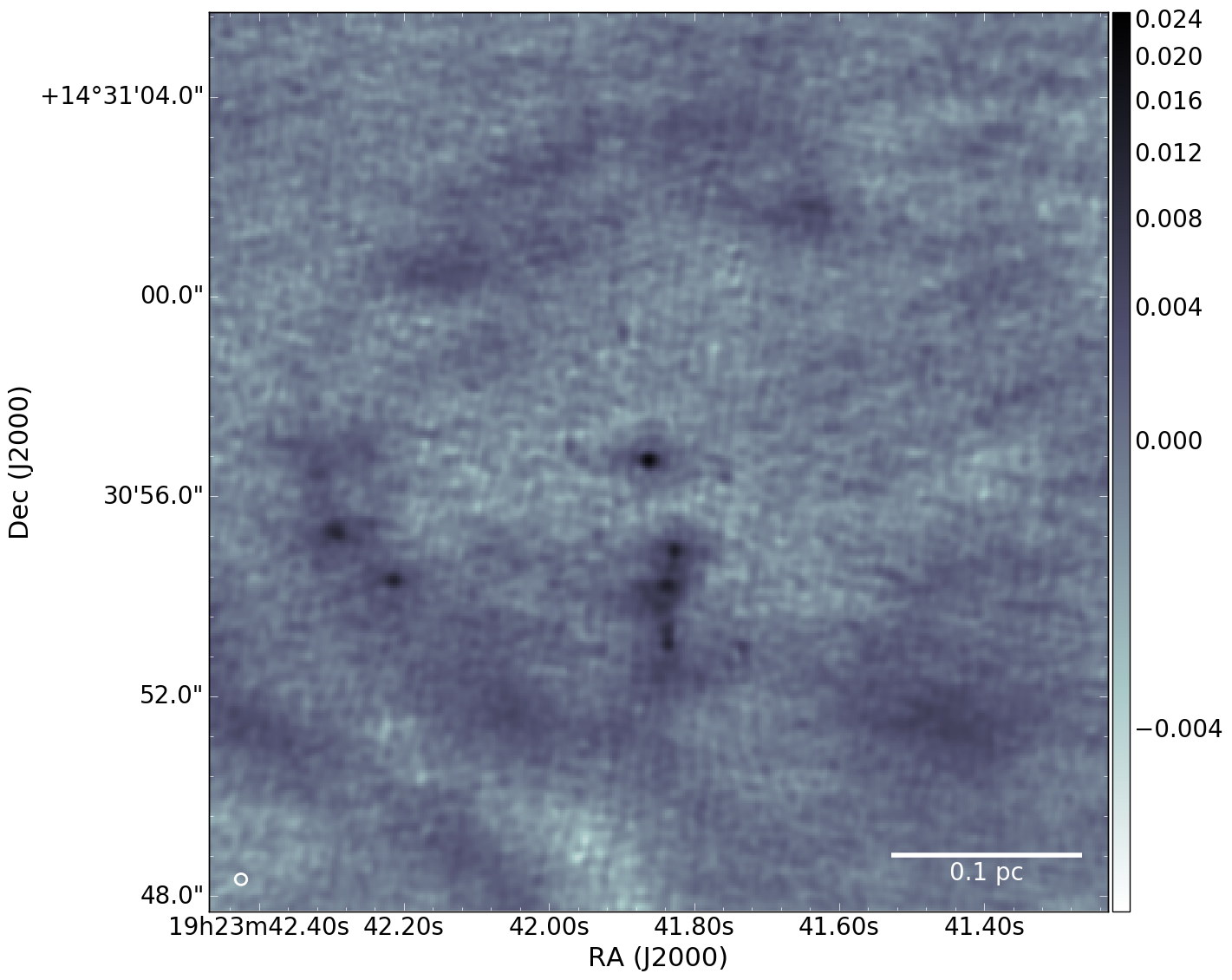}{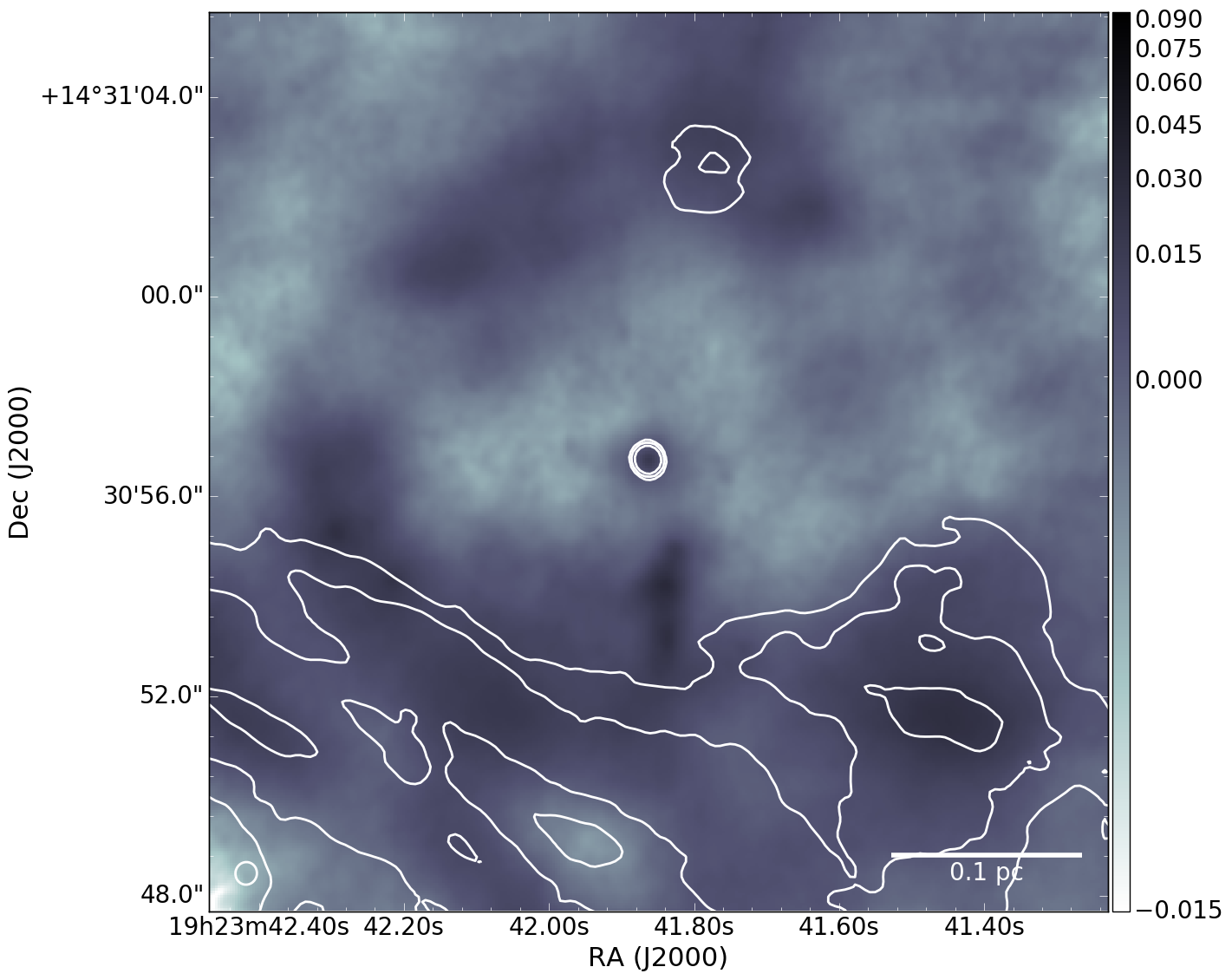}
{The bubble around source e5.  The bubble interior shows no sign of centimeter
emission, though the lower-left region of the shell - just south of the
``cores'' - coincides with part of the W51 Main ionized shell.  The source of
the ionization is not obvious.
({\it Left}): A robust -2.0 image with a small (0.2\arcsec) beam and poor
recovery of large angular scale emission.  This image highlights the presence
of protostellar cores on the left edge of the bubble and along a filament just
south of the central source.
({\it Right}): A robust +2.0 image with a larger (0.4\arcsec) beam and better
recovery of large angular scales.  The contours show radio continuum (14.5 GHz)
emission at 1.5, 3, and 6 mJy/beam.  While some of the detected 1.4 mm emission
in the south could be free-free emission, the eastern and northern parts of the
shell show no emission down to the 50 $\mu$Jy noise level of the Ku-band map,
confirming that they consist only of dust emission.
}{fig:e5bubble}{1}{8cm}

\clearpage
\section{Data Release}
The reduced images and data cubes used in this paper are publicly available at
\dataset[doi:10.7910/DVN/8QJT3K]{https://dataverse.harvard.edu/dataset.xhtml?persistentId=doi:10.7910/DVN/8QJT3K}.
The reduction and analysis scripts are at
\url{https://github.com/adamginsburg/W51_ALMA_2013.1.00308.S}.

The continuum image used for source identification and analysis is
\texttt{W51\_te\_continuum\_best.fits}.  Spectral cubes covering individual
important lines are specified with filenames
\texttt{w51\_\{linename\}\_contsub.image.pbcor.fits}, where \texttt{linename}
is one of the lines imaged in this paper.  

The data also include cutout images covering the full spectral range but a
limited spatial area.  These cutouts are useful for spectral line searches and
morphological comparison; most of the lines detected in our observations are
only visible in the cutout regions.  These data are in files fitting the
template\\
\texttt{\{sourcename\}cutout\_full\_W51\_7m12m\_spw\{number\}\{optional
suffix\}\_lines.fits}, where \texttt{sourcename} is one of the sources
discussed in this paper (\texttt{e2}, \texttt{e8}, or \texttt{north}),
\texttt{number} is one of the spectral windows \texttt{0-3} listed in Tables
\ref{tab:linesspw0} - \ref{tab:linesspw3}, and \texttt{optional\_suffix} is
either empty or \texttt{hires}; the hires cubes used robust=-0.5 and the others
used robust=1.0 in \texttt{tclean}.  The higher robust parameter results
in lower spatial resolution but better sensitivity.

\end{document}